\documentclass[11pt,a4paper]{article}
\pdfoutput=1
\usepackage{jheppub}
\usepackage[T1]{fontenc}
\usepackage[english]{babel}
\usepackage{lmodern}
\usepackage[colorlinks = true]{hyperref}
\usepackage{amsfonts}
\usepackage{amsmath} 
\usepackage{bbm}
\usepackage{bm}
\usepackage{color}
\usepackage{slashed}
\usepackage{graphicx}
\usepackage{dsfont}
\usepackage{bbold}
\usepackage{braket}
\usepackage{caption}
\usepackage{lineno}
\usepackage{soul}

\makeatletter \@addtoreset{equation}{section} \makeatother

\allowdisplaybreaks

\title{Effective field theories for dark matter pairs in the early universe: center-of-mass recoil effects}

\author[a]{S.~Biondini,}
\author[b,c,d]{N.~Brambilla,}
\author[b]{G.~Qerimi}
\author[b]{and A. Vairo}

\affiliation[a]{Department of Physics, University of Basel,\\
Klingelbergstr. 82, CH-4056 Basel, Switzerland}
\affiliation[b]{Technical University of Munich, TUM School of Natural Sciences, Physics Department,\\
James-Franck-Str.  1, 85748 Garching, Germany}
\affiliation[c]{Institute for Advanced Study, Technical University Munich, \\
Lichtenbergstrasse 2 a, 85748 Garching, Germany}
\affiliation[d]{Munich Data Science Institute, Technical University Munich, \\
Walther-von-Dyck-Strasse 10, 85748 Garching, Germany}

\emailAdd{simone.biondini@unibas.ch}
\emailAdd{nora.brambilla@tum.de}
\emailAdd{gramos.qerimi@tum.de}
\emailAdd{antonio.vairo@tum.de}

\abstract{
  For non-relativistic thermal dark matter, close-to-threshold effects largely dominate the evolution of the number density for most of the times after thermal freeze-out, and hence affect the cosmological relic density.
  A precise evaluation of the relevant interaction rates in a thermal medium representing the early universe includes accounting for the relative motion of the dark matter particles and the thermal medium.
  We consider a model of dark fermions interacting with a plasma of dark gauge bosons, which is equivalent to thermal QED. 
  The temperature is taken to be smaller than the dark fermion mass and the inverse of the typical size of the dark fermion-antifermion bound states,
  which allows for the use of non-relativistic effective field theories.
  For the annihilation cross section, bound-state formation cross section, bound-state dissociation width and bound-state transition width of dark matter fermion-antifermion pairs,
  we compute the leading recoil effects in the reference frame of both the plasma and the center-of-mass of the fermion-antifermion pair. We explicitly verify the Lorentz transformations among these quantities.
  We evaluate the impact of the recoil corrections on the dark matter energy density.
  Our results can be directly applied to account for the relative motion of quarkonia in the quark-gluon plasma formed in heavy-ion collisions.
  They may be also used to precisely assess thermal effects in atomic clocks based on atomic transitions;
  the present work provides a first field theory derivation of time dilation for these processes in vacuum and in a medium.
}

\begin{document}
\maketitle

\section{Introduction}
The dark matter (DM) problem together with the origin of the matter-antimatter asymmetry in the universe poses a challenging quest across particle physics, astrophysics and cosmology.
Historically, there has been a tight connection between dark matter and unresolved issues in the realm of the Standard Model of particle physics.
A popular option consists in postulating the existence of novel particles at the weak scale as a solution of the hierarchy problem
and, at the same time, interpret some of them as viable dark matter candidates~\cite{Ellis:1983ew,Jungman:1995df,Catena:2013pka}.
More recently, and at variance with ultraviolet complete theories, complementary simplified model approaches have been used to explore additional dark matter candidates and their phenomenological signatures~\cite{deSimone:2014pda,Garny:2015wea,Albert:2016osu}.
In both cases, dark matter comes in the form of a weakly interacting massive particle with sizeable couplings with the Standard Model degrees of freedom to accommodate the observed dark matter energy density.
The dark matter energy density is typically computed via the \emph{thermal freeze-out} mechanism.
There, one assumes that, at a temperature of the universe, $T$, much larger than the dark matter mass $M$, the weak interactions between dark matter and Standard Model particles are sufficient to keep the dark matter in thermal equilibrium.
As the universe expands and cools down, dark matter pair annihilation becomes ineffective at a typical temperature of about $T \approx M/25$, which is the freeze-out temperature that determines eventually the observed relic density.
Since the freeze-out temperature is much smaller than the dark matter mass, dark matter pairs are non-relativistic at the time of the freeze-out.    

The increasingly stringent experimental constraints on novel weakly interacting particles have triggered a renewed interest for different classes of dark matter models \cite{Arcadi:2017kky}.
Most notably, it has been shown that thermal freeze-out may occur entirely within an extended dark sector,
and that the observed dark matter energy density $\Omega_{\hbox{\tiny DM}} h^2 = 0.1200 \pm 0.0012$~\cite{Planck:2018nkj} can be reproduced without the need for sizeable couplings between the dark and the visible sector~\cite{Pospelov:2007mp,Duerr:2016tmh,Evans:2017kti}.
Dark matter particles annihilate into lighter particles of the dark sector, such as dark gauge bosons or dark scalars, so that the relevant cross sections are controlled by couplings of the hidden sector.
These lighter dark particles may decay at a later stage into Standard Model particles via interactions encoded in the portal of the dark sector in the Standard Model or, if very light and stable, be part of a remnant dark radiation.

Irrespective of the specific nature of the dark matter candidate, thermal freeze-out stands out as a unifying production mechanism in the early universe across a wide variety of dark matter models.
A central quantity entering the evolution equations of the dark matter particle densities is the thermally averaged annihilation cross section $\langle \sigma_{\hbox{\scriptsize ann}} v_{\hbox{\scriptsize rel} } \rangle$ in the non-relativistic regime.
Whenever heavy dark matter pairs interact with lighter states of the dark (or visible) sector near threshold, i.e. at small relative velocity, 
one needs to account for the repeated exchange of soft degrees of freedom, i.e. degrees of freedom with energy and momentum smaller than $M$.\footnote{
The existence of lighter particles is typically a necessary condition for the freeze-out, so that the two-body annihilation process can occur in the first place.
The annihilation products can also be the same particles responsible for long-range interactions, such as dark photons or dark scalars.}
Near-threshold effects modify the annihilation cross section in many ways, which go from the Sommerfeld enhancement for above-threshold scattering states \cite{Sommerfeld,Hisano:2004ds} to the formation and decay of meta-stable bound states~\cite{Feng:2009mn,vonHarling:2014kha}.
Therefore, the actual input in the Boltzmann equation governing the evolution of the dark matter particle densities is rather an effective cross section $ \langle  \sigma_{\textrm{eff}} \, v_{\textrm{rel}} \rangle$ \cite{Ellis:2015vaa} 
that entails as much as possible of the near-threshold dynamics of the heavy dark matter pairs in a thermal environment (cf.~eq.~\eqref{Cross_section_eff}).
In the last few years, there has been a significant effort in the inclusion of near-threshold effects when extracting the relic density of various dark matter
models~\cite{vonHarling:2014kha,Beneke:2014hja,Petraki:2016cnz,Kim:2016kxt,Beneke:2016ync,Biondini:2017ufr,Harz:2018csl,Biondini:2018pwp,Binder:2018znk,Oncala:2018bvl,Oncala:2019yvj,Harz:2019rro,Binder:2019erp,Biondini:2019zdo,Binder:2020efn,Biondini:2021ycj,Biondini:2021ccr,Binder:2023ckj},
together with an assessment of their phenomenological impact~\cite{March-Russell:2008klu,Laha:2015yoa,Pearce:2015zca,Cirelli:2016rnw,An:2016kie,Mitridate:2017izz,Biondini:2018ovz,Baldes:2020hwx,Garny:2021qsr,Becker:2022iso,Baumgart:2023pwn,Biondini:2023ksj,Biondini:2023yxt}. 

Dealing with heavy particle pairs near threshold in a thermal environment is complicated by the presence of several energy scales.
There are the scales that are generated by the non-relativistic relative motion, namely the momentum transfer and the kinetic/binding energy of the pair.
These scales are hierarchically ordered as $M \gg M v_{\hbox{\scriptsize rel} } \gg Mv_{\hbox{\scriptsize rel} }^2$, $v_{\hbox{\scriptsize rel} }$ being the relative velocity of the particles in the pair.
The corresponding hierarchy of energy scales for bound states can be easily inferred for Coulombic states, where the relative velocity of the pair is $v_{\hbox{\scriptsize rel} } \sim \alpha$, and hence one finds $M \gg M \alpha \gg M \alpha^2$,
where $\alpha=g^2/(4\pi)$ and $g$ is the coupling between the heavy dark matter particle and the mediator responsible for the binding.
Then, there are the thermodynamical scales that include the plasma temperature $T$ and possibly, depending on the model details, a Debye mass for the force mediator.
The latter corresponds to the inverse of the chromoelectric screening length and, at weak coupling, $m_{\hbox{\scriptsize D} } \sim gT$.
Finally, if the heavy dark matter is in kinetic equilibrium with the plasma, then its typical momentum is of order $\sqrt{MT}$.
Depending on the plasma temperature, $\sqrt{MT}$ may be larger or smaller than the scales $M\alpha$ or $M\alpha^2$. 
Contributions coming from the different energy scales may be disentangled and computed in a systematic framework by means of non-relativistic effective field theories~\cite{Brambilla:2004jw}, which  are the method that we adopt in this work.

Non-relativistic effective field theories have been used to compute annihilation and bound-state formation cross sections, and annihilation, bound-state dissociation and transition widths in~\cite{Biondini:2023zcz}.
Here we follow up and compute in the same setup the leading-order corrections to these quantities, and ultimately to the effective cross section, due to the relative motion between the plasma and the center-of-mass of the heavy dark matter pair.
In the laboratory frame, where the plasma is at rest, we may interpret these effects as due to the center-of-mass of the heavy dark matter pair recoiling with a momentum $\bm{P}$ of order $\sqrt{MT}$ with respect to the plasma.
The energy and momentum scale $\sqrt{MT}$ is generated by the Brownian motion of the heavy pair in the plasma of temperature $T$, which eventually leads the heavy particles to be in kinetic equilibrium.
The complete leading-order recoil corrections for bound-state formation and dissociation in a thermal bath are computed here for the first time.

In most of the paper, we consider a QED-like dark sector made of Dirac fermions and photons decoupled from the Standard Model.
The corresponding non-relativistic effective field theories are non-relativistic QED~\cite{Caswell:1985ui} for dark matter, $\textrm{NRQED}_{\textrm{DM}}$, and potential NRQED~\cite{Pineda:1998kn} for dark matter, $\textrm{pNRQED}_{\textrm{DM}}$.
Center-of-mass momentum-dependent operators in the Lagrangians of these effective field theories have been considered in~\cite{Brambilla:2003nt,Brambilla:2008zg,Berwein:2018fos}.
The non-relativistic effective field theories in a thermal bath have been developed in~\cite{Brambilla:2008cx,Escobedo:2008sy}.
Furthermore, in the paper, we also discuss the case of a dark sector made of Dirac fermions coupled to SU($N$) gauge bosons.
For $N=3$ this is the QCD case that describes heavy quarkonium and heavy quark-antiquark pairs near threshold in a quark gluon plasma.
Quarkonium dissociation in a medium by gluon absorption has been studied in the framework of non-relativistic effective field theories in~\cite{Brambilla:2011sg}. 
The effect of a moving thermal bath in the center-of-mass frame has been considered in~\cite{Escobedo:2011ie,Escobedo:2013tca} (see~\cite{Song:2007gm} for an earlier reference not based on effective field theories). 
Our results complement those calculations from the point of view of the laboratory frame and complete them by adding some missing contributions.

Recoil corrections are relevant both for precision measurements and, if large, to describe macroscopic effects.
They affect atomic clocks based on atomic transitions.
The results presented here are novel for atomic clocks moving in a thermal medium.
They also affect quarkonium production and suppression in heavy-ion collisions, as the quarkonium is moving with respect to the quark-gluon plasma.
There, the effect may be large if the relative velocity with respect to the plasma is large.
For the dark matter relic energy density the recoil effect depends on the temperature, and may become a sizeable fraction at high temperature close to the freeze-out.
Anyway, as we shall see, it is larger than the uncertainty on the measured dark matter energy density for a large time of the universe evolution.

The outline of the paper is the following.
In section~\ref{sec:NREFTs}, we set up the abelian dark matter model and the hierarchy of energy scales that we consider throughout most of this work.
Then, in section~\ref{sec:ann}, we inspect the annihilation of dark matter pairs in $\textrm{NRQED}_{\textrm{DM}}$ and $\textrm{pNRQED}_{\textrm{DM}}$.
We derive the order $\bm{P}^2/M^2$ recoil corrections.
Section~\ref{sec:bsf} and~\ref{sec:bsd} are respectively devoted to bound-state formation and dissociation processes.
We compute the leading recoil effects in the reference frame of both the plasma (laboratory frame) and the center-of-mass of the fermion-antifermion pair.
Bound-state to bound-state transitions are studied in section~\ref{sec:bound_bound}, where we compute the leading recoil corrections in the laboratory frame.
Atomic transitions are at the base of atomic clocks.
Relativistic effects have been studied in this context and we connect to some of these studies in the section.
 In section \ref{sec:non_abelian_model}, we present the leading center-of-mass recoil corrections for fermionic dark matter charged under a non-abelian SU($N$) model.
In section~\ref{sec:numerics}, we assess the size of the different recoil corrections and solve numerically the Boltzmann equations for the dark matter densities,
using as input the cross sections and widths with recoil corrections computed in the previous sections.
We eventually also assess the relative impact of the recoil correction on the dark matter relic density.
Conclusions can be found in section~\ref{sec:concl}.
Complementing material about Lorentz transformations, thermal averages and dipole matrix elements are collected in the appendices.

\section{The model: dark QED}
\label{sec:NREFTs}
We consider a simple model where the dark sector consists of a dark Dirac fermion $X$ that is charged under an abelian gauge group (QED$_{\textrm{DM}}$)~\cite{Feldman:2006wd,Fayet:2007ua,Goodsell:2009xc,Morrissey:2009ur,Andreas:2011in}.
We call $\gamma$ the dark photon associated to the gauge group.
The general form of the Lagrangian density is 
\begin{equation}
\mathcal{L}=\bar{X} (i \slashed {D} -M) X -\frac{1}{4} F_{\mu \nu} F^{\mu \nu} + \mathcal{L}_{\textrm{portal}} \, ,
\label{lag_mod_0}
\end{equation}
where $D_\mu=\partial_\mu + i g A_\mu$ is the gauge covariant derivative, $A_\mu$ the dark photon field and $F_{\mu \nu} = \partial_\mu A_\nu - \partial_\nu A_\mu$ the field strength tensor;
$g$ is the gauge coupling and $\alpha \equiv g^2/(4 \pi)$ the fine structure constant.
The term $\mathcal{L}_{\textrm{portal}}$ comprises all possible couplings of the dark photon with the Standard Model degrees of freedom.
For the purpose of this work, we do not consider the effect of portal interactions when computing the cross sections and decay widths of dark matter particles, however we still assume they are responsible for keeping the Standard Model and dark sector at the same temperature.\footnote{
A popular interaction at a renormalizable level is a mixing with the neutral components of the Standard Model gauge fields \cite{Holdom:1985ag,Foot:1991kb}, also called kinetic mixing.
The dark photon acquires a non-vanishing coupling with the Standard Model fermions. Such interactions are responsible for the eventual decay of the
dark photons and maintain the dark and Standard Model sectors in thermal equilibrium even for very small values of the mixing parameter, see e.g.~\cite{Evans:2017kti}.
In our work, we assume the dark gauge coupling to be much larger than the mixing-induced coupling, hence we practically neglect portal interactions when computing the cross sections and widths of dark matter particles.} 
 
We assume the following hierarchy of energy scales~\cite{Biondini:2023zcz}:
\begin{equation}
M \gg M\alpha \gtrsim \sqrt{MT} \gg M\alpha^2 \gtrsim T \,.
\label{scale_arrang}
\end{equation}
This hierarchy is realized for most of the time after chemical decoupling.

In the model that we consider, only three formation/dissociation processes turn out to be relevant: dark fermion pair \textit{annihilation} (ann), discussed in section~\ref{sec:ann},
\textit{bound-state formation} (bsf), i.e. the formation of a dark fermion-antifermion bound state through the emission of a dark photon, discussed in section~\ref{sec:bsf},
and \textit{bound-state dissociation} (bsd), i.e. the dissociation of a dark fermion-antifermion bound state through the absorption of a dark photon, discussed in section~\ref{sec:bsd}.
In~\cite{Biondini:2023zcz}, we have computed the cross sections and widths for these processes in the rest frame of the thermal bath assuming that the center of mass of the dark fermion pair is comoving with it.
In this paper, we account for \textit{recoil corrections}, i.e. for the fact that the center of mass of the dark fermion pair is moving relative to the thermal bath with a momentum $\bm{P}$ of magnitude of the order of $\sqrt{MT}$.
We compute recoil corrections at leading order in the non-relativistic expansion, i.e. at order $\bm{P}^2/M^2$.

\section{Annihilation}
\label{sec:ann}
Annihilation of a dark fermion pair releases dark photons of an energy of order $2M$.
Threshold effects are due to the exchange of dark photons of momentum of order $M\alpha$.
The effective field theories that best describe fermion-antifermion annihilation near threshold are 
\textit{non-relativistic QED}~\cite{Caswell:1985ui} for dark fermions and photons (NRQED$_{\textrm{DM}}$),
which follows from QED$_{\textrm{DM}}$ by integrating out \textit{hard} modes associated with the scale $M$, 
and \textit{potential NRQED}~\cite{Pineda:1998kn} for dark fermions and photons (pNRQED$_{\textrm{DM}}$),
which follows from NRQED$_{\textrm{DM}}$ by integrating out \textit{soft} photons of momentum or energy larger than $M\alpha^2$.
At leading order, annihilation happens through the emission of two dark photons: $X\bar{X}\to \gamma \gamma$.
In this case, the dark fermion pair, $X\bar{X}$, is in a spin-singlet configuration.

\subsection{Annihilation in NRQED$_{\textrm{DM}}$}
\label{sec:NRQED}
The sector of the NRQED$_{\textrm{DM}}$ Lagrangian density responsible for dark fermion-antifermion annihilation is the four-fermion sector.
We consider the following four-fermion operators~\cite{Brambilla:2008zg,Berwein:2018fos}
\begin{equation}
\begin{aligned}
  & \mathcal{L}_{\textrm{NRQED}_{\textrm{DM}}} ~\supset~
       \frac{d_s}{M^2} \psi^\dagger \chi \, \chi^\dagger \psi
       +\frac{d_v}{M^2} \psi^\dagger \, \bm{\sigma} \, \chi \cdot \chi^\dagger \, \bm{\sigma} \, \psi \\
       &  
       + \frac{g_{\textrm{a\,cm}}}{M^4} \nabla^i(\psi^\dagger \sigma^j \chi) \nabla^i(\chi^\dagger \sigma^j \psi)
         + \frac{g_{\textrm{b\,cm}}}{M^4} \bm{\nabla}\cdot(\psi^\dagger \bm{\sigma} \chi) \bm{\nabla}\cdot(\chi^\dagger \bm{\sigma} \psi)
                  + \frac{g_{\textrm{c\,cm}}}{M^4} \bm{\nabla}(\psi^\dagger \chi) \cdot \bm{\nabla}(\chi^\dagger \psi) \, ,
\label{local_op_total_mom}
\end{aligned}
\end{equation}
where $\psi$ is the two-component Pauli spinor that annihilates a dark fermion, $\chi^\dagger$ is the Pauli spinor that annihilates a dark antifermion
and $\sigma^i$ are the Pauli matrices.
The first line of eq.~\eqref{local_op_total_mom} contains all dimension 6 four-fermion operators.
They project only on fermion-antifermion pairs in an S-wave configuration.
In the second line, we show the dimension 8 operators projecting on S waves that depend on the center-of-mass momentum;
they provide the leading center-of-mass momentum corrections to the S-wave annihilation widths~\cite{Brambilla:2008zg,Berwein:2018fos}.
The annihilation cross sections and widths depend on the imaginary parts of the matching coefficients $d_s$, $d_v$, $g_{\textrm{a\,cm}}$, $g_{\textrm{b\,cm}}$ and $g_{\textrm{c\,cm}}$.
Two photon annihilations induce the following imaginary parts at order $\alpha^2$~\cite{Barbieri:1979be,Hagiwara:1980nv}: 
\begin{eqnarray}
  &&  {\rm{Im}} \, d_s = \pi \alpha^2 ,
    \label{Im_ds_LO}
      \\
               &&  {\rm{Im}} \, d_v = 0 \,,
    \label{Im_dv_LO}
\end{eqnarray}
and, for the imaginary parts of the matching coefficients associated with the center-of-mass dependent operators,~\cite{Brambilla:2008zg,Berwein:2018fos}:
\begin{eqnarray}
&&  {\rm{Im}}\, g_{\textrm{a\,cm}} = {\rm{Im}}\, g_{\textrm{b\,cm}} = 0 \,, \label{gabcm}\\
&&  {\rm{Im}}\, g_{\textrm{c\,cm}} = -\frac{1}{4}{\rm{Im}} \, d_s = -\frac{\pi \alpha^2}{4} \,. \label{gccm}
\end{eqnarray}
The relation $\displaystyle {\rm{Im}}\, g_{\textrm{c\,cm}} = -\frac{1}{4}{\rm{Im}} \, d_s$ is exact, i.e. valid at all orders, and follows from the Poincar\'e invariance of QED.
The vanishing of ${\rm{Im}} \, d_v$, ${\rm{Im}}\, g_{\textrm{a\,cm}}$  and ${\rm{Im}}\, g_{\textrm{b\,cm}}$ at order $\alpha^2$ reflects the fact that only spin-singlet fermion-antifermion pairs can decay into two photons.

From the optical theorem, it follows that the spin-averaged annihilation cross section, $\sigma_{\hbox{\scriptsize ann}}$, can be written in full generality as~\cite{Gondolo:1990dk,ParticleDataGroup:2022pth}
\begin{equation}
  \sigma_{\hbox{\scriptsize ann}} v_{\hbox{\tiny M\o l}}  = \frac{{\rm{Im}}[\mathcal{M}_{\hbox{\tiny NR}}(\psi\chi \to \psi\chi)]}{2}\,,
\label{optical_cross_section}
\end{equation}
where $\mathcal{M}_{\hbox{\tiny NR}}(\psi\chi \to \psi\chi)$ is the $2 \to 2$ scattering amplitude with initial and final states normalized non-relativistically,\footnote{
  The relation between a relativistically normalized state, $|\textrm{R}\rangle$, and a non-relativistically normalized one, $|\textrm{NR}\rangle$, is
  $|\textrm{R}\rangle = \sqrt{2E}\,|\textrm{NR}\rangle$, $E$ being the energy of the state.
  }  
and $v_{\hbox{\tiny M\o l}}$ is the so-called M\o ller velocity, which is the flux of incoming particles divided by the energies of the two colliding particles carrying four-momenta $p_i = (E_i,\bm{p}_i)$,
\begin{equation}
v_{\hbox{\tiny M\o l}} = \frac{\sqrt{(p_1\cdot p_2)^2-M^4}}{E_1E_2}\,.
\label{def:Moller}
\end{equation}
The M\o ller velocity has a simple expression in terms of the particle velocities $\bm{v}_i \equiv \bm{p}_i/E_i$:
\begin{equation}
v_{\hbox{\tiny M\o l}}^2 = v_{\hbox{\tiny rel}}^2 - (\bm{v}_1 \times \bm{v}_2)^2 \,,
\end{equation}
where $v_{\hbox{\tiny rel}} \equiv |\bm{v}_1 - \bm{v}_2|$ is the relative velocity of the colliding particles.
Note that in the non-relativistic limit, the relative velocity is of order $1/M$, whereas $|\bm{v}_1 \times \bm{v}_2|$ is of order~$1/M^2$.

We can write $\bm{p}_1 = \bm{p} + \bm{P}/2$ and $\bm{p}_2 = -\bm{p} + \bm{P}/2$, where $\bm{p}$ is the relative and $\bm{P}$ the center-of-mass momentum.
From this it follows that in the \textit{center-of-mass frame} (cm) of the dark fermion-antifermion pair ($\bm{P} = \bm{0}$) it holds that 
\begin{align}
  &\bm{p}_1 = - \bm{p}_2 = \bm{p}, \qquad E_1=E_2=\sqrt{\bm{p}^2+M^2}\,, \label{pEcm}\\
  &\bm{v}_1 = -\bm{v}_2 = \frac{\bm{p}}{\sqrt{\bm{p}^2+M^2}}\,, \label{vcm}\\ 
  & \left(v_{\hbox{\tiny M\o l}}\right)_{\textrm{cm}} = \left(v_{\hbox{\tiny rel}}\right)_{\textrm{cm}}
  = 2 \frac{|\bm{p}|}{M} \,\left[1 + \mathcal{O}\left(\frac{\bm{p}^2}{M^2}\right)\right] \,. \label{vMolcm}
\end{align}
We see that $v^{(0)}_{\hbox{\tiny rel}} \equiv 2 |\bm{p}|/M$ is the relative velocity at leading order in the non-relativistic expansion.
In the center-of-mass frame of the fermion-antifermion pair, the thermal bath is moving with velocity  about $-\bm{P}/(2M)$,  where $\bm{P}$ is the center-of-mass momentum in the thermal bath frame (see appendix~\ref{sec:app_A}).

\begin{figure}[t!]
    \centering
    \includegraphics[scale=0.47]{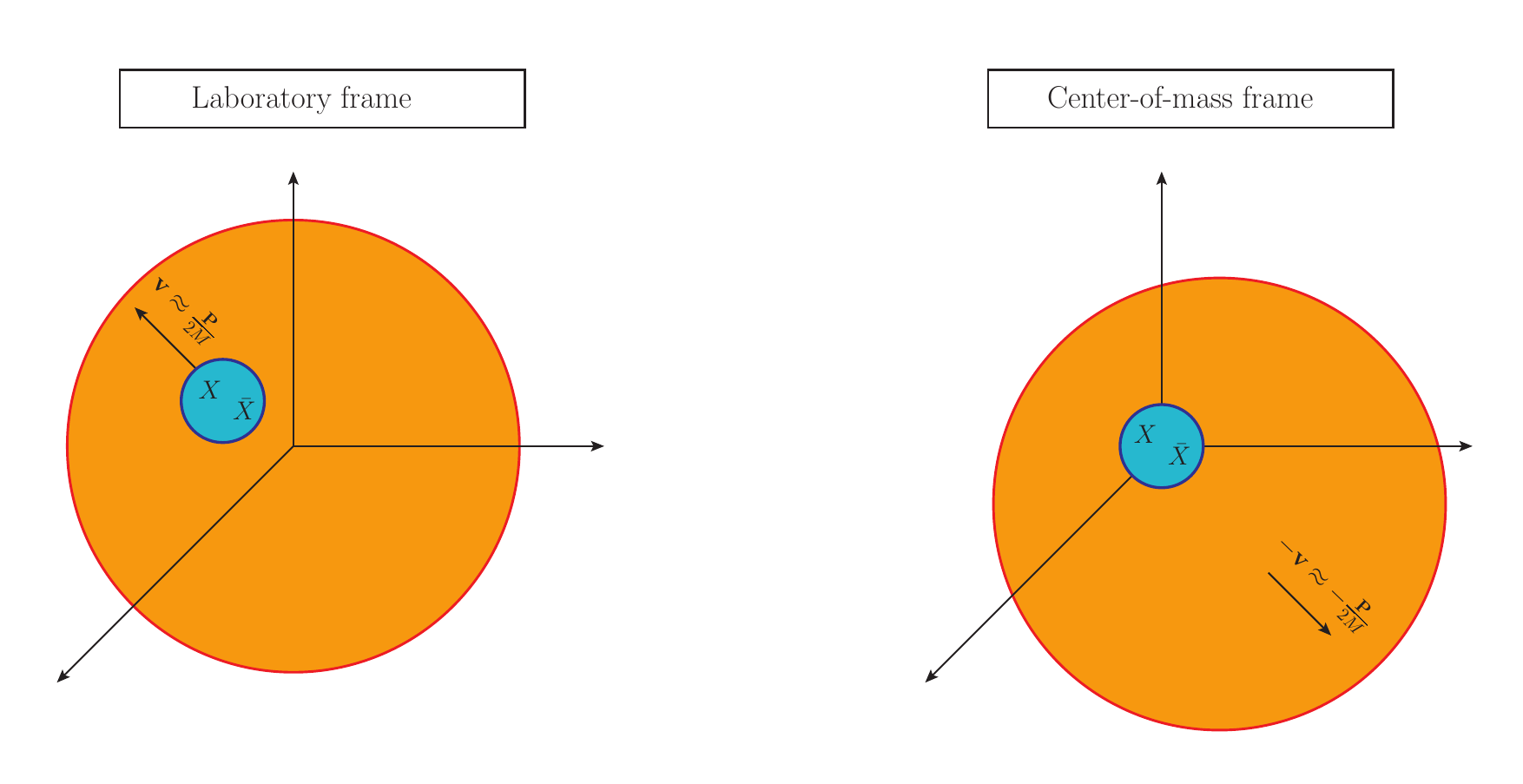}
    \caption{Schematic representation of the laboratory frame vs the center-of-mass frame.}
    \label{fig:reference_frames}
\end{figure}

We call \textit{laboratory frame} (lab) the reference frame where the thermal bath is at rest.
In this case, the center of mass of the fermion-antifermion pair is moving with velocity $\bm{v}  \approx \bm{P}/(2M)$.
The laboratory frame is sometimes also called \textit{cosmic comoving frame}.
The center-of-mass and laboratory frame are shown in figure~\ref{fig:reference_frames} for the case of a dark matter fermion-antifermion pair moving in a thermal bath.

At leading order in the non-relativistic expansion and in the coupling we get~\cite{Dirac:1930bga} 
\begin{equation}
  \sigma_{\hbox{\scriptsize ann}} v_{\hbox{\tiny M\o l}}
  = \frac{  {{\rm{Im}}\,d_s}+3 {{\rm{Im}}\,d_v} }{M^2} = \frac{\pi \alpha^2}{M^2}  \equiv \sigma^{\hbox{\tiny NR}}_{\hbox{\scriptsize ann}} v^{(0)}_{\hbox{\tiny rel}}\,,
    \label{NR_hard_cross_section}
\end{equation}
which corresponds to an S-wave spin-singlet dark fermion-antifermion pair annihilating into two dark photons: $X\bar{X}\to \gamma \gamma$.
Equation~\eqref{NR_hard_cross_section} follows straightforwardly from computing the amplitude $\mathcal{M}_{\hbox{\tiny NR}}(\psi\chi \to \psi\chi)$ generated by the dimension 6 four-fermion operators
listed in the first line of eq.~\eqref{local_op_total_mom}.
If we add the contributions coming from the dimension 8 four-fermion operators listed in the second line of eq.~\eqref{local_op_total_mom},
then we get the spin-averaged S-wave annihilation cross section in the laboratory frame at leading order in $\alpha$ and at order $\bm{P}^2/M^2$ in the center-of-mass momentum 
\begin{equation}
  \left(\sigma_{\hbox{\scriptsize ann}} v_{\hbox{\tiny M\o l}}\right)_{\textrm{lab}}(\bm{P})
  = \frac{  {{\rm{Im}}\,d_s}}{M^2} \left(1-\frac{\bm{P}^2}{4M^2}\right)  
  = \sigma^{\hbox{\tiny NR}}_{\hbox{\scriptsize ann}} v^{(0)}_{\hbox{\scriptsize rel}}\left(1- \frac{\bm{P}^2}{4M^2}\right).
 \label{NR_hard_cross_section_lab_frame}
\end{equation}

The result in eq.~\eqref{NR_hard_cross_section_lab_frame} could also have been derived by Lorentz-boosting $\sigma_{\hbox{\scriptsize ann}} v_{\hbox{\tiny M\o l}}$ from the center-of-mass frame to the laboratory frame. 
In particle physics, the cross section is defined in such a way to be Lorentz invariant, hence boosting $\sigma_{\hbox{\scriptsize ann}} v_{\hbox{\tiny M\o l}}$ just means boosting $v_{\hbox{\tiny M\o l}}$.
According to its definition \eqref{def:Moller}, the M\o ller velocity transforms under a Lorentz transformation as the inverse of an energy square since the flux is Lorentz invariant.
In particular, transforming from the center-of-mass to the laboratory frame we get 
\begin{eqnarray}
  \left(v_{\hbox{\tiny M\o l}}\right)_{\textrm{lab}} 
&=& \frac{\sqrt{(p_1\cdot p_2)^2-M^4}}{\gamma(\sqrt{\bm{p}^2+M^2} - \bm{p}\cdot\bm{v}) \, \gamma(\sqrt{\bm{p}^2+M^2} + \bm{p}\cdot\bm{v})}
  = \frac{\left(v_{\hbox{\tiny M\o l}}\right)_{\textrm{cm}}}{\gamma^2\left(1 - (\bm{p}\cdot\bm{v})^2/({\bm{p}^2+M^2})\right)}
\nonumber\\
&=& \left(v_{\hbox{\tiny M\o l}}\right)_{\textrm{cm}} \left(1- \frac{\bm{P}^2}{4M^2} + \dots\right),
\label{MolLor}  
\end{eqnarray}
where $\gamma = 1/\sqrt{1-\bm{v}^2}$ is the Lorentz factor, $\bm{v}$ is the center-of-mass velocity, $\bm{p}$ the relative momentum in the center-of-mass frame,  $\bm{P}$ the center-of-mass momentum in the laboratory frame
and the dots stand for higher-order terms in the $1/M$ expansion.
Therefore, Lorentz-boosting eq.~\eqref{NR_hard_cross_section} to the laboratory frame leads precisely to eq.~\eqref{NR_hard_cross_section_lab_frame}.

\subsection{Annihilation in pNRQED$_{\textrm{DM}}$}
\label{sec:pNRQED}
Threshold effects are accounted for in pNRQED$_{\textrm{DM}}$ by integrating out from NRQED$_{\textrm{DM}}$ dark photons of momentum or energy larger than $M\alpha^2$
and casting the information of those soft modes into a potential~\cite{Biondini:2023zcz}.
In the effective field theory, the equation of motion for the dark fermion-antifermion field  is at leading order a Schr\"odinger equation governed by the above potential.
According to the scale arrangement \eqref{scale_arrang}, thermal photons are not integrated out, and therefore we can set $T=0$ when matching to pNRQED$_{\textrm{DM}}$.
The resulting potential is at leading order the Coulomb potential $-\alpha/r$, where $\bm{r}$ is the dark fermion-antifermion distance,
while we denote $\bm{R}$ the center-of-mass coordinate.

In the center-of-mass frame, the spectrum of bound states below the mass threshold is given at order $\alpha^2$ by
\begin{align}
  &(E_n)_{\textrm{cm}}  = 2M + E^b_n\,,
  \label{Encm}\\
  &\textrm{with binding energy}\, E^b_n = -\frac{M\alpha^2}{4n^2} = -\frac{1}{Ma_0^2n^2} \; \textrm{and Bohr radius}\, a_0= \frac{2}{M\alpha}. \nonumber
\end{align}
The corresponding Coulombic bound-state wavefunctions are $(\Psi_{n\ell m}(\bm{r}))_{\textrm{cm}} \!= R_{n\ell}(r)\,Y^m_\ell(\hat{\bm{r}})$,
with $n$ the principal quantum number, $\ell$ the orbital angular momentum quantum number, $m$ the magnetic quantum number, $R_{n\ell}(r)$ the radial part of the wavefunction and $Y^m_\ell(\hat{\bm{r}})$ spherical harmonics.
We call a bound state made of a dark fermion-antifermion pair \textit{darkonium}; in the S-wave case, we may further distinguish 
between a spin-singlet \textit{paradarkonium} state, and a spin-triplet \textit{orthodarkonium} state in analogy to the positronium case in QED.
In the center-of-mass frame, the continuum spectrum of scattering states above the mass threshold is given at leading order  in the relative momentum by
\begin{equation}
(E_p)_{\textrm{cm}} = 2M + \frac{\bm{p}^2}{M}.
\end{equation}  
The corresponding Coulombic scattering wavefunctions are $(\Psi_{\bm{p}}(\bm{r}))_{\textrm{cm}}$, whose partial waves we denote $(\Psi_{\bm{p}\ell}(\bm{r}))_{\textrm{cm}}$.
In the laboratory frame, the spectrum and wavefunctions get corrections that depend on the center-of-mass momentum $\bm{P}$.
The leading-order correction to the spectrum is the center-of-mass kinetic energy $\displaystyle \frac{\bm{P}^2}{4M}$,
which is of the same order as $E^b_n$ or $\bm{p}^2/M$ if $P \sim p \sim \sqrt{MT}$ and $M\alpha^2 \sim T$.
Higher order corrections are computed in appendix~\ref{sec:app_B}.

In the laboratory frame, the interaction between the dark fermion-antifermion field $\phi (t,\bm{r},\bm{R})$ at a time $t$ and the dark photon field is described by the pNRQED$_{\textrm{DM}}$ Lagrangian density
\begin{eqnarray}
  \mathcal{L}_{\textrm{pNRQED}_{\textrm{DM}}} \supset   \int d^3r \; \phi^\dagger(t,\bm{r},\bm{R})
              \left[ \bm{r} \cdot g\bm{E}(t,\bm{R})  + \frac{\bm{r}}{2} \cdot  \left\lbrace \frac{\bm{P}}{2M}\times \, , \, g\bm{B}(t,\bm{R}) \right\rbrace + \dots \right] \phi (t,\bm{r},\bm{R})  \,,
              \nonumber
              \\
\label{pNREFT_1}
\end{eqnarray}
where $E^i=F^{i0}$ is the (dark) electric field, $B^i=-\epsilon_{ijk} F^{jk}/2$ is the (dark) magnetic field,  
and $\{\dots,\dots\}$ stands for the anticommutator.
The term proportional to the electric field is an electric-dipole interaction term; it provides the leading interaction between fermion and photon fields in pNRQED$_{\textrm{DM}}$.
The term proportional to the magnetic field provides the leading interaction between fermion and photon fields in pNRQED$_{\textrm{DM}}$ that is proportional to the center-of-mass momentum~$\bm{P}$.
It is sometimes also called \textit{R\"ontgen term}~\cite{James_D_Cresser_2003}.
The R\"ontgen term originates form the \textit{Lorentz force} $\bm{F} = \bm{v}\times g\bm{B}$ and it shows up as a manifestation of the Poincar\'e invariance of QED~\cite{Brambilla:2003nt}.
It is suppressed in the center-of-mass velocity $\bm{v}$ compared to the electric-dipole term.
The dots in \eqref{pNREFT_1} stand for irrelevant operators that are subleading with respect to the electric-dipole one, if they do not depend on the center-of-mass momentum,
or to the R\"ontgen term, if they do depend on the center-of-mass momentum.

The four-fermion operators responsible for annihilation in NRQED$_{\textrm{DM}}$ give rise to contact potentials in pNRQED$_{\textrm{DM}}$,
\begin{equation}
  \mathcal{L}_{\textrm{pNRQED}_{\textrm{DM}}} ~\supset~    - \int d^3r \; \phi^\dagger(t,\bm{r},\bm{R}) \; {\rm{Im}}\,\delta V^{\textrm{ann}} \; \phi (t,\bm{r},\bm{R})  \,,
\label{pNREFT_2}
\end{equation}
with 
\begin{equation}
\begin{aligned}
& {\rm{Im}} \,\delta V^{\textrm{ann}}
  = - \frac{1}{M^2} \, \delta^3(\bm{r})\, \left[ 2 {\rm{Im}}\,d_s - \bm{S}^2 \left( {\rm{Im}}\,d_s- {\rm{Im}}\,d_v \right) \right] \\
  & \; + \frac{1}{M^4} \, \delta^3(\bm{r})\, \nabla_{\bm{R}}^i \nabla_{\bm{R}}^j
  \left[ {2\rm{Im}}\, g_{\textrm{c\,cm}}\,\delta_{ij}
    - \bm{S}^2 \left( {\rm{Im}}\, g_{\textrm{c\,cm}} - {\rm{Im}}\,g_{\textrm{a\,cm}}\right)\delta_{ij}
    + S^i S^j \,{\rm{Im}}\,g_{\textrm{b\,cm}}  \right],
\label{annpNRQED}
\end{aligned}
\end{equation}  
if we match to the imaginary parts of the four-fermion operators listed in eq.~\eqref{local_op_total_mom}.
The operator $\bm{S}=\bm{S}_1+\bm{S}_2$ is the total spin of the dark fermion-antifermion pair, 
with $\bm{S}_1=\bm{\sigma}_1/2$ and $\bm{S}_2=\bm{\sigma}_2/2$ the spin operators acting on the fermion and antifermion, respectively.

The resummation of multiple soft photon exchanges within the dark fermion-antifer\-mion pair leads to a modification of the pair wavefunction close to threshold from free to either a bound-state wavefunction or a scattering wavefunction.
This modification ultimately alters the annihilation cross section and decay width.
The spin averaged annihilation cross section may be computed from the optical theorem analogously to eq.~\eqref{optical_cross_section}:
\begin{equation}
\sigma_{\hbox{\scriptsize ann}} v_{\hbox{\tiny M\o l}}  = \frac{{\rm{Im}}[\mathcal{M}_{\hbox{\tiny NR}}(\phi \to \phi)]}{2}\,,
\label{optical_cross_section2}
\end{equation}
where the amplitude $\mathcal{M}_{\hbox{\tiny NR}}(\phi \to \phi)$ describes the propagation of the fermion-antifermion field $\phi$ projected on scattering states.
The amplitude is given by the expectation value of $-{\rm{Im}} \,\delta V^{\textrm{ann}}$ on the fermion-antifermion wavefunction for scattering states.
The spin-averaged S-wave annihilation cross section in the laboratory frame at leading order in $\alpha$ and at order $\bm{P}^2/M^2$ in the center-of-mass momentum reads, therefore, 
\begin{equation}
  \left(\sigma_{\hbox{\scriptsize ann}} v_{\hbox{\tiny M\o l}}\right)_{\textrm{lab}}(\bm{p},\bm{P})
  =   \sigma^{\hbox{\tiny NR}}_{\hbox{\scriptsize ann}} v^{(0)}_{\hbox{\scriptsize rel}} \,
  \left(\left|\Psi_{\bm{p} 0}(\bm{0})\right|^2\right)_{\textrm{lab}}(\bm{P})\,\left(1- \frac{\bm{P}^2}{4M^2}\right),
\label{ann_fact_scatwflab}
\end{equation}
where we have used eqs.~\eqref{gabcm} and \eqref{gccm}, and $\sigma^{\hbox{\tiny NR}}_{\hbox{\scriptsize ann}} v^{(0)}_{\hbox{\scriptsize rel}}$ has been defined in eq.~\eqref{NR_hard_cross_section}.
According to eq.~\eqref{Psiporigin}, we can replace $\left(\left|\Psi_{\bm{p} 0}(\bm{0})\right|^2\right)_{\textrm{lab}}(\bm{P})$ with
the corresponding quantity in the center-of-mass frame, $\left(\left|\Psi_{\bm{p} 0}(\bm{0})\right|^2\right)_{\textrm{cm}} \equiv S_{\hbox{\scriptsize ann}}(\zeta)$, 
where $S_{\hbox{\scriptsize ann}}(\zeta)$ is called \emph{Sommerfeld factor}~\cite{Sommerfeld} and reads (see e.g.~\cite{Iengo:2009ni,Cassel:2009wt}) 
\begin{equation}
  S_{\hbox{\scriptsize ann}}(\zeta)=\frac{2 \pi \zeta}{1-e^{-2 \pi \zeta}} \, , \qquad\qquad \zeta \equiv \frac{\alpha}{v^{ (0)}_{\hbox{\scriptsize rel}}} = \frac{1}{a_0|\bm{p}_{\textrm{cm}}|}\, .
\label{Somme_0}
\end{equation}
Hence, the spin-averaged S-wave annihilation cross section in the laboratory frame at leading order in $\alpha$ and at order $\bm{P}^2/M^2$ in the center-of-mass momentum can be written as
\begin{equation}
  \left(\sigma_{\hbox{\scriptsize ann}} v_{\hbox{\tiny M\o l}}\right)_{\textrm{lab}}(\bm{p},\bm{P})
=  \sigma^{\hbox{\tiny NR}}_{\hbox{\scriptsize ann}} v^{(0)}_{\hbox{\scriptsize rel}}  \, S_{\hbox{\scriptsize ann}}(\zeta)\left(1- \frac{\bm{P}^2}{4M^2}\right) \,.
\label{ann_fact_scat}
\end{equation}
In the above expression, the center-of-mass relative momentum $\bm{p}_{\textrm{cm}}$ is expressed in terms of the relative momentum in the laboratory frame through eq.~\eqref{LT_prel_COM_lab_v2}.

Below threshold, spin-singlet bound states, paradarkonia, decay via annihilation into two dark photons.\footnote{
The annihilation of spin-triplet orthodarkonia requires a final state made of three dark photons.
It happens, therefore, at order $\alpha^3$, which is beyond our present accuracy.}
The decay width can be computed from
\begin{equation}
\Gamma_{\hbox{\scriptsize ann}} = 2 \,{\rm{Im}}[\mathcal{M}_{\hbox{\tiny NR}}(\phi \to \phi)]\,,
\label{optical_cross_section3}
\end{equation}
which is analogous to eq.~\eqref{optical_cross_section2}, but now we do not average over the spin of the initial states as we project $\phi$ onto the specific bound state that is decaying.
Proceeding like in the case of the annihilation cross section,
it follows that the paradarkonium S-wave annihilation width in the laboratory frame at order $\bm{P}^2/M^2$ in the center-of-mass momentum is given by 
\begin{equation}
  \left(\Gamma^{n,\hbox{\scriptsize para}}_{\textrm{ann}}\right)_{\textrm{lab}}(\bm{P}) =
  \frac{4 {\rm{Im}}\,d_s}{M^2} \,\left(\left|\Psi_{n00}(\bm{0})\right|^2\right)_{\textrm{lab}}(\bm{P})\, \left(1- \frac{\bm{P}^2}{4M^2}\right). 
\label{ann_paralab}
\end{equation}
Using eq.~\eqref{Psinorigin}, we can replace $\left(\left|\Psi_{n00}(\bm{0})\right|^2\right)_{\textrm{lab}}(\bm{P})$ with
$\gamma \left(\left|\Psi_{n00}(\bm{0})\right|^2\right)_{\textrm{cm}}$ $\approx$ $(1 + \bm{v}^2/2)$  $\left(\left|\Psi_{n00}(\bm{0})\right|^2\right)_{\textrm{cm}}$  $=$  $(1 + \bm{P}^2/(8M^2))\left(\left|\Psi_{n00}(\bm{0})\right|^2\right)_{\textrm{cm}}$. 
We then get for the paradarkonium S-wave annihilation width in the laboratory frame at leading order in $\alpha$ and at order $\bm{P}^2/M^2$ in the center-of-mass momentum 
\begin{equation}
  \left(\Gamma^{n,\hbox{\scriptsize para}}_{\textrm{ann}}\right)_{\textrm{lab}}(\bm{P}) =
  \frac{4 \pi \alpha^2}{M^2} \left(\left|\Psi_{n00}(\bm{0})\right|^2\right)_{\textrm{cm}} \left(1- \frac{\bm{P}^2}{8M^2}\right),
\label{ann_para}
\end{equation}
which is now expressed in terms of the square of the bound-state wavefunction in the center-of-mass frame at the origin, $\left(\left|\Psi_{n00}(\bm{0})\right|^2\right)_{\textrm{cm}} = \left|R_{n0}(\bm{0})\right|^2/(4\pi)$.\footnote{
For the ground state it holds $\left|R_{10}(\bm{0})\right|^2 = 4/a_0^3 = (M\alpha)^3/2$, which leads to
$ \displaystyle  \left(\Gamma^{1,\hbox{\tiny para}}_{\textrm{ann}}\right)_{\textrm{lab}}(\bm{P})  = \frac{M \alpha^5}{2}\left(1- \frac{\bm{P}^2}{8M^2}\right)$,
where $M \alpha^5/2$ is the 1S paradarkonium annihilation width in the center-of-mass frame.
}
Since $\displaystyle \frac{4 \pi \alpha^2}{M^2} \left(\left|\Psi_{n00}(\bm{0})\right|^2\right)_{\textrm{cm}}$ is the annihilation width in the center-of-mass frame, $\left(\Gamma^{n,\hbox{\scriptsize para}}_{\textrm{ann}}\right)_{\textrm{cm}}$, 
eq.~\eqref{ann_para} simply states the expected Lorentz dilation of time intervals:
\begin{equation}
  \left(\Gamma^{n,\hbox{\scriptsize para}}_{\textrm{ann}}\right)_{\textrm{lab}}(\bm{P})  = \frac{\left(\Gamma^{n,\hbox{\scriptsize para}}_{\textrm{ann}}\right)_{\textrm{cm}}}{\gamma}
  \approx \left(\Gamma^{n,\hbox{\scriptsize para}}_{\textrm{ann}}\right)_{\textrm{cm}}\left(1- \frac{\bm{P}^2}{8M^2}\right).
\label{Lorentz_dilation}
\end{equation}
Our calculation is restricted to the paradarkonium case, but clearly the same relation also holds for the orthodarkonium decay.

In computing annihilation cross sections and widths we have neglected  the thermal distribution of the photons in the final state.
This is justified by the fact that the energy of the final state photons is of order $M$.
Therefore, according to the hierarchy \eqref{scale_arrang}, the corresponding Bose--Einstein distribution is exponentially suppressed: $n_{\textrm{B}}(M) \approx e^{-M/T}$.

\section{Bound-state formation}
\label{sec:bsf}
\begin{figure}[ht]
    \centering
    \includegraphics[scale=0.7]{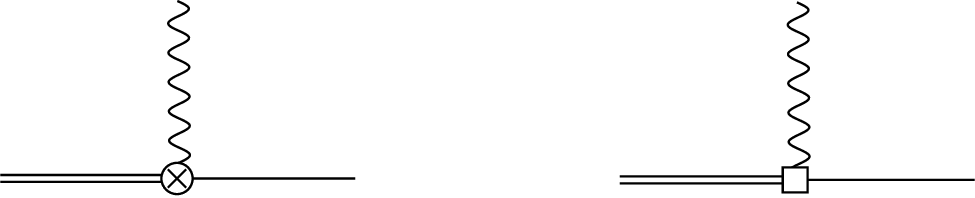}
    \caption{Feynman diagrams for electric and magnetic-dipole transitions between fermion-antifermion scattering states (double line) and bound states (single line).
   The wavy lines stand for the photon external fields.
    The circle with a cross denotes the electric-dipole vertex $i\bm{r}\cdot g\bm{E}$ 
      and the square denotes the magnetic-dipole vertex $i\bm{r} \cdot  \left\lbrace \bm{P}\times \, , \, g\bm{B} \right\rbrace/(4M)$ due to the R\"ontgen term~\cite{Brambilla:2003nt}.
     The momentum $\bm{P}$ is the sum of the center-of-mass momentum of the incoming 
     fermion-antifermion pair and the momentum of the incoming photon.     
    The vertices follow from the Lagrangian~\eqref{pNREFT_1}.}
    \label{fig:el_magn_vertex}
\end{figure}

In the laboratory frame, i.e. in the reference frame where the thermal bath is at rest and the center of mass of the dark fermion-antifermion pair is moving,
near-threshold processes such as the formation of bound states or their dissociation into scattering states are due at order $\bm{r}$ in pNRQED$_{\textrm{DM}}$ 
to the two dipole interaction operators in the right-hand side of eq.~\eqref{pNREFT_1}.
The corresponding vertices are shown in figure~\ref{fig:el_magn_vertex}.
The photons emitted or absorbed in the bound-state formation and bound-state dissociation processes respectively are \textit{ultrasoft}, i.e. they carry energy and momentum of order $M \alpha^2$ or $T$,
which justifies the multipole expansion for a system that fulfills the hierarchy of energy scales~\eqref{scale_arrang}.

Under the hierarchy of energy scales~\eqref{scale_arrang}, the electric-dipole operator is the leading operator responsible for bound-state formation and bound-state dissociation.
Its effect has been computed in ref.~\cite{Biondini:2023zcz}.
Together with kinetic energy corrections to the electric-dipole vertex,  
the magnetic-dipole vertex accounts for the leading correction to bound-state formation and bound-state dissociation due to the center-of-mass motion of the dark fermion-antifermion pair relative to the thermal bath.
Its effect is suppressed by $P/M \sim v \sim \sqrt{T/M}$ (if $P \sim \sqrt{MT}$) with respect to the effect of the electric-dipole vertex. 

The laboratory frame may be a convenient frame where to compute recoil effects, because thermal  distributions have there a particularly simple form. 
For instance, the thermal distribution of photons in a thermal bath at rest is the Bose--Einstein distribution
\begin{equation}
 n_{\text{B}}(E) = \frac{1}{e^{E/T}-1}\,.
\label{BoseEinstein}
\end{equation}  
The Bose--Einstein distribution for a moving thermal bath is given in eq.~\eqref{bose_fkt} and requires the introduction of a velocity-dependent effective temperature (see discussion following eq.~\eqref{bose_fkt}).

In this section, we compute the bound-state formation cross section in pNRQED$_{\textrm{DM}}$ and in section~\ref{sec:bsd} the bound-state dissociation width in pNRQED$_{\textrm{DM}}$ including order $\bm{P}^2/M^2$ recoil effects.
In both cases, we make use of the optical theorem and express the rates in terms of self-energy diagrams whose vertices are shown in figure~\ref{fig:el_magn_vertex}.
We check explicitly that the cross section and width obtained in the laboratory frame agree with the ones derived by boosting the cross section and width obtained in the center-of-mass frame.

\subsection{Formation of bound states in the laboratory frame}
\label{sec:bsf_lab}
\begin{figure}[ht]
    \centering
    \includegraphics[scale=0.9]{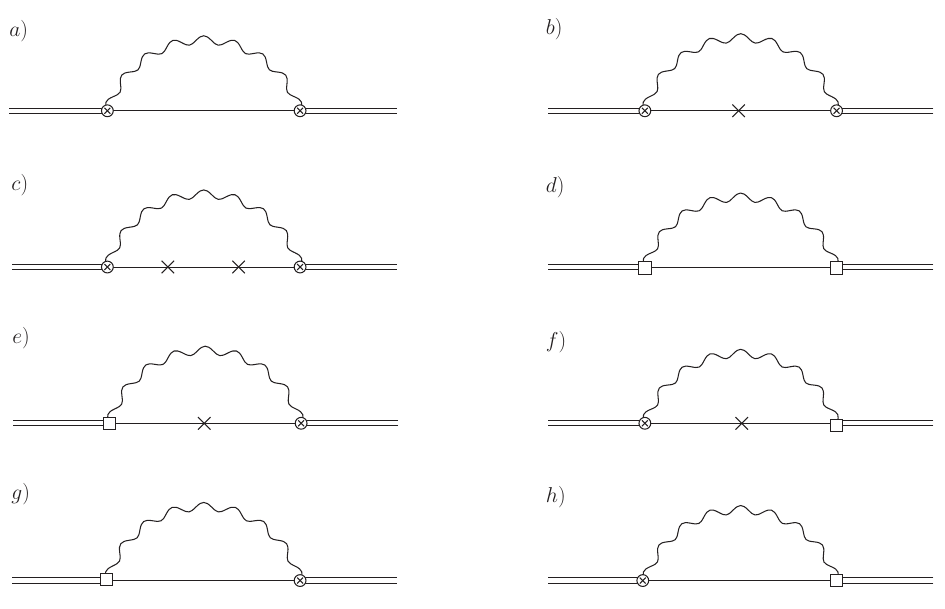}
    \caption{Self-energy diagrams in  pNRQED$_{\textrm{DM}}$ with an initial scattering state (solid double line) and an intermediate bound state (solid line) contributing up to relative order $T/M$ in the laboratory frame.
      Diagram $a)$ is the leading-order diagram. Diagrams $b)$ to $h)$ are suppressed by $T/M$ or $\Delta E^p_n/M$.
      Since the dark photon in the loop carries a spatial momentum $\bm{k}$, the bound fermion-antifermion pair recoils by a spatial momentum $\bm{P}-\bm{k}$.
      Crosses in the bound-state propagator stand for insertions of the recoil correction $-(2\bm{P}\cdot \bm{k}-\bm{k}^2)/(4M)$ in the non-recoiling propagator ${i}/{(\Delta E^p_n-k_0+i\epsilon)}$.
      Electric and magnetic couplings are as in figure~\ref{fig:el_magn_vertex}.}
    \label{fig:pnEFT_DM_self}
\end{figure}

At leading order  in the coupling expansion and first order in the temperature and recoil energy over $M$ ratio, the cross section for the process
\begin{equation}
(X\bar{X})_p\to \gamma + (X\bar{X})_n\,,
\label{ptogamman}
\end{equation}
where a bound state $(X\bar{X})_n$ is formed from a scattering state $(X\bar{X})_p$ via the emission of an ultrasoft dark photon,
can be determined by cutting at finite $T$ the time-ordered self-energy diagrams shown in figure~\ref{fig:pnEFT_DM_self}.
They depend on four correlators: \\
{\it a)} the electric-electric correlator
\begin{equation}
\langle E_i(t,\bm{R})E_j(0,\bm{R}') \rangle  = \int \frac{d^4k}{(2\pi)^4}e^{-ik^0t+i\bm{k}\cdot(\bm{R}-\bm{R}')}\left[k_0^2D_{ij}(k)+k_ik_jD_{00}(k)\right] \, ,
\label{elelcorr}
\end{equation}
{\it b)} the magnetic-magnetic correlator
\begin{equation}
\langle B_i(t,\bm{R})B_j(0,\bm{R}') \rangle = \epsilon_{ilm}\epsilon_{jnr}\int \frac{d^4k}{(2\pi)^4}e^{-ik^0t+i\bm{k}\cdot(\bm{R}-\bm{R}')}k_lk_n D_{mr}(k) \, ,
\label{magnmagncorr}
\end{equation}
{\it c)} the electric-magnetic correlator
\begin{equation}
\langle B_i(t,\bm{R})E_j(0,\bm{R}') \rangle = -\epsilon_{ilm}\int \frac{d^4k}{(2\pi)^4}e^{-ik^0t+i\bm{k}\cdot(\bm{R}-\bm{R}')}k_lk_0 D_{mj}(k) \, ,
\label{elmagncorr}
\end{equation}
{\it d)} the magnetic-electric correlator
\begin{equation}
\langle E_i(t,\bm{R})B_j(0,\bm{R}') \rangle = -\epsilon_{jlm}\int \frac{d^4k}{(2\pi)^4}e^{-ik^0t+i\bm{k}\cdot(\bm{R}-\bm{R}')}k_lk_0 D_{im}(k) \, ,
\label{magnelcorr}
\end{equation}
where $D_{\mu \nu}$ is the dark photon propagator in real-time formalism.
The correlators are gauge invariant and therefore may be evaluated in any gauge.
The explicit expression of the photon propagator in Coulomb gauge at leading order can be found in~\cite{Landshoff:1992ne}, 
whereas the expression of the dark fermion-antifermion pair propagator can be found in~\cite{Biondini:2023zcz}.\footnote{
In real-time formalism, propagators are $2\times 2$ matrices.
However, the fermion-antifermion propagator gets a particularly simple form in the heavy-fermion limit,
as thermal corrections are exponentially suppressed and the $12$-component vanishes~\cite{Brambilla:2008cx}.
The bound-state formation cross section can be computed from the imaginary part of the $11$-component of the self energy, which is the approach that we follow in this section, 
or alternatively from the $21$-component of the self energy~\cite{Biondini:2023zcz}.
}

In the laboratory frame, the dark fermion-antifermion pair moves with momentum $\bm{P}$.
After emitting a dark photon of spatial momentum $\bm{k}$, the fermion-antifermion pair recoils by a spatial momentum $\bm{P-k}$.
Therefore, the propagator of the recoiling fermion-antifermion pair is in pNRQED$_{\textrm{DM}}$
\begin{equation}
\begin{aligned}
 & \frac{i}{E_p + \bm{P}^2/(4M) - (E_n + (\bm{P}-\bm{k})^2/(4M)) -k_0+i\epsilon} =\\
 &\hspace{7cm}  \frac{i}{\Delta E^p_n + {(2\bm{P}\cdot \bm{k}-\bm{k}^2)}/{(4M)}-k_0+i\epsilon}\,,
\label{recoilprop}
\end{aligned}
\end{equation}
where we have defined 
\begin{equation}
  \Delta E^{p}_n \equiv (\Delta E^{p}_n)_{\textrm{lab}} \equiv E_p - E_n >0
\end{equation}
the energy difference between the scattering state and the bound state in the laboratory frame.
The term $(2\bm{P}\cdot \bm{k}-\bm{k}^2)/(4M)$ is  a recoil correction to the kinetic energy.
It is indeed a correction, since the term $\bm{P}\cdot \bm{k}/(2M)$ is suppressed by $\sqrt{T/M}$ and the term $ - \bm{k}^2/(4M)$ by $T/M$ with respect to $\Delta E^{p}_n$ and $T$.
Hence, the expansion of the propagator \eqref{recoilprop} up to terms of relative order $T/M$ reads
\begin{equation}
\begin{aligned}
  &\frac{i}{\Delta E^p_n + {(2\bm{P}\cdot \bm{k}-\bm{k}^2)}/{(4M)} -k_0+i\epsilon} =  \frac{i}{\Delta E^p_n-k_0+i\epsilon}\\
  &\hspace{1cm} + \frac{i}{(\Delta E^p_n-k_0+i\epsilon)^2}\left( - \frac{2\bm{P}\cdot \bm{k}-\bm{k}^2}{4M} \right)
  +\frac{i}{(\Delta E^p_n-k_0+i\epsilon)^3}\left(\frac{\bm{P}\cdot \bm{k}}{2M}\right)^2 + \dots \, ,
\label{den_expansion}
\end{aligned}
\end{equation}
where the dots stand for higher-order terms.
In Feynman diagrams, we indicate with a cross the insertion of a correction $-(2\bm{P}\cdot \bm{k}-\bm{k}^2)/(4M)$  in the fermion-antifermion propagator.\footnote{
In eq.~\eqref{recoilprop}, we have added to the propagator the $1/M$ center-of-mass recoil corrections that 
  stem from the kinetic energy.
  The fermion-antifermion potential gets also affected by center-of-mass-momentum dependent corrections, $\delta V(\bm{r},\bm{P})$, at order $1/M^2$~\cite{Barchielli:1986zs,Barchielli:1988zp,Brambilla:2001xk,Brambilla:2003nt}:
  $$
  \delta V(\bm{r},\bm{P})
  = -\frac{1}{8M^2}\bm{P}^2V^{(0)}(r) + \frac{1}{8M^2}(\bm{r}\cdot\bm{P})^2\frac{V^{(0)\prime}(r)}{r}
  - \frac{1}{8M^2}(\bm{r}\times \bm{P})\cdot (\bm{\sigma}_1 -\bm{\sigma}_2) \frac{V^{(0)\prime}(r)}{r} , 
  $$
  where $V^{(0)}(r)$ is the static potential, $V^{(0)\prime}(r)$ its derivative, $\bm{\sigma}_1$ the Pauli matrix acting on the fermion and $\bm{\sigma}_2$ the Pauli matrix acting on the antifermion.
  Of the same order is also the kinetic energy correction $- \bm{P}^4/(64 M^3)-(\bm{p}\cdot\bm{P})^2/(4M^3) - \bm{p}^2\bm{P}^2/(8M^3)$.
   These corrections may contribute at order  $T/M$ and $\Delta E^p_n/M$ only through diagram $a)$ of figure~\ref{fig:pnEFT_DM_self}, as insertions of $\delta V(\bm{r},\bm{P})$ into diagrams $b)$-$h)$
  are  suppressed.
  In diagram $a)$, the leading recoil correction due to $\delta V(\bm{r},\bm{P})$ is proportional to the energy shift 
  $\langle \bm{p}| \delta V(\bm{r},\bm{P})|\bm{p}\rangle  - \langle n| \delta V(\bm{r},\bm{P}-\bm{k}) | n \rangle$.
  Terms that do not depend on $\bm{k}$ are absorbed into $\Delta E^p_n$, 
  terms linear in $\bm{k}$ give rise to odd integrands in $\bm{k}$, which vanish in the integral,
  while terms proportional to $\bm{k}^2$ give rise to corrections of order  $(\Delta E^p_n/M)^2$, which are beyond our accuracy.
  A similar reasoning holds for the kinetic energy correction.
\label{footnotepotential}
}

The leading-order cross section can be computed from the imaginary part of the self-energy diagram with two electric-dipole vertices.
This is diagram~\textit{a)} of figure~\ref{fig:pnEFT_DM_self}.
The time-ordered $11$-component of diagram~\textit{a)}, 
projected onto a scattering state with relative momentum $\bm{p}$, reads~\cite{Biondini:2023zcz}
\begin{equation}
\begin{aligned}
  &\left( -i\Sigma^{11}_{\textrm{diag.}\,\,\textit{a)}}\right)(\bm{p}) = 
  -g^2 \sum \limits_{n} \int \frac{d^4k}{(2\pi)^4} \, \frac{i}{\Delta E^p_n-k_0+i\epsilon}\\
  & \hspace{2cm} \times \left[|\langle n|\bm{r}|\bm{p}\rangle|^2k_0^2 - \langle n|\bm{r}\cdot \bm{k}|\bm{p}\rangle|^2\right] \left[\frac{i}{k^2+i\epsilon}+2\pi \delta{(k^2)}n_{\textrm{B}}(|k_0|)\right] \,,
\label{ee_diagram_LO}
\end{aligned}
\end{equation}
where we have summed over all intermediate bound states.

In this work, we add to \eqref{ee_diagram_LO} all corrections of relative order $\bm{P}^2/M^2 \sim T/M$ and $\Delta E^p_n/M$.
They are encoded in the self-energy diagrams~\textit{b)}~-~\textit{h)} shown in figure~\ref{fig:pnEFT_DM_self}.
They are computed projecting onto a scattering state with relative momentum $\bm{p}$ and center-of-mass momentum $\bm{P}$ in the laboratory frame.

The time-ordered $11$-component of the electric-electric self-energy diagram~\textit{b)} reads
\begin{equation}
\begin{aligned}
  &\left( -i\Sigma^{11}_{\textrm{diag.}\,\textit{b)}}\right) (\bm{p}) = 
  -g^2 \sum \limits_{n} \int \frac{d^4k}{(2\pi)^4} \,\frac{i}{(\Delta E^p_n-k_0+i\epsilon)^2}\,\frac{\bm{k}^2}{4M} \\
  & \hspace{2cm} \times \left[|\langle n|\bm{r}|\bm{p}\rangle|^2k_0^2 - \langle n|\bm{r}\cdot \bm{k}|\bm{p}\rangle|^2\right] \left[\frac{i}{k^2+i\epsilon}+2\pi \delta{(k^2)}n_{\textrm{B}}(|k_0|)\right] \, .
\label{ee1_diagram}
\end{aligned}
\end{equation}
It comes from considering one recoil correction insertion in the bound-state propagator.
We have dropped the term proportional to $-\bm{P}\cdot \bm{k}/(2M)$, since it generates an integrand that is odd in $\bm{k}$ and, therefore, gives rise to a vanishing integral. 
Note that, since $\left( -i\Sigma^{11}_{\textrm{diag.}\,\textit{b)}}\right)(\bm{p})$ is not proportional to the center-of-mass momentum, this contribution is present also in the center-of-mass frame.

The time-ordered $11$-component of the electric-electric self-energy diagram~\textit{c)} reads
\begin{equation}
\begin{aligned}
  &\left( -i\Sigma^{11}_{\textrm{diag.}\,\textit{c)}}\right)
  (\bm{p},\bm{P}) = 
  -g^2 \sum \limits_{n} \int \frac{d^4k}{(2\pi)^4} \, \frac{i}{(\Delta E^p_n-k_0+i\epsilon)^3}\left(\frac{\bm{P}\cdot \bm{k}}{2M}\right)^2\\
  & \hspace{2cm} \times \left[|\langle n|\bm{r}|\bm{p}\rangle|^2k_0^2 - \langle n|\bm{r}\cdot \bm{k}|\bm{p}\rangle|^2\right] \left[\frac{i}{k^2+i\epsilon}+2\pi \delta{(k^2)}n_{\textrm{B}}(|k_0|)\right] \,.
\label{ee_diagram}
\end{aligned}
\end{equation}
It comes from considering two recoil correction insertions in the bound-state propagator.
Recoil corrections proportional to $[\bm{k}^2/(4M)]^2$ are of higher order and have been neglected.

The time-ordered $11$-component of the magnetic-magnetic self-energy diagram~\textit{d)} reads
\begin{equation}
\begin{aligned}
  &\left( -i\Sigma^{11}_{\textrm{diag.}\,\textit{d)}}\right)
  (\bm{p},\bm{P}) = 
  -\frac{g^2}{4M^2}\epsilon_{ijk}\epsilon_{lmn}\epsilon_{krs}\epsilon_{ntu} \sum \limits_{n} \int \frac{d^4k}{(2\pi)^4} \left(P_i - \frac{k_i}{2} \right) \frac{i}{\Delta E^p_n -k_0+i\epsilon} \\
  & \hspace{1.3cm}  \times  \left(P_l - \frac{k_l}{2} \right) \langle \bm{p}|r^j|n\rangle \langle n|r^m|\bm{p}\rangle k_r k_t \left(\delta_{su}-\frac{k_s k_u}{|\bm{k}|^2}\right)\left[\frac{i}{k^2+i\epsilon}+2\pi \delta{(k^2)}n_{\textrm{B}}(|k_0|)\right] \, .
\label{mm_diagram}
\end{aligned}
\end{equation}
We have taken the propagator~\eqref{den_expansion} at leading order in $T/M$ as the two magnetic vertices already provide a relative suppression of order $T/M$ with respect to $\left( -i\Sigma^{11}_{\textrm{diag.}\,\,\textit{a)}}\right)(\bm{p})$. 
Recoil corrections proportional to $k_ik_lk_rk_t/(4M)^2$ are of higher order and have been neglected.

The time-ordered $11$-component of the magnetic-electric self-energy diagram~\textit{e)} of figure~\ref{fig:pnEFT_DM_self} reads at relative order $T/M$
\begin{equation}
\begin{aligned}
  &\left( -i\Sigma^{11}_{\textrm{diag.}\, e)}\right)
  (\bm{p},\bm{P}) = 
  \frac{g^2}{2M}\epsilon_{ijk}\epsilon_{klm} \sum \limits_{n} \int \frac{d^4k}{(2\pi)^4} \frac{i}{(\Delta E^p_n-k_0+i\epsilon)^2}\left( - \frac{\bm{P}\cdot \bm{k}}{2M} \right)\left(P_i-\frac{k_i}{2}\right) \\
   & \hspace{2cm} \times \langle \bm{p}|r^s|n\rangle \langle n|r^j|\bm{p}\rangle k_l k_0 \left(\delta_{sm}-\frac{k_s k_m}{|\bm{k}|^2}\right)\left[\frac{i}{k^2+i\epsilon}+2\pi \delta{(k^2)}n_{\textrm{B}}(|k_0|)\right] \,.
\label{em_diagram}
\end{aligned}
\end{equation}
We have dropped the term proportional to $\bm{k}^2/(4M)$ in the recoil correction insertion in the bound-state propagator,  
because for that term either the integrand is odd in $\bm{k}$, and hence the integral vanishes, or the contribution is of higher order.
Similarly, for the symmetric electric-magnetic self-energy diagram~\textit{f)} we obtain 
\begin{equation}
\begin{aligned}
  &\left( -i\Sigma^{11}_{\textrm{diag.}\,f)}\right)
  (\bm{p},\bm{P}) = 
  \frac{g^2}{2M}\epsilon_{ijk}\epsilon_{klm} \sum \limits_{n} \int \frac{d^4k}{(2\pi)^4} \left(P_i-\frac{k_i}{2}\right) \frac{i}{(\Delta E^p_n-k_0+i\epsilon)^2}\left( - \frac{\bm{P}\cdot \bm{k}}{2M} \right) \\
  & \hspace{2cm}  \times \langle \bm{p}|r^j|n\rangle \langle n|r^s|\bm{p}\rangle k_l k_0 \left(\delta_{ms}-\frac{k_m k_s}{|\bm{k}|^2}\right)\left[\frac{i}{k^2+i\epsilon}+2\pi \delta{(k^2)}n_{\textrm{B}}(|k_0|)\right] \, ,
\label{me_diagram}
\end{aligned}
\end{equation}
which is equal to eq.~\eqref{em_diagram}, as the term proportional to $-k_i/2$ in the magnetic-dipole vertex gives rise to an integrand that is odd in $\bm{k}$, which leads to a vanishing integral.

The term proportional to $-k_i/2$, however, contributes if we consider the diagrams without recoil correction insertion, i.e. diagram~\textit{g)},
\begin{equation}
\begin{aligned}
  &\left( -i\Sigma^{11}_{\textrm{diag.}\,\textit{g)}}\right) (\bm{p}) = 
  \frac{g^2}{2M}\epsilon_{ijk}\epsilon_{klm} \sum \limits_{n} \int \frac{d^4k}{(2\pi)^4} \frac{i}{\Delta E^p_n-k_0+i\epsilon}\left(P_i-\frac{k_i}{2}\right)\\
   & \hspace{2cm} \times \langle \bm{p}|r^s|n\rangle  \langle n|r^j|\bm{p}\rangle k_l k_0 \left(\delta_{sm}-\frac{k_s k_m}{|\bm{k}|^2}\right)\left[\frac{i}{k^2+i\epsilon}+2\pi \delta{(k^2)}n_{\textrm{B}}(|k_0|)\right] \,,
\label{me_diagram2}
\end{aligned}
\end{equation}
and diagram~\textit{h)}
\begin{equation}
\begin{aligned}
  &\left( -i\Sigma^{11}_{\textrm{diag.}\,\textit{h)}}\right) (\bm{p}) = 
  \frac{g^2}{2M}\epsilon_{ijk}\epsilon_{klm} \sum \limits_{n} \int \frac{d^4k}{(2\pi)^4} \left(P_i-\frac{k_i}{2}\right) \frac{i}{\Delta E^p_n-k_0+i\epsilon}\\
   & \hspace{2cm} \times \langle \bm{p}|r^j|n\rangle \langle n|r^s|\bm{p}\rangle k_l k_0 \left(\delta_{sm}-\frac{k_s k_m}{|\bm{k}|^2}\right)\left[\frac{i}{k^2+i\epsilon}+2\pi \delta{(k^2)}n_{\textrm{B}}(|k_0|)\right] \,.
\label{me_diagram2h}
\end{aligned}
\end{equation}
In this case,  it is the term proportional to $P_i$ that gives rise to a vanishing integral, whereas the term proportional to $-k_i/2$ gives a contribution of relative order $T/M$ or $\Delta E^p_n/M$.
For these diagrams, it holds what we wrote about diagram~\textit{b)}, i.e. that they are not proportional to the center-of-mass momentum and, therefore, they contribute also in the center-of-mass frame.
This may, at first, be surprising as the diagrams involve the magnetic-dipole vertex originating from the R\"ontgen term.
However, the physical reason is that, once a photon of momentum $\bm{k}$ is emitted, the fermion-antifermion pair recoils by a momentum $-\bm{k}$ also in the center-of-mass frame.

The sum of the imaginary parts of the self-energies~\eqref{ee_diagram_LO}-\eqref{me_diagram2h} gives the bound-state formation cross section up to relative order $T/M$ and $\Delta E^p_n/M$:
\begin{equation}
\begin{aligned}
  &(\sigma_{\hbox{\scriptsize bsf}} \, v_{\hbox{\scriptsize M\o l}})_{\textrm{lab}}(\bm{p},\bm{P}) \equiv  \sum \limits_n   (\sigma^n_{\hbox{\scriptsize bsf}} \, v_{\hbox{\scriptsize M\o l}})_{\textrm{lab}}(\bm{p},\bm{P})
  = \\
  & \hspace{0.8cm}  -2\,\textrm{Im}\left[ \left(\Sigma^{11}_{\textrm{diag.}\,\textit{a)}}\right)(\bm{p}) + \left(\Sigma^{11}_{\textrm{diag.}\,\textit{b)}}\right)(\bm{p})
    +\left(\Sigma^{11}_{\textrm{diag.}\,\textit{c)}}\right)
     (\bm{p},\bm{P})+ \left(\Sigma^{11}_{\textrm{diag.}\,\textit{d)}}\right)
     (\bm{p},\bm{P})\right.\\
    &\hspace{1.8cm} \left. + \left(\Sigma^{11}_{\textrm{diag.}\,\textit{e)}}\right)
     (\bm{p},\bm{P})
    + \left(\Sigma^{11}_{\textrm{diag.}\,\textit{f)}}\right)
     (\bm{p},\bm{P}) + \left(\Sigma^{11}_{\textrm{diag.}\,\textit{g)}}\right)(\bm{p})
     + \left(\Sigma^{11}_{\textrm{diag.}\,\textit{h)}}\right)(\bm{p})
    \right] .
\label{bsf_xsection}
\end{aligned}
\end{equation}
The equation follows from the optical theorem and eq.~\eqref{optical_cross_section2}, by noticing that $\mathcal{M}_{\hbox{\tiny NR}} = - \Sigma^{11}$
and that in $\Sigma^{11}$ we have already implicitly averaged over the spins of the initial states.\footnote{
The electric-dipole interaction and the R\"ontgen term are both spin independent.
Computing their matrix elements on the spin-independent wavefunction is equivalent to averaging over the possible spin configurations.}
 
Including all corrections of order $\bm{P}^2/M^2 \sim T/M$ and $\Delta E^p_n/M$, the bound-state formation cross section in the laboratory frame reads
\begin{equation}
\begin{aligned}
 & (\sigma_{\hbox{\scriptsize bsf}} \, v_{\hbox{\scriptsize M\o l}})_{\textrm{lab}}(\bm{p},\bm{P}) = \\
 & \hspace{2mm} \frac{4}{3}\alpha \sum \limits_n  (\Delta E^p_n)^3 \left(1+n_{\textrm{B}}( \Delta E^p_n)\right)
 \left( |\langle n|\bm{r}|\bm{p}\rangle_{\textrm{lab}}|^2 F^n_1(p,P) + \left|\langle n|\bm{r}\cdot \frac{\bm{P}}{2M}|\bm{p}\rangle_{\textrm{lab}}\right|^2 F^n_2(p,P)\right) ,
\label{bsf_xsection_final_result}
\end{aligned}
\end{equation}
with
\begin{equation}
\begin{aligned}
  F^n_1(p,P) =&\,  1 - \frac{3}{4}\frac{\Delta E^p_n }{M}+\frac{\bm{P}^2}{4M^2}\\
  &\, + n_{\textrm{B}}( \Delta E^p_n) \,\frac{\Delta E^p_n}{4M} \, \frac{\Delta E^p_n}{T}\\
  &\, - n_{\textrm{B}}( \Delta E^p_n) \, \frac{\bm{P}^2}{4M^2}
     \frac{\Delta E^p_n}{T} \left[ 1 - \frac{\Delta  E^p_n}{5T}  - \frac{2}{5} n_{\textrm{B}}( \Delta E^p_n) \frac{\Delta E^p_n}{T} \right],
  \label{bsf_xsection_final_result_F1}
\end{aligned}
\end{equation}
and
\begin{equation}
  F^n_2(p,P) =  1 - \frac{1}{10} \, n_{\textrm{B}}( \Delta E^p_n) \, \frac{(\Delta E^p_n)^2}{T^2} \, \left( 1+ 2 \, n_{\textrm{B}}( \Delta E^p_n) \right).
\label{bsf_xsection_final_result_F2}
\end{equation}
Note that $\Delta E^p_n/T$ may be of order one according to our hierarchy of energy scales \eqref{scale_arrang}.
The statistical factor $1+n_{\textrm{B}}( \Delta E^p_n)$ in \eqref{bsf_xsection_final_result} reflects the fact the the dark photon is emitted into the thermal bath.

\subsection{Formation of bound states in the center-of-mass frame}
\label{sec:bsf_cm}
In this section, we consider the dark-matter pair at rest, while the thermal bath moves with constant velocity $-\bm{v}$.\footnote{
A general formula for the center-of-mass velocity of the heavy pair with respect to the moving thermal medium as seen from a  generic laboratory frame is given by~\cite{Escobedo:2013tca}
$$
\bm{v} = \frac{-P^0\bm{w}+\frac{\bm{P}\cdot \bm{w}}{\bm{w}^2}\bm{w}+\left(\bm{P}-\frac{\bm{P}\cdot \bm{w}}{\bm{w}^2}\bm{w}\right)\sqrt{1-\bm{w}^2}}{P^0-\bm{P}\cdot\bm{w}} \,,
$$
where $P^0$ and $\bm{P}$ are the total energy and center-of-mass momentum of the dark-matter pair with respect to the laboratory frame, respectively,
and $\bm{w}$ is the velocity of the thermal medium with respect to the laboratory frame.
If the laboratory frame coincides with the center-of-mass frame of the pair, then $\bm{P}=0$ and $\bm{v}=-\bm{w}$.
Instead, if the laboratory frame coincides with the frame where the medium is at rest, then $\bm{w}=0$ and $\bm{v}=\bm{P}/P^0$.
\label{velocities}}
We provide the bound-state formation cross section in the center-of-mass frame of the dark-matter fermion-antifermion pair.

First, we consider the vacuum case. 
The bound-state formation cross section in the laboratory frame is given by eq.~\eqref{bsf_xsection_final_result} with the Bose--Einstein distribution,
$n_{\textrm{B}}$, set to zero; in particular, in the vacuum case we have that 
$\displaystyle F^n_1(p,P) =  1 - \frac{3}{4}\frac{\Delta E^p_n }{M}+\frac{\bm{P}^2}{4M^2}$ and $F^n_2(p,P)~=~1$.
The cross section in the center-of-mass frame follows from setting $\bm{P}=\bm{0}$, 
\begin{equation}
(\sigma_{\hbox{\scriptsize bsf}} \, v_{\hbox{\scriptsize M\o l}})_{\textrm{cm}}(\bm{p}) = \frac{4}{3}\alpha \sum \limits_n  (\Delta E^p_n)^3_{\textrm{cm}} 
|\langle n|\bm{r}|\bm{p}\rangle_{\textrm{cm}}|^2 \left(  1 - \frac{3}{4}\frac{(\Delta E^p_n)_{\textrm{cm}} }{M} \right),
\label{bsf_xsection_final_result_cm_vacuum}
\end{equation}
where we have made explicit in which reference frame the matrix element and the energy difference are computed. 
Here and in the following, the relative momentum, $\bm{p}$, and distance, $\bm{r}$, in quantities marked with the subscript cm are to be understood as measured in the center-of-mass frame.
From eq.~\eqref{dipolenp} it follows that the dipole matrix elements transform from the center-of-mass reference frame to the laboratory frame as 
\begin{equation}
\begin{aligned}  
  &\left|\langle n|\bm{r}|\bm{p}\rangle_{\textrm{cm}}\right|^2 = \left|\langle n|\bm{r}|\bm{p}\rangle_{\textrm{lab}}\right|^2\left(1+\frac{\bm{v}^2}{2} \right) + \left|\langle n|\bm{r}\cdot\bm{v}|\bm{p}\rangle_{\textrm{lab}}\right|^2 , \\
  &\left|\langle n|\bm{r}\cdot\bm{v}|\bm{p}\rangle_{\textrm{cm}}\right|^2 = \left|\langle n|\bm{r}\cdot  \bm{v} |\bm{p}\rangle_{\textrm{lab}}\right|^2,
\label{Lorentz_trafo_matrix_elements}
\end{aligned}
\end{equation}
when keeping only corrections up to order $\bm{v}^2$.
From eq.~\eqref{DeltaEgamma} it follows that the energy difference transforms up to order $\bm{v}^2$ as
\begin{equation}
\begin{aligned}
      (\Delta E^{p}_n)_{\textrm{cm}} = \gamma (\Delta E^p_n)_{\textrm{lab}} = (\Delta E^p_n)_{\textrm{lab}} \left(1+\frac{\bm{v}^2}{2}\right) .
\label{Lorentz_trafo_energy}
\end{aligned}
\end{equation}
Plugging eqs.~\eqref{Lorentz_trafo_matrix_elements} and~\eqref{Lorentz_trafo_energy} into eq.~\eqref{bsf_xsection_final_result_cm_vacuum},
setting $\bm{v} = \bm{P}/(2M)$, and keeping only terms up to order $\bm{P}^2/M^2$, we get 
\begin{equation}
 (\sigma_{\hbox{\scriptsize bsf}} \, v_{\hbox{\scriptsize M\o l}})_{\textrm{lab}}(\bm{p},\bm{P}) = \left(1 - \frac{\bm{P}^2}{4M^2} \right) (\sigma_{\hbox{\scriptsize bsf}} \, v_{\hbox{\scriptsize M\o l}})_{\textrm{cm}}(\bm{p}) \,.
\label{Lorentz_boost}
\end{equation}
This relation is consistent with the discussion in section~\ref{sec:ann}: the in-vacuum cross section is Lorentz invariant
and the transformation \eqref{Lorentz_boost} just reflects the Lorentz transformation \eqref{MolLor} of the M\o ller velocity.

We consider now the  case of a moving thermal bath.
The Bose--Einstein distribution for thermal dark photons in the moving bath reads~\cite{Weldon:1982aq,Escobedo:2011ie}
\begin{equation}
\begin{aligned}
    n_{\textrm{B}}(|k^{\mu}u_{\mu}|) = \frac{1}{e^{|k^{\mu}u_{\mu}|/T}-1} \, ,
\label{bose_fkt}
\end{aligned}
\end{equation}
where $u^\mu = (1,-\bm{v})\,\gamma$, and, as in the rest of the paper, $\gamma = 1/\sqrt{1-\bm{v}^2}$ is the Lorentz factor.
In the laboratory frame, where the bath is at rest ($\bm{v}=0$), we have $k^{\mu}u_{\mu} = k^0$, and the distribution \eqref{bose_fkt} reduces to \eqref{BoseEinstein}.
For on-shell thermal dark photons from the bath, we can write
\begin{equation}
\begin{aligned}
     \frac{k^{\mu}u_{\mu}}{T} = \frac{k^0 -|\bm{v}||\bm{k}|\cos{\theta}}{T\sqrt{1-\bm{v}^2}} = \frac{|\bm{k}|}{T}\frac{1-|\bm{v}|\cos{\theta}}{\sqrt{1-\bm{v}^2}} \equiv \frac{|\bm{k}|}{T_{\textrm{eff}}}    \, ,
\end{aligned}
\end{equation}
where $\theta$ is the angle between the medium velocity $-\bm{v}$ and the dark photon momentum $\bm{k}$.
The \textit{effective temperature}, $T_{\textrm{eff}}$, is defined as~\cite{Costa:1995yv}
\begin{equation}
\begin{aligned}
     T_{\textrm{eff}}(|\bm{v}|,\theta) = \frac{T\sqrt{1-\bm{v}^2}}{1-|\bm{v}|\cos{\theta}} \,.
\label{bose_fkt_argument}
\end{aligned}
\end{equation}
It may be understood as the temperature experienced by an observer at rest;
it is different from $T$ because of the Doppler effect.\footnote{
Depending on the angle $\theta$, i.e. whether the medium moves towards the observer $(0\leq \theta < \pi/2)$ or away from the observer $(\pi/2 < \theta \leq \pi)$,
the temperature measured by the observer is larger or smaller than $T$, $T$ being the temperature of the thermal bath in the medium rest frame.
The maximum and minimum temperature is for $\theta=0$ and $\theta=\pi$, respectively,
$$
T_{\textrm{max}} = T\sqrt{\frac{1+|\bm{v}|}{1-|\bm{v}|}} \, ,  \qquad\quad T_{\textrm{min}} = T\sqrt{\frac{1-|\bm{v}|}{1+|\bm{v}|}} \, .
$$}
Expanding the distribution function \eqref{bose_fkt} for small medium velocities $|\bm{v}|\ll 1$ up to order $\bm{v}^2$, we get\footnote{
Because of the hierarchy \eqref{scale_arrang}, we consider here only the case of a thermal bath moving at small velocity.
For the case of a thermal bath of photons moving at high velocity, see ref.~\cite{Escobedo:2011ie}.} 
\begin{equation}
   n_{\textrm{B}}(|k^{\mu}u_{\mu}|) = n_{\textrm{B}}(|\bm{k}|)
  \left[ 1+ \left(1+n_{\textrm{B}}(|\bm{k}|)\right) \left(   -\frac{\bm{v}\cdot \bm{k}}{T}-\frac{|\bm{k}|}{T}\frac{\bm{v}^2}{2}  + \frac{(\bm{k} \cdot \bm{v})^2}{2 T^2} (2n_{\textrm{B}}(|\bm{k}|) +1) \right) \right].
\label{bose_fkt_expanded}
\end{equation}

In the center-of-mass frame, the dark fermion-antifermion pair recoils by a spatial momentum $-\bm{k}$ when emitting a photon of spatial momentum $\bm{k}$.
The resulting propagator, when the incoming pair is in a scattering state and the outcoming one in a bound state, can be expanded in the center-of-mass kinetic energy $\bm{k}^2/(4M)$,
which is suppressed by $T/M$ with respect to $\Delta E^{p}_n$ and $T$, leading to
\begin{equation}
    \frac{i}{(\Delta E^p_n)_{\textrm{cm}} - \bm{k}^2/(4M) -k_0+i\epsilon}=\frac{i}{(\Delta E^p_n)_{\textrm{cm}} -k_0+i\epsilon} +\frac{i}{((\Delta E^p_n)_{\textrm{cm}} - k_0+i\epsilon)^2}\frac{\bm{k}^2}{4M} + \dots \, .
\end{equation} 
Higher-order terms are beyond our accuracy. 
The bound-state formation cross section in the center-of-mass frame up to relative order $\bm{v}^2$, which in our case is about $T/M$, and $\Delta E^p_n/M$ depends on the self-energy diagrams $a)$, $b)$, $g)$ and $h)$ of figure~\ref{fig:pnEFT_DM_self}.
Crosses stand for insertions of the recoil correction $\bm{k}^2/(4M)$ in the non-recoiling propagator $i/((\Delta E^p_n)_{\textrm{cm}} -k_0+i\epsilon)$.
Hence, from the optical theorem it follows that
\begin{equation}
\begin{aligned}
&  (\sigma_{\hbox{\scriptsize bsf}} \, v_{\hbox{\scriptsize M\o l}})_{\textrm{cm}}(\bm{p},\bm{v}) \equiv  \sum \limits_n   (\sigma^n_{\hbox{\scriptsize bsf}} \, v_{\hbox{\scriptsize M\o l}})_{\textrm{cm}}(\bm{p},\bm{v}) = \\
& \hspace{0.8cm} -2\,\textrm{Im}\left[ \left(\Sigma^{11}_{\textrm{diag.}\,\textit{a)}}\right)(\bm{p},\bm{v}) + \left(\Sigma^{11}_{\textrm{diag.}\,\textit{b)}}\right)(\bm{p},\bm{v}) + \left(\Sigma^{11}_{\textrm{diag.}\,\textit{g)}}\right)(\bm{p},\bm{v}) + \left(\Sigma^{11}_{\textrm{diag.}\,\textit{h)}}\right)(\bm{p},\bm{v})
  \right] .
\end{aligned}
\end{equation}
After explicit calculation we get
\begin{equation}
\begin{aligned}
 & (\sigma_{\hbox{\scriptsize bsf}} \, v_{\hbox{\scriptsize M\o l}})_{\textrm{cm}}(\bm{p},\bm{v}) = \\
 & \hspace{2mm} \frac{4}{3}\alpha \sum \limits_n  (\Delta E^p_n)^3_{\textrm{cm}} \left(1+n_{\textrm{B}}( (\Delta E^p_n)_{\textrm{cm}})\right)
 \left( |\langle n|\bm{r}|\bm{p}\rangle_{\textrm{cm}}|^2 \tilde{F}^{n}_1(p,v) + \left|\langle n|\bm{r}\cdot \bm{v}|\bm{p}\rangle_{\textrm{cm}}\right|^2 \tilde{F}^{n}_2(p,v)\right) ,
\label{bsf_xsection_bath_rest}
\end{aligned}
\end{equation}
with
\begin{equation}
\begin{aligned}
  \tilde{F}^{n}_1(p,v) =&\,  1 - \frac{3}{4}\frac{(\Delta E^p_n )_{\textrm{cm}}}{M}\\
  & + n_{\textrm{B}}((\Delta E^p_n)_{\textrm{cm}}) \,\frac{(\Delta E^p_n)_{\textrm{cm}}}{4M} \, \frac{(\Delta E^p_n)_{\textrm{cm}}}{T}\\
  & - n_{\textrm{B}}((\Delta E^p_n)_{\textrm{cm}}) \, \bm{v}^2 \frac{(\Delta E^p_n)_{\textrm{cm}}}{T}
\left[ \frac{1}{2} - \frac{(\Delta  E^p_n)_{\textrm{cm}}}{5T}  - \frac{2}{5} n_{\textrm{B}}((\Delta E^p_n)_{\textrm{cm}}) \frac{(\Delta E^p_n)_{\textrm{cm}}}{T} \right],
\label{bsf_xsection_bath_rest_F1}
\end{aligned}
\end{equation}
and
\begin{equation}
  \tilde{F}^{n}_2(p,v) =  - \frac{1}{10} \, n_{\textrm{B}}((\Delta E^p_n)_{\textrm{cm}}) \, \frac{(\Delta E^p_n)^2_{\textrm{cm}}}{T^2} \, \left( 1+ 2 \, n_{\textrm{B}}((\Delta E^p_n)_{\textrm{cm}}) \right),
\label{bsf_xsection_bath_rest_F2}
\end{equation}
where we have made explicit in which reference frame the matrix elements and the energy difference are computed,
and we have neglected relative corrections smaller than $\bm{v}^2$ and $\Delta E^p_n/M$.

The relation between the bound-state formation cross section in the laboratory frame \eqref{bsf_xsection_final_result} and the bound-state formation cross section in the center-of-mass frame \eqref{bsf_xsection_bath_rest} is 
\begin{equation}
  (\sigma_{\hbox{\scriptsize bsf}} \, v_{\hbox{\scriptsize M\o l}})_{\textrm{lab}}(\bm{p},\bm{P}) = 
  (\sigma_{\hbox{\scriptsize bsf}} \, v_{\hbox{\scriptsize M\o l}})_{\textrm{cm}}(\bm{p},\bm{P}/(2M)) \left(1 - \frac{\bm{P}^2}{4M^2} \right)\,,
\label{Lorentz_boost2}
\end{equation}
if we transform the matrix elements according to eqs.~\eqref{Lorentz_trafo_matrix_elements}, the energy difference according to eq.~\eqref{Lorentz_trafo_energy}, set $\bm{v} = \bm{P}/(2M)$ and keep only terms up to order $\bm{P}^2/M^2$.
This is the same relation as in the vacuum case and the same comments apply.  We remind that the relative momentum in $(\sigma_{\hbox{\scriptsize
bsf}} \, v_{\hbox{\scriptsize M\o l}})_{\textrm{lab}}(\bm{p},\bm{P})$ is measured
in the laboratory frame, the one in $ (\sigma_{\hbox{\scriptsize bsf}} \, v_{\hbox{\scriptsize M\o
l}})_{\textrm{cm}}(\bm{p},\bm{P}/(2M))$ is measured in the center-of-mass
frame and the center-of-mass momentum $\bm{P}$ is measured in the
laboratory frame.

\section{Bound-state dissociation}
\label{sec:bsd}
Bound-state dissociation happens when a bound state $(X\bar{X})_n$ absorbs a thermal dark photon from the bath and dissociates into a scattering state  $(X\bar{X})_p$ through the reaction
\begin{equation}
\gamma + (X\bar{X})_n\to (X\bar{X})_p\,.
\label{thermaldiss}
\end{equation}
The bound-state dissociation width
can be determined from the imaginary parts of the self-energy diagrams shown in figure~\ref{fig:pnEFT_DM_self} with the propagators of the scattering states (double line) and bound states (single line) exchanged,
because now the incoming and outgoing pair is bound, while the pair in the loop is unbound.
We project the self energies onto bound states with quantum numbers $n$ and center-of-mass momentum $\bm{P}$ in the laboratory frame, and label them accordingly.
The dissociation width can then be computed in the laboratory frame, up to corrections of relative order $T/M$ and $\Delta E^p_n/M$, as 
\begin{equation}
\begin{aligned}
  & (\Gamma^{n}_{\textrm{bsd}})_{\textrm{lab}}(\bm{P}) = \\
  & \hspace{0.8cm} -2\,\textrm{Im}\left[ \left(\Sigma^{11}_{\textrm{diag.}\,\textit{a)}}\right)(n) + \left(\Sigma^{11}_{\textrm{diag.}\,\textit{b)}}\right)(n)
    +\left(\Sigma^{11}_{\textrm{diag.}\,\textit{c)}}\right)
    (n,\bm{P})  + \left(\Sigma^{11}_{\textrm{diag.}\,\textit{d)}}\right)
    (n,\bm{P})  \right.\\
   &\hspace{1.8cm} \left. + \left(\Sigma^{11}_{\textrm{diag.}\,\textit{e)}}\right)
    (n,\bm{P})
    + \left(\Sigma^{11}_{\textrm{diag.}\,\textit{f)}}\right)
    (n,\bm{P})  + \left(\Sigma^{11}_{\textrm{diag.}\,\textit{g)}}\right)(n) + \left(\Sigma^{11}_{\textrm{diag.}\,\textit{h)}}\right) (n)
        \right] 
    .
\label{bsd_width}
\end{aligned}
\end{equation}
The propagator of the recoiling unbound fermion-antifermion pair in the loop reads
\begin{equation}
\begin{aligned}
 & \frac{i}{E_n + \bm{P}^2/(4M) - (E_p + (\bm{P}-\bm{k})^2/(4M)) -k_0+i\epsilon} =\\
 &\hspace{7cm}  \frac{i}{-\Delta E^p_n + {(2\bm{P}\cdot \bm{k}-\bm{k}^2)}/{(4M)}-k_0+i\epsilon}\,,
\label{propagator_loop}
\end{aligned}
\end{equation}
where we notice the sign difference in front of $\Delta E^p_n$ with respect to eq.~\eqref{recoilprop}.
The propagator is expanded in the recoil correction to the kinetic energy, which generates the diagrams with crosses;
the non-recoiling propagator is $i/(-\Delta E^p_n -k_0+i\epsilon)$.

The computation of \eqref{bsd_width} goes like the one of \eqref{bsf_xsection}, and may be derived from that one by considering the different initial and intermediate states, 
and keeping track of the different sign in front of the energy difference $\Delta E^p_n$.
Including all corrections of order $\bm{P}^2/M^2 \sim T/M$ and $\Delta E^p_n/M$, the bound-state dissociation width in the laboratory frame reads
\begin{equation}
\begin{aligned}
  & (\Gamma^{n}_{\textrm{bsd}})_{\textrm{lab}}(\bm{P})
  = \\
 & \hspace{2mm} \frac{4}{3}\alpha  \int \frac{d^3p}{(2\pi)^3}  (\Delta E^p_n)^3 n_{\textrm{B}}( \Delta E^p_n)
 \left( |\langle n|\bm{r}|\bm{p}\rangle_{\textrm{lab}}|^2 D^n_1(p,P) + \left|\langle n|\bm{r}\cdot \frac{\bm{P}}{2M}|\bm{p}\rangle_{\textrm{lab}}\right|^2 D^n_2(p,P)\right) ,
\label{bsd_width_final_result}
\end{aligned}
\end{equation}
with
\begin{equation}
\begin{aligned}
  D^n_1(p,P) =&\,  1 + \frac{3}{4}\frac{\Delta E^p_n }{M}+\frac{\bm{P}^2}{4M^2}\\
  & - (1+n_{\textrm{B}}( \Delta E^p_n)) \,\frac{\Delta E^p_n}{4M} \, \frac{\Delta E^p_n}{T}\\
  & - (1+n_{\textrm{B}}( \Delta E^p_n)) \, \frac{\bm{P}^2}{4M^2}
  \frac{\Delta E^p_n}{T} \left[ 1 + \frac{\Delta  E^p_n}{5T}  - \frac{2}{5} (1+n_{\textrm{B}}( \Delta E^p_n)) \frac{\Delta E^p_n}{T} \right],
  \label{bsd_width_final_result_D1}
\end{aligned}
\end{equation}
and
\begin{equation}
  D^n_2(p,P) =  1 - \frac{1}{10} \, (1+n_{\textrm{B}}( \Delta E^p_n)) \, \frac{(\Delta E^p_n)^2}{T^2} \, \left( 1+ 2 \, n_{\textrm{B}}( \Delta E^p_n) \right).
\label{bsd_width_final_result_D2}
\end{equation}
The statistical factor $n_{\textrm{B}}( \Delta E^p_n)$ in \eqref{bsd_width_final_result} reflects the fact the the dark photon is absorbed from the thermal bath.
The dissociation width does not contain a vacuum part because 
bound-state dissociation is kinematically forbidden in vacuum.
Hence, the bound-state dissociation width is a purely thermal width.

Proceeding like in section~\ref{sec:bsf_cm}, we compute from the diagrams in figure~\ref{fig:pnEFT_DM_self} that do not vanish for $\bm{P}=\bm{0}$ (diagrams $a)$, $b)$, $g)$ and $h)$) 
the bound-state dissociation width in the center-of-mass frame.
There, the center of mass of the dark fermion-antifermion pair is at rest and the thermal bath is moving with velocity $-\bm{v}$.
At relative order $\bm{v}^2$ and $\Delta E^p_n/M$, we get
\begin{equation}
\begin{aligned}
 & (\Gamma^{n}_{\textrm{bsd}})_{\textrm{cm}}(\bm{v}) = \\
 & \hspace{2mm} \frac{4}{3}\alpha \int \frac{d^3 p }{(2\pi)^3} (\Delta E^p_n)^3_{\textrm{cm}} \,n_{\textrm{B}}( (\Delta E^p_n)_{\textrm{cm}})
 \left( |\langle n|\bm{r}|\bm{p}\rangle_{\textrm{cm}}|^2 \tilde{D}^{n}_1(p,v) + \left|\langle n|\bm{r}\cdot \bm{v}|\bm{p}\rangle_{\textrm{cm}}\right|^2 \tilde{D}^{n}_2(p,v)\right) ,
\label{bsd_width_bath_rest}
\end{aligned}
\end{equation}
with
\begin{equation}
\begin{aligned}
  \tilde{D}^{n}_1(p,v) =&\,  1 + \frac{3}{4}\frac{(\Delta E^p_n )_{\textrm{cm}}}{M}\\
  & -(1+n_{\textrm{B}}((\Delta E^p_n)_{\textrm{cm}})) \,\frac{(\Delta E^p_n)_{\textrm{cm}}}{4M} \, \frac{(\Delta E^p_n)_{\textrm{cm}}}{T}\\
  & -(1+n_{\textrm{B}}((\Delta E^p_n)_{\textrm{cm}})) \, \bm{v}^2 \frac{(\Delta E^p_n)_{\textrm{cm}}}{T}
\left[ \frac{1}{2} + \frac{(\Delta  E^p_n)_{\textrm{cm}}}{5T}  \right.\\
&\hspace{6cm}
\left.
- \frac{2}{5}  \left( 1 + n_{\textrm{B}}((\Delta E^p_n)_{\textrm{cm}})\right) \frac{(\Delta E^p_n)_{\textrm{cm}}}{T} \right]\!,
\label{bsd_width_bath_rest_D1}
\end{aligned}
\end{equation}
and
\begin{equation}
  \tilde{D}^{n}_2(p,v) =  - \frac{1}{10} \, (1+n_{\textrm{B}}((\Delta E^p_n)_{\textrm{cm}})) \, \frac{(\Delta E^p_n)^2_{\textrm{cm}}}{T^2} \, \left( 1+ 2 \, n_{\textrm{B}}((\Delta E^p_n)_{\textrm{cm}}) \right),
\label{bsd_width_bath_rest_D2}
\end{equation}
where we have made explicit in which reference frame the matrix elements and the energy difference are computed. 
The momentum integral in~\eqref{bsd_width_bath_rest} is over the relative momentum in the center-of-mass frame.

The relation between the bound-state dissociation width in the laboratory frame \eqref{bsd_width_final_result} and the bound-state dissociation width in the center-of-mass frame \eqref{bsd_width_bath_rest} is 
\begin{equation}
 (\Gamma^{n}_{\textrm{bsd}})_{\textrm{lab}}(\bm{P}) = \frac{(\Gamma^{n}_{\textrm{bsd}})_{\textrm{cm}}(\bm{v})}{\gamma} \approx (\Gamma^{n}_{\textrm{bsd}})_{\textrm{cm}}(\bm{P}/(2M)) \left(1 - \frac{\bm{P}^2}{8M^2} \right)\,,
\label{Lorentz_boost3}
\end{equation}
if we transform the matrix elements according to eqs.~\eqref{Lorentz_trafo_matrix_elements}, the energy difference according to eq.~\eqref{Lorentz_trafo_energy},
the momentum-space volume as
$d^3p_{\textrm{cm}} = d^3p_{\textrm{lab}}/\gamma$ (see eq.~\eqref{d3pgamma}), set $\bm{v} = \bm{P}/(2M)$ and keep only terms up to order $\bm{P}^2/M^2$.
Equation~\eqref{Lorentz_boost3} expresses the Lorentz dilation of time intervals.

\subsection{Bound-state dissociation for $T\gg \Delta E^p_n$}
Our result for the bound-state dissociation width in the rest frame of the pair, cf.~\eqref{bsd_width_bath_rest}, can be compared with the result obtained in ref.~\cite{Escobedo:2013tca}.
We compare in the case of temperatures much larger than the exchanged energy.
In this case, we can expand the right-hand side of~\eqref{bsd_width_bath_rest} in $(\Delta E^p_n)_{\textrm{cm}}/T \ll 1$ and approximate
\begin{equation}
    1 + n_{\textrm{B}}((\Delta E^p_n)_{\textrm{cm}}) \approx n_{\textrm{B}}((\Delta E^p_n)_{\textrm{cm}}) \approx \frac{T}{(\Delta E^p_n)_{\textrm{cm}}} \,.
\end{equation}
At leading order in $(\Delta E^p_n)_{\textrm{cm}}/T$ and at relative order $\bm{v}^2$ and $\Delta E^p_n/M$, we obtain 
\begin{equation}
  \begin{aligned}
 (\Gamma^{n}_{\textrm{bsd}})_{\textrm{cm}}(\bm{v})\Big|_{T\gg \Delta E^p_n} 
    &= \frac{4}{3}\alpha T \int \frac{d^3p}{(2\pi)^3} ((\Delta E^p_n)_{\textrm{cm}})^2 \\
&\times \left( |\langle n|\bm{r}|\bm{p}\rangle_{\textrm{cm}}|^2 \left[1+\frac{(\Delta E^p_n)_{\textrm{cm}}}{2M}-\frac{\bm{v}^2}{10}\right] -\frac{1}{5}|\langle n|\bm{r}\cdot \bm{v}|\bm{p}\rangle_{\textrm{cm}}|^2 \right)\,.
    \label{bsd_width_bath_rest_approx}
\end{aligned}
\end{equation}

For what concerns the result of ref.~\cite{Escobedo:2013tca}, 
we expand the abelian analogue of the bound-state self-energy expression in the case $T\gg \Delta E^p_n$ for small plasma velocities $\bm{v}$ and rewrite the matrix element of $(\bm{p}\cdot \bm{v})^2/M^2$ on bound states,
upon using the commutator $[H,r^i]=-2ip^i/M$  with $H= \bm{p}^2/M -\alpha/r$ and inserting a complete set of scattering eigenstates $\displaystyle \int \frac{d^3p}{(2\pi)^3}|\bm{p}\rangle \langle \bm{p}| = 1$, as
\begin{equation}
\begin{aligned}
    \frac{1}{M^2}\langle n|(\bm{p}\cdot\bm{v})^2|n\rangle = -\frac{1}{4}\langle n|([H,\bm{r}\cdot \bm{v}])^2|n\rangle = \frac{1}{4}\int \frac{d^3p}{(2\pi)^3} (\Delta E^p_n)^2|\langle n|\bm{r}\cdot \bm{v}|\bm{p}\rangle|^2 \, .
\label{expression1}
\end{aligned}
\end{equation}
Moreover, we rewrite the matrix element of $\bm{p}^2/M^2$ on bound states as
\begin{equation}
\begin{aligned}
    \frac{1}{M^2}\langle n|\bm{p}^2|n\rangle = \frac{1}{4}\langle n|r^i(E_n - H)^2 r^i|n\rangle = \frac{1}{4}\int \frac{d^3p}{(2\pi)^3} (\Delta E^p_n)^2|\langle n|\bm{r}|\bm{p}\rangle|^2  \, .
\label{expression2}
\end{aligned}
\end{equation}
In this way, we recover from~\cite{Escobedo:2013tca} our result~\eqref{bsd_width_bath_rest_approx}, up to the velocity-independent term proportional to $(\Delta E^p_n)_{\textrm{cm}}/(2M)$.

The reason for the discrepancy can be traced back to the fact that in~\cite{Escobedo:2013tca} only self-energy contributions involving electric-dipole vertices were considered.
They correspond to our diagrams $a)$, $b)$ in figure~\ref{fig:pnEFT_DM_self}.
However, as we have shown in this work, also diagrams $g)$ and $h)$ of figure~\ref{fig:pnEFT_DM_self}, involving a magnetic-dipole vertex, contribute at the same order.
This is due to the recoiling of the dark fermion-antifermion pair against the absorbed dark photon from the bath, which happens also in the center-of-mass frame.
Diagrams $g)$ and $h)$ provide exactly the missing contribution proportional to $(\Delta E^p_n)_{\textrm{cm}}/(2M)$.
This, after being added to the result from ref.~\cite{Escobedo:2013tca}, eventually gives back eq.~\eqref{bsd_width_bath_rest_approx}.

\section{Bound-state to bound-state transitions}
\label{sec:bound_bound}
Bound-state to bound-state transitions include \textit{de-excitations} of excited bound states into bound states of lower energy by emission of a dark photon, $(X\bar{X})_n\to \gamma + (X\bar{X})_{n'}$,
and \textit{excitations} of bound states into bound states of higher energy due to the absorption of a dark photon from the bath, $\gamma + (X\bar{X})_n\to (X\bar{X})_{n'}$.
The computation of the de-excitation transition width goes like the computation of the bound-state formation cross section done in section~\ref{sec:bsf},
whereas the computation of the excitation transition width goes like the computation of the bound-state dissociation width done in section~\ref{sec:bsd}.
The results may be read directly from the results listed in those sections by replacing the scattering state $|\bm{p}\rangle$ with the bound state $|n\rangle$ in the case of the de-excitation transition width
and with the bound state $|n'\rangle$ in the case of the excitation transition width.

The total de-excitation width in the laboratory frame, where the incoming bound state moves with center-of-mass momentum $\bm{P}$, reads up to relative order $T/M$ and $\Delta E^p_n/M$ 
\begin{eqnarray}
(\Gamma^{n}_{\textrm{de-ex.}})_{\textrm{lab}}(\bm{P}) \; &=& \!\! \sum_{n', E_{n'}<E_n} \!\!  (\Gamma^{n\rightarrow n'}_{\textrm{de-ex.}})_{\textrm{lab}}(\bm{P}) \nonumber\\
&=& \frac{4}{3}\alpha \!\! \sum \limits_{n', E_{n'}<E_n} (\Delta E^n_{n'})^3 \left(1+n_{\textrm{B}}(\Delta E^n_{n'})\right) \label{de-excitation_width}\\
&& \hspace{1cm}\times \left( |\langle n'|\bm{r}|n\rangle_{\textrm{lab}}|^2 F^{n'}_1(n,P) + \left|\langle n'|\bm{r}\cdot \frac{\bm{P}}{2M}|\bm{p}\rangle_{\textrm{lab}}\right|^2 F^{n'}_2(n,P)\right),
\nonumber
\end{eqnarray}
where the form factors $F^{n'}_1(n,P)$ and $F^{n'}_2(n,P)$ are defined as in eqs.~\eqref{bsf_xsection_final_result_F1} and~\eqref{bsf_xsection_final_result_F2}, respectively,
but with the energy difference  $\Delta E^p_{n}$ replaced by $\Delta E^n_{n'} = E_n-E_{n'}$, which in the center-of-mass frame is $(M\alpha^2/4)\left(1/{n'}^2-1/n^2\right)$.
In the center-of-mass frame, where the bath is moving with velocity $-\bm{v}$, the de-excitation width, $(\Gamma^{n}_{\textrm{de-ex.}})_{\textrm{cm}}(\bm{v})$,
has the same expression as \eqref{de-excitation_width}, but with $\bm{P}/(2M)$ replaced by $\bm{v}$ and form factors $\tilde{F}^{n'}_1(n,v)$ and $\tilde{F}^{n'}_2(n,v)$
defined as in eqs.~\eqref{bsf_xsection_bath_rest_F1} and~\eqref{bsf_xsection_bath_rest_F2}, respectively, but in terms of $(\Delta E^n_{n'})_{\textrm{cm}}$.

The  total excitation width in the laboratory frame reads up to relative order $T/M$ and $\Delta E^p_n/M$ 
\begin{eqnarray}
(\Gamma^{n}_{\textrm{ex.}})_{\textrm{lab}}(\bm{P}) \; &=& \!\! \sum_{n', E_{n'}>E_n} \!\! (\Gamma^{n\rightarrow n'}_{\textrm{ex.}})_{\textrm{lab}}(\bm{P}) \nonumber\\
&=&\frac{4}{3}\alpha \!\! \sum \limits_{n', E_{n'}>E_n} \, (\Delta E^{n'}_n)^3 \, n_{\textrm{B}}(\Delta E^{n'}_n) \, \label{excitation_width} \\
&& \hspace{1cm} \times \left( |\langle n'|\bm{r}|n\rangle_{\textrm{lab}}|^2 D^n_1(n',P) + \left|\langle n'|\bm{r}\cdot \frac{\bm{P}}{2M}|n \rangle_{\textrm{lab}}\right|^2 D^n_2(n',P)\right) ,
\nonumber
\end{eqnarray}
where the form factors $D^{n}_1(n',P)$ and $D^{n}_2(n',P)$ are defined as in eqs.~\eqref{bsd_width_final_result_D1} and~\eqref{bsd_width_final_result_D2}, respectively,
but with the energy difference  $\Delta E^p_{n}$ replaced by $\Delta E^{n'}_n = E_{n'}-E_n$.
In the center-of-mass frame, the excitation width, $(\Gamma^{n}_{\textrm{ex.}})_{\textrm{cm}}(\bm{v})$,
has the same expression as \eqref{excitation_width}, but with $\bm{P}/(2M)$ replaced by $\bm{v}$ and form factors $\tilde{D}^{n'}_1(n,v)$ and $\tilde{D}^{n'}_2(n,v)$
defined as in eqs.~\eqref{bsd_width_bath_rest_D1} and~\eqref{bsd_width_bath_rest_D2}, respectively, but in terms of $(\Delta E^{n'}_n)_{\textrm{cm}}$.

\subsection{Atomic clock}
The obtained results apply immediately to hydrogen-like atoms moving in a hot plasma after replacing the dark matter mass $M$ and coupling $\alpha$ 
with $2m_{e}$, where $m_e$ is the electron mass, and $\alpha_\text{em}$, the fine structure constant, respectively.
The results are specifically relevant for performing precision measurements or "calibrate" the atom in a thermal environment
using as an \textit{atomic clock} the time measured for a de-excitation or excitation transition between the ground state and an excited state. 
The lifetime, $\tau^n$,  of an excited state $n$ of an hydrogen-like atom
is inversely proportional to the de-excitation width given in eq.~\eqref{de-excitation_width},
hence the atomic clock is sensitive to the medium and, in particular, to its temperature and motion (see also~\cite{PhysRevLett.94.050404}).
Time dilation may be explicitly checked at hand of our expressions:
\begin{equation}
\begin{aligned}
  (\tau^{n})_{\textrm{lab}}(\bm{P})
  &= \frac{1}{(\Gamma^{n}_{\textrm{de-ex.}})_{\textrm{lab}}(\bm{P})} 
  = \frac{\gamma}{(\Gamma^{n}_{\textrm{de-ex.}})_{\textrm{cm}}(\bm{v})}
  = \gamma\,(\tau^{n})_{\textrm{cm}}(\bm{v})\\
  &\approx (\tau^{n})_{\textrm{cm}}(\bm{P}/(2M)) \left(1 + \frac{\bm{P}^2}{8M^2} \right)\,.
\label{Lorentz_boost4}
\end{aligned}
\end{equation}
In vacuum, time dilation due to the center-of-mass motion has been derived and studied in a quantum-mechanical framework in several recent papers~\cite{PhysRevLett.89.123001,James_D_Cresser_2003,Giacosa:2015mpm,PhysRevResearch.3.023053}.
To our knowledge, the derivation presented here is the first one done in a field theory framework.
The results for the in-medium effects are original.
Non-relativistic effective field theories at finite temperature may help in understanding how moving atoms behave in thermal environments,
aiding in the design of more robust and accurate atomic clocks~\cite{doi:10.1126/science.1192720,Lorek:2015rua,Nicholson_2015}.

\begin{figure}[ht]
    \centering
    \includegraphics[scale=0.75]{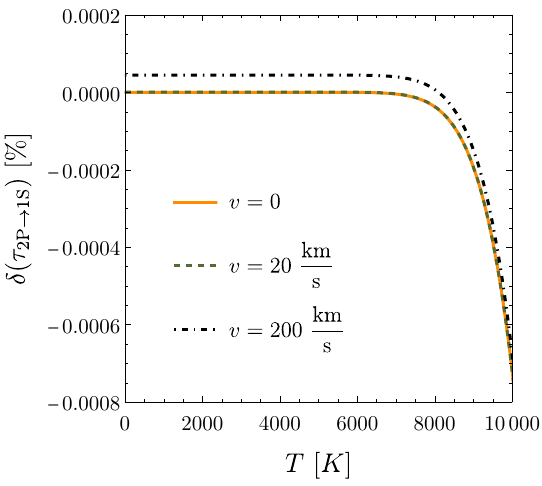}
        \hspace{1cm}
    \includegraphics[scale=0.7]{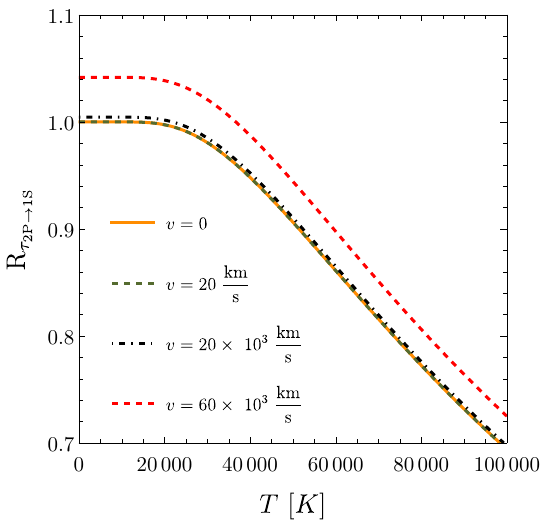}
    \caption{(Left) Relative correction (in percent) to the 2P lifetime due to the thermal medium and center-of-mass motion in the laboratory frame, plotted as a function of the thermal bath temperature in Kelvin. 
    The orange solid line is for the  2P center of mass at rest, the green dashed and black dash-dotted lines for center-of-mass velocities of 20 km/s and 200 km/s, respectively.
    (Right) The ratio $R_{\tau_{\textrm{2P}\rightarrow \textrm{1S}}} = (\tau^{\textrm{2P}\rightarrow \textrm{1S}})_{\textrm{lab}}(\bm{P})/(\tau^{\textrm{2P}\rightarrow \textrm{1S}})_{\textrm{vac}}$ as a function of the temperature up to $10^5$ Kelvin.}
    \label{plotatomicclock}
\end{figure}

As an example in atomic physics that may also be of some relevance in  astrophysics, we consider the Lyman-$\alpha$ (Ly-$\alpha$) transition in a neutral hydrogen
atom in thermal equilibrium with a photon gas and moving with respect to it with velocity $\bm{v} = \bm{P}/(2M)$.
The Ly-$\alpha$ transition is the (de-)excitation transition between the bound state 2P and the lowest energy configuration 1S \cite{Weinberg:2003eg,McQuinn:2015icp}. 
We would like to determine quantitatively the corrections to the lifetime due to the thermal bath and the center-of-mass recoil. 
In figure~\ref{plotatomicclock}, left, we plot the relative correction $\delta(\tau_{\textrm{2P}\rightarrow \textrm{1S}}) \equiv ((\tau^{\textrm{2P}\rightarrow \textrm{1S}})_{\textrm{lab}}(\bm{P})-(\tau^{\textrm{2P}\rightarrow \textrm{1S}})_{\textrm{vac}})$ $/$ $(\tau^{\textrm{2P}\rightarrow \textrm{1S}})_{\textrm{vac}}$ (in percent),
where $(\tau^{\textrm{2P}\rightarrow \textrm{1S}})_{\textrm{lab}}(\bm{P})$ is defined in~\eqref{Lorentz_boost4} and $(\tau^{\textrm{2P}\rightarrow \textrm{1S}})_{\textrm{vac}}$ is the lifetime of a 2P state at rest in vacuum,\footnote{
In vacuum and at rest, the lifetime of the 2P state is $(\tau^{\textrm{2P}\rightarrow \textrm{1S}})_{\textrm{vac}} = 1/(\Gamma^{2\textrm{P}\rightarrow 1\textrm{S}}_{\textrm{de-ex.}})_{\textrm{vac}} \approx 1.6 \cdot 10^{-9}$ s,
where at leading order $(\Gamma^{2\textrm{P}\rightarrow 1\textrm{S}}_{\textrm{de-ex.}})_{\textrm{vac}}=2^8 m_e\alpha_\text{em}^5/3^8$, $m_e=511$~keV, and $\alpha_\text{em} \approx 1/137$ is the fine-structure constant in QED.
In our plots, $(\Gamma^{2\textrm{P}\rightarrow 1\textrm{S}}_{\textrm{de-ex.}})_{\textrm{vac}}$ also accounts for recoil corrections affecting the 1S state in the center-of-mass frame of the 2P state.} for a 2P state at rest (orange solid line),
moving with center-of-mass velocity 20 km/s (green dashed line) and with center-of-mass velocity 200 km/s (black dash-dotted line) as a function of the medium temperature (in Kelvin). 
We observe that up to temperatures of the order of 8000 Kelvin the relative correction is a positive constant with a value up to $0.5 \times 10^{-4}$\% and increases with increasing center-of-mass velocity, reflecting the time dilation phenomenon.
With increasing temperature up to $10^5$ K ($\approx$ 8.6 eV), the finite temperature effects cancel and eventually dominate over the recoil corrections and decrease the lifetime, since the de-excitation width increases  due to stimulated emission. 

In the right panel of figure~\ref{plotatomicclock}, it can be seen that for center-of-mass velocities up to $20 \times 10^3$ km/s (black dash-dotted line)
the ratio $R_{\tau_{\textrm{2P}\rightarrow \textrm{1S}}} \equiv (\tau^{\textrm{2P}\rightarrow \textrm{1S}})_{\textrm{lab}}(\bm{P})/(\tau^{\textrm{2P}\rightarrow \textrm{1S}})_{\textrm{vac}}$ reduces significantly up to 30\% with temperature, but is rather insensitive to the recoil corrections. 
The ratio increases significantly for larger center-of-mass velocities; see, for instance, the red dashed line, which  shows the ratio for $v=60\times 10^3$ km/s $\approx$ 20\% of the velocity of light. 
Velocities close to the velocity of light, however, eventually slow/break down the non-relativistic expansion on which the calculation is based.

Similarly, one can study the opposite case, namely a hydrogen atom that absorbs Ly-$\alpha$ radiation and hence  becomes excited. 
Corrections from the center-of-mass recoil and the thermal medium are of comparable size as in  the de-excitation process.

\section{Recoil corrections in a non-abelian SU($N$) model}
\label{sec:non_abelian_model}
In this section, we consider dark matter fermions, $X$, that are in the fundamental representation of a non-abelian group SU($N$) with $N\ge 2$.
The SU($N$) gauge invariant Lagrangian density with dark fermions $X$ has the form  
\begin{equation}
    \mathcal{L}=\bar{X} (i \slashed {D} -M) X -\frac{1}{4} G^a_{\mu \nu} G^{a\,\mu \nu} + \mathcal{L}_{\textrm{portal}} \, ,
    \label{non_ab_model}
\end{equation}
where $D_\mu=\partial_\mu+i g A_\mu^a T^a$, $T^a$ are the SU($N$) generators, $A_\mu^a$ the SU($N$) dark gauge fields and $G^a_{\mu \nu}$ the corresponding field strength tensor. 
As in the abelian case, we neglect for the present discussion the portal interactions.\footnote{
At variance with the abelian model of section~\ref{sec:NREFTs}, the kinetic mixing with the Standard Model is not a viable portal interaction because of gauge invariance.
The common practice encompasses different solutions, such as the inclusion of non-renormalizable operators \cite{Juknevich:2009ji,Juknevich:2009gg,Boddy:2014yra}, a Higgs portal \cite{Cline:2013zca}, or the introduction of vectorlike fermions \cite{Juknevich:2009gg}.
We assume that these interactions allow for keeping the dark sector and the Standard Model at the same temperature and that, at the same time, the corresponding couplings are much smaller than the dark gauge coupling.
We also assume that the additional dark-sector degrees of freedom are much heavier than the dark matter particles and do not participate in the freeze-out process.} 
The extension of the abelian results to the dark non-abelian model based on the gauge group SU($N$) is straightforward if we assume that the multipole expansion holds
and the coupling constant is sufficiently small to allow for a perturbative treatment also at the ultrasoft scale~\cite{Biondini:2023zcz}.
This happens if we are in the energy regime $M \gg M \alpha \gtrsim \sqrt{MT} \gg M \alpha^2 \gtrsim T \gg \Lambda$, where $\Lambda$ denotes the non perturbative scale at which a weak-coupling expansion in $\alpha \equiv g^2/(4\pi)$ breaks down.
The SU(3) case, which corresponds to QCD, is relevant to describe the effects of a moving quark-gluon plasma on quarkonium formation and dissociation in heavy-ion collisions~\cite{Song:2007gm,Escobedo:2013tca}.

The spin- and color-averaged annihilation cross section in the laboratory frame up to second order in the center-of-mass velocity of the annihilating pair is given by
\begin{eqnarray}
(\sigma_{\hbox{\scriptsize ann}} \, v_{\hbox{\scriptsize M\o l}})_{\textrm{lab}}(\bm{p},\bm{P}) &=&
  \frac{1}{N^2} \left[(\sigma^{\hbox{\tiny NR}}_{\hbox{\scriptsize ann}} v^{(0)}_{\hbox{\scriptsize rel}})^{[\mathds{1}]} \, \left(|\Psi^{[\mathds{1}]}_{\bm{p}}(\boldsymbol{0})|^2\right)_{\textrm{lab}}(\bm{P}) \right.\nonumber\\
&&\hspace{1cm}
\left. + (N^2-1) (\sigma^{\hbox{\tiny NR}}_{\hbox{\scriptsize ann}} v^{(0)}_{\hbox{\scriptsize rel}})^{[\textbf{adj}]} \,  \left(|\Psi^{[\textbf{adj}]}_{\bm{p}}(\boldsymbol{0})|^2\right)_{\textrm{lab}}(\bm{P})\right]\left(1- \frac{\bm{P}^2}{4M^2}\right) \nonumber\\
&=& (\sigma_{\hbox{\scriptsize ann}} \, v_{\hbox{\scriptsize M\o l}})_{\textrm{cm}}(\bm{p})\left(1- \frac{\bm{P}^2}{4M^2}\right),
\label{total_non_abelian_ann_colored_av}
\end{eqnarray}
where the free singlet and adjoint annihilation cross sections up to next-to-leading order in the coupling are given in~\cite{Biondini:2023zcz}.
In the last equality we have replaced the squared SU($N$)-singlet and SU($N$)-adjoint scattering wavefunctions at the origin in the laboratory frame with the ones in the center-of-mass frame
using the Lorentz transformations derived in appendix~\ref{sec:app_A} in analogy to the abelian case.
Furthermore, in the non-abelian SU($N$) model the annihilation width in the laboratory frame can be inferred from the annihilation width in the center-of-mass frame using the Lorentz contraction formula~\eqref{Lorentz_dilation},
which is valid for paradarkonia and orthodarkonia.

The bound-state formation cross section in the laboratory frame from dark heavy fermion-antifermion pairs charged under an SU($N$) gauge group without additional light degrees of freedom is given up to relative order $\bm{P}^2/M^2 \sim T/M$ and $\Delta E^p_n/M$ by
\begin{eqnarray}
  (\sigma_{\hbox{\scriptsize bsf}} \, v_{\hbox{\scriptsize M\o l}})_{\textrm{lab}}(\bm{p},\bm{P}) &=& 
  \frac{4C_F}{3N^2}\,\alpha(\mu_{\textrm{us}}) \sum \limits_n (\Delta E^p_n)^3 \left(1+n_{\textrm{B}}(\Delta E^p_n)\right)\nonumber\\
&\times&
  \left(\left|\langle n|\bm{r}|\bm{p}\rangle_{\textrm{lab}}^{[\textbf{adj}]}\right|^2 F^n_1(p,P) + \left|\langle n|\bm{r}\cdot \frac{\bm{P}}{2M}|\bm{p}\rangle_{\textrm{lab}}^{[\textbf{adj}]}\right|^2 F^n_2(p,P)\right) ,
\label{bsf_xsection_final_result_non_abelian}
\end{eqnarray}
where $C_F \equiv (N^2-1)/(2N)$, $|\boldsymbol{p} \rangle^{[\textbf{adj}]}$ is the energy eigenstate made of an adjoint heavy fermion-antifermion pair of relative momentum $\bm{p}$,
and the overall coupling $\alpha$ is evaluated at an ultrasoft scale $\mu_{\textrm{us}}$ of the order of $M \alpha^2$ or $T$, with $\alpha(\mu_{\textrm{us}}) \gg \alpha(2M)$ because of asymptotic freedom.
The energy difference between the color-adjoint unbound pair and the color-singlet bound pair in the laboratory frame is $\Delta E_{n}^{p}$
and the functions $F^n_1(p,P)$ and $F^n_2(p,P)$ are defined in~\eqref{bsf_xsection_final_result_F1} and \eqref{bsf_xsection_final_result_F2}, respectively. 

Similarly, the thermal bound-state dissociation width in the laboratory frame is given up to relative order $\bm{P}^2/M^2 \sim T/M$ and $\Delta E^p_n/M$ by
\begin{eqnarray}
  (\Gamma^{n}_{\textrm{bsd}})_{\textrm{lab}}(\bm{P}) &=& \frac{4}{3}C_F\alpha(\mu_{\textrm{us}}) \int \frac{d^3p}{(2\pi)^3} \, (\Delta E^p_n)^3 \,  n_{\textrm{B}}(\Delta E^p_n)\nonumber\\
&&\times
\left( \left|\langle n|\bm{r}|\bm{p}\rangle_{\textrm{lab}}^{[\textbf{adj}]}\right|^2 D^n_1(p,P) + \left|\langle n|\bm{r}\cdot \frac{\bm{P}}{2M}|\bm{p}\rangle_{\textrm{lab}}^{[\textbf{adj}]}\right|^2 D^n_2(p,P)\right).
\label{bsd_width_final_result_non_abelian}
\end{eqnarray}
The functions $D^n_1(p,P)$, $D^n_2(p,P)$ are defined in~\eqref{bsd_width_final_result_D1} and~\eqref{bsd_width_final_result_D2}, respectively.

\section{Dark matter energy density}
\label{sec:numerics}
In the following two sections, we evaluate the numerical impact of the recoil corrections to the cross sections and widths, section~\ref{sec:numerics_A},
and to the Boltzmann equation for the determination of the dark matter relic abundance, section~\ref{sec:numerics_B}.
Cross sections, widths and thermal averages are computed in the laboratory frame, see appendix~\ref{sec:app_B}.
Nevertheless, it is convenient to express, first, the quantum-mechanical matrix elements, which include wavefunctions at the origin and dipole matrix elements,
and the energy differences in the center-of-mass frame according to eqs.~\eqref{DeltaEgamma} and~\eqref{Psinorigin}-\eqref{dipolenp},
since in the center-of-mass frame the spectrum and wavefunctions are the usual Coulombic energy levels and eigenfunctions, see ref.~\cite{Biondini:2023zcz}.
Then the center-of-mass momenta are boosted back in the laboratory frame according to eq.~\eqref{pcmplab}.

\subsection{Numerical impact of recoil corrections on cross sections and widths}
\label{sec:numerics_A}
In order to quantify the effect of the recoil corrections due to the center-of-mass motion on cross-sections and widths, 
we compare the results for the bound-state formation cross section, eq.~\eqref{bsf_xsection_final_result},
bound-state dissociation width, eq.~\eqref{bsd_width_final_result}, bound-state de-excitation and excitation transition widths, eqs.~\eqref{de-excitation_width} and~\eqref{excitation_width} respectively,
with the corresponding leading-order expressions without recoil corrections given in~\cite{Biondini:2023zcz,Binder:2020efn}.
These read (at leading order in the non-relativistic expansion $v_{\hbox{\tiny M\o l}}(\bm{P}=\bm{0}) = v_{\hbox{\tiny rel}} \approx  v^{(0)}_{\hbox{\tiny rel}}$)
\begin{equation}
\begin{aligned}
  (\sigma_{\hbox{\scriptsize bsf}} \, v^{(0)}_{\hbox{\tiny rel}})(\bm{p}) =
  \frac{4}{3}\alpha &\sum \limits_n (\Delta E^p_n)^3 \, \left(1+n_{\textrm{B}}(\Delta E^p_n)\right) \, \left|\langle n|\bm{r}|\bm{p}\rangle\right|^2  \, ,
\label{bsf_xsection_old}
\end{aligned}
\end{equation}
\begin{equation}
\Gamma^{n}_{\textrm{bsd}} =
\frac{4}{3}\alpha \int \frac{d^3p}{(2\pi)^3} \, (\Delta E^p_n)^3 \, n_{\textrm{B}}(\Delta E^p_n) \, |\langle n|\bm{r}|\bm{p}\rangle|^2 \, ,
\label{bsd_width_LO}
\end{equation} 
\begin{eqnarray}
\Gamma^{n}_{\textrm{de-ex.}} =\frac{4}{3}\alpha \sum_{n', E_{n'}<E_n}\left(\Delta E_{n'}^{n}\right)^3 \, \left(1+n_{\text{B}}\left(\Delta E_{n'}^{n}\right)\right) \, \left| \langle n'|  \bm{r}  | n \rangle \right|^2 \, ,
\label{gamma_de-excitation1}
\end{eqnarray}
and
\begin{eqnarray}
\Gamma^{n}_{\textrm{ex.}} =\frac{4}{3}\alpha \sum_{n', E_{n'}>E_n}\, (\Delta E_{n}^{n'})^3  \, n_{\text{B}}(\Delta E_{n}^{n'}) \, \left| \langle n' | \bm{r} | n \rangle \right|^2 \, .
\label{gamma_excitation1}
\end{eqnarray}
Processes happening at the hard scale, like annihilations, depend weakly on the thermal medium.\footnote{ 
For the Coulombic bound states considered in this work, the leading thermal correction to the annihilation cross section or width comes from a loop diagram with two electric dipole vertices and the insertion of an imaginary contact potential~\eqref{annpNRQED}.
This is suppressed by at least $\alpha (a_0T)^2 \lesssim \alpha T/M$ with respect to the leading width.   \label{foot:annthermal}
}
Recoil corrections due to the center-of-mass motion may be incorporated using the Lorentz invariance of the annihilation cross section, see eq.~\eqref{ann_fact_scat},
or time dilation in the case of the annihilation width, see eq.~\eqref{ann_para}.

\begin{figure}[ht]
    \centering
    \includegraphics[scale=0.75]{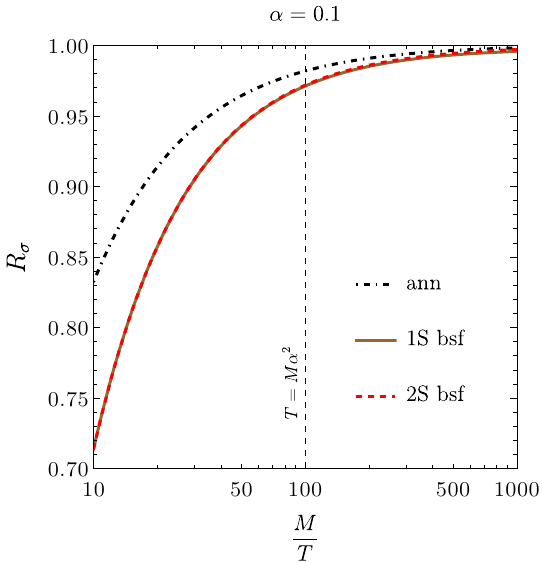}
    \hspace{1cm}
    \includegraphics[scale=0.74]{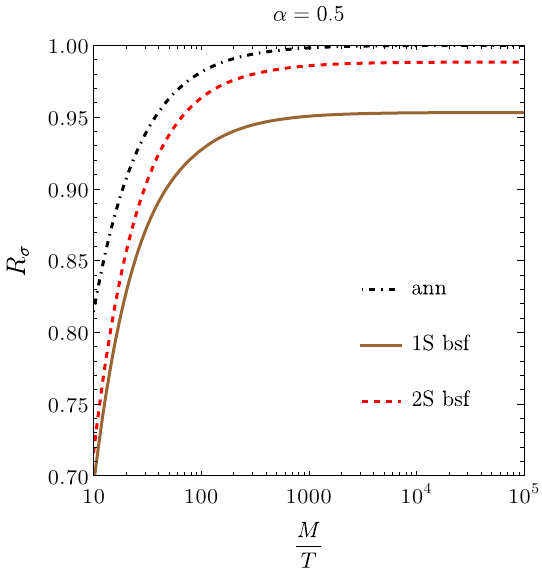}
    \caption{(Left) Ratios of thermally averaged cross sections in the laboratory frame with recoil corrections
      and the corresponding thermally averaged cross sections without recoil corrections plotted as a function of $M/T$ for coupling $\alpha=0.1$.
      The black dash-dotted line follows from the annihilation cross section~\eqref{ann_fact_scat},
      and the brown solid and red dashed lines from the bound-state formation cross section~\eqref{bsf_xsection_final_result} for the 1S and 2S state, respectively.
      (Right) Ratios for coupling $\alpha=0.5$.
      The vertical line in the left plot marks the position where $T=M\alpha^2$ for $\alpha=0.1$; the same line is outside the right plot.}
    \label{fig:bsf_abelian_plots}
\end{figure}

We consider the formation and dissociation of the 1S ground state and first excited 2S state, and the bound-state to bound-state transitions between 1S and 2P states.
The explicit expressions of the dipole matrix elements in the center-of-mass frame can be found in~\cite{Biondini:2023zcz} in the case of transitions among bound states, $\langle n' | \bm{r} | n \rangle$,
and in appendix~\ref{sec:app_C} in the case of transitions between bound and scattering states, $\langle n|\bm{r}|\bm{p}\rangle$.
Concerning the formation of bound states, in figure~\ref{fig:bsf_abelian_plots} we plot the ratio of the cross section~\eqref{bsf_xsection_final_result} thermally averaged in the laboratory frame 
and the corresponding thermally averaged cross section~\eqref{bsf_xsection_old}. 
The thermal average has been defined in appendix~\ref{sec:app_B}.
Similarly, for dark matter fermion pair annihilation, we take the ratio of the thermal average of~\eqref{ann_fact_scat} in the laboratory frame and the corresponding thermally averaged annihilation cross section without recoil corrections. 
The left and right panels of figure~\ref{fig:bsf_abelian_plots} show the ratios for the couplings $\alpha=0.1$ and 0.5, respectively, as functions of $M/T$.
For the smaller coupling, the effect of the center-of-mass recoil corrections to the bound-state formation cross section is up to 3\% for the 1S state and 2S state at temperatures such that $M/T \gtrsim 1/\alpha^2 = 100$.
For the larger coupling $\alpha=0.5$, recoil corrections may become as large as 20-25\% around thermal freeze-out; for this choice of the coupling, the condition $M \alpha^2 \gtrsim T$ is fulfilled even at freeze-out temperature.
For the whole range of considered couplings, recoil corrections are larger for bound-state formation cross sections than for annihilation.
In general, it holds that the ratio is $R_\sigma < 1$, since the annihilation and bound-state formation cross sections are both Lorentz contracted in the laboratory frame, although to a different degree. 

\begin{figure}[ht]
    \centering
    \includegraphics[scale=0.75]{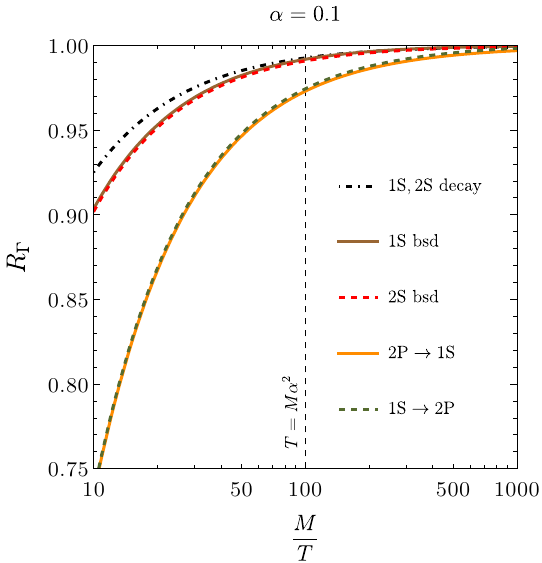}
    \caption{Ratios of thermally averaged widths in the laboratory frame with recoil corrections
      and the corresponding thermally averaged widths without recoil corrections plotted as a function of $M/T$ for coupling $\alpha=0.1$.
      The black dash-dotted line denotes the ratio when taking the thermal average of~\eqref{ann_para},
      the brown solid and red dashed lines when taking the thermal average of~\eqref{bsd_width_final_result} for the 1S and 2S state, respectively, 
      the orange solid line when taking the thermal average of~\eqref{de-excitation_width} for the transition 2P $\rightarrow$ 1S,
      and the green dashed line when taking the thermal average of~\eqref{excitation_width} for the transition 1S $\rightarrow$ 2P. 
      The vertical line marks the position where $T=M\alpha^2$.
      }
    \label{fig:bsd_abelian_plots}
\end{figure}
  
Next, we consider the effect of recoil corrections on the dissociation of 1S and 2S bound-states and the (de-)excitation transitions between 1S and 2P states.
In figure~\ref{fig:bsd_abelian_plots}, we plot the ratios of the widths in~\eqref{bsd_width_final_result},~\eqref{de-excitation_width} and~\eqref{excitation_width} thermally averaged in the laboratory frame 
and the corresponding thermally averaged widths in~\eqref{bsd_width_LO},~\eqref{gamma_de-excitation1} and~\eqref{gamma_excitation1} as functions of $M/T$ with $\alpha=0.1$.
We also plot the ratios of the thermally averaged 1S and 2S paradarkonium decay widths in the laboratory frame with recoil corrections, cf.~\eqref{ann_para}, and the corresponding ones without recoil corrections.
The most significant recoil corrections are for the 1S $\leftrightarrow$ 2P transition widths at large temperatures.
Hence, specially for bound-state to bound-state transitions
the contribution from the motion of the center of mass should be taken into account whenever doing precision computations.

\begin{figure}[ht]
    \centering
    \includegraphics[scale=0.75]{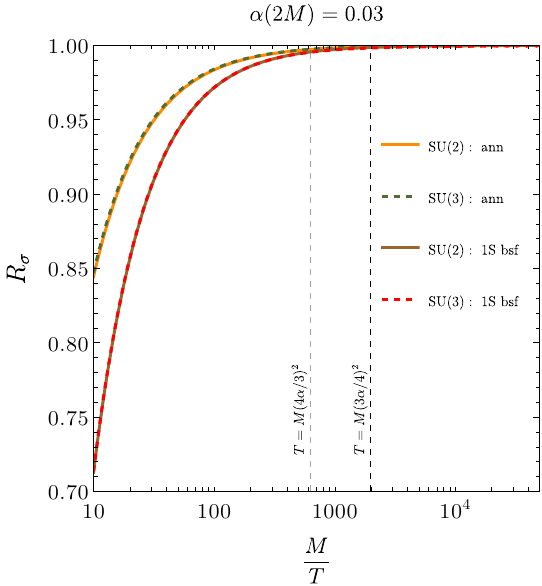}
    \hspace{1cm}
   \includegraphics[scale=0.74]{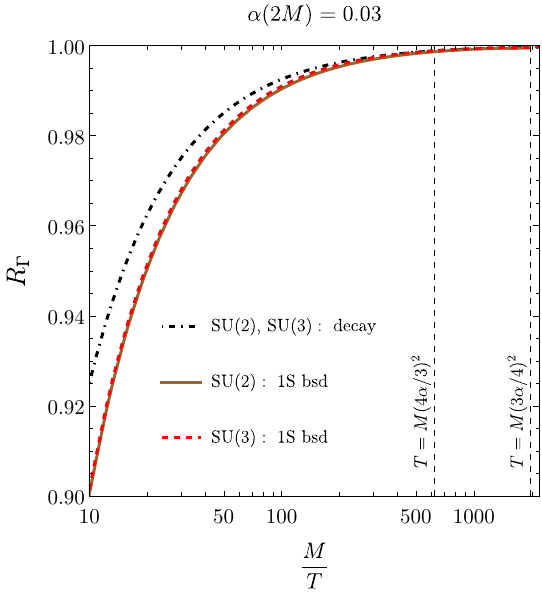}
   \caption{(Left) Ratios of thermally averaged cross sections in the laboratory frame with recoil corrections
     and the corresponding thermally averaged cross sections without recoil corrections plotted as a function of $M/T$ for the non-abelian model~\eqref{non_ab_model}.
     The orange solid and green dashed lines denote the ratios when taking the thermal average of~\eqref{total_non_abelian_ann_colored_av} in SU(2) and SU(3), respectively,
     the brown solid and red dashed lines when taking the thermal average of~\eqref{bsf_xsection_final_result_non_abelian} for the 1S state in SU(2) and SU(3), respectively.
     (Right) Ratios of thermally averaged widths in the laboratory frame with recoil corrections 
     and the corresponding thermally averaged widths without recoil corrections plotted as a function of $M/T$ for the non-abelian model~\eqref{non_ab_model}.
     The black dash-dotted line is for the para- and orthodarkonium decay width, which is valid in any SU(N) model,
     the brown solid and red dashed lines are for taking the thermal average of~\eqref{bsd_width_final_result_non_abelian} for the 1S state in SU(2) and SU(3), respectively.
     The coupling runs at one loop, starting from $\alpha(2M)=0.03$.
     The vertical lines mark the positions where $T=M(C_F\alpha)^2$.}
    \label{fig:bsf_plotsSU3}
\end{figure}

In the non-abelian model~\eqref{non_ab_model}, we obtain results similar to the abelian case for the annihilation cross section~\eqref{total_non_abelian_ann_colored_av},  
bound-state formation cross section~\eqref{bsf_xsection_final_result_non_abelian}, which are shown in the left panel of figure~\ref{fig:bsf_plotsSU3},
and for the bound-state annihilation and dissociation width~\eqref{bsd_width_final_result_non_abelian} shown in the right panel of figure~\ref{fig:bsf_plotsSU3}.
There are no bound-state to bound-state transitions in a pure SU(N) theory due to charge conservation.
The coupling is taken to be $\alpha=0.03$ at the hard scale $2M$ and runs down to the lower energy scales at one loop.\footnote{
With this choice of the coupling at the scale $2M$, the weak-coupling expansion is secure up to the lowest considered temperature $T \approx 10^{-5}M$~\cite{Biondini:2023zcz}.}
We observe that the size of the recoil corrections does not change much from SU(2) to SU(3).
At high temperatures, the relative effect of the recoil corrections is largest for the bound-state formation cross section.

\subsection{Boltzmann equations with recoil corrections}
\label{sec:numerics_B}
The coupled Boltzmann equations for the 1S paradarkonium number density $n_{1\textrm{S}}^{\textrm{para}}\equiv n_{1\textrm{S}}$ and the sum of the dark matter particle and antiparticle number densities $n = n_X + n_{\bar{X}} = 2 n_X$ in the laboratory frame,
neglecting (de-)excitations between bound states, are~\cite{Gondolo:1990dk,vonHarling:2014kha, Biondini:2023zcz}\footnote{We neglect the contribution from the spin-triplet orthodarkonium, since we include only two-photon annihilations in the annihilation cross section and decay width.}
\begin{equation}
\begin{aligned}
  (\partial_t + 3H) n &\!=\! -
  \frac{1}{2}\langle \left(\sigma_{\hbox{\scriptsize ann}} v_{\hbox{\tiny M\o l}}\right)_{\textrm{lab}} \rangle_{\textrm{lab}} (n^2-n^2_{\textrm{eq}})\!
  - \!\frac{1}{2}\langle (\sigma^{1\text{S}}_{\hbox{\scriptsize bsf}} \, v_{\hbox{\scriptsize M\o l}})_{\textrm{lab}} \rangle_{\textrm{lab}} n^2
  + 2 \langle (\Gamma^{1\text{S}}_{\textrm{bsd}})_{\textrm{lab}} \rangle_{\textrm{lab}} n_{1\textrm{S}} , \\
  (\partial_t + 3H)n_{1\textrm{S}} &\!=\!
  -\langle (\Gamma^{1\text{S},\hbox{\scriptsize para}}_{\textrm{ann}})_{\textrm{lab}} \rangle_{\textrm{lab}} (n_{1\textrm{S}} - n_{1\textrm{S},\textrm{eq}})
  \! - \! \langle (\Gamma^{1\text{S}}_{\textrm{bsd}})_{\textrm{lab}} \rangle_{\textrm{lab}} n_{1\textrm{S}}
  +  \frac{1}{16}\langle (\sigma^{1\text{S}}_{\hbox{\scriptsize bsf}} \, v_{\hbox{\scriptsize M\o l}})_{\textrm{lab}} \rangle_{\textrm{lab}} n^2 \!,
\label{coupled_boltzmann_eqs}
\end{aligned}
\end{equation}
where $H$ is the Hubble rate, i.e. the expansion rate of the universe.
In contrast with the expressions in~\cite{Biondini:2023zcz}, the particle production and decay rates depend now on the total momentum $\bm{P}$
and, therefore, all cross-sections and widths, including the annihilation and bound-state dissociation widths, are taken thermally averaged over the total momentum in the laboratory frame. 
The coupled equations~\eqref{coupled_boltzmann_eqs} follow from integrating the evolution equations for the dark matter bound-state and scattering distribution functions.
For kinetically equilibrated pairs, this leads to factorized thermally averaged rates and number densities.
For instance, we obtain the annihilation decay part of the collision term, $\langle (\Gamma^{\text{1S}}_{\textrm{ann}})_{\text{lab}}(\bm{P}) \rangle_{\text{lab}} n_{1\textrm{S}}$, from
\begin{equation}
\begin{aligned}
  \int \frac{d^3 P}{(2\pi)^3} (\Gamma^{\text{1S}}_{\textrm{ann}})_{\text{lab}}(\bm{P})f_{\text{1S}}(\bm{P},t) &= \frac{n_{1\textrm{S}}(t)}{n_{1\textrm{S},\textrm{eq}}}
  \int \frac{d^3 P}{(2\pi)^3} (\Gamma^{\text{1S}}_{\textrm{ann}})_{\text{lab}}(\bm{P})f_{\text{1S},\textrm{eq}}(\bm{P}) \\
  &= \langle (\Gamma^{\text{1S}}_{\textrm{ann}})_{\text{lab}}(\bm{P}) \rangle_{\text{lab}} n_{1\textrm{S}} \, ,
\end{aligned}
\end{equation}
where the first equality is a consequence of $f_{\text{1S}}(\bm{P},t) = f_{\text{1S},\textrm{eq}}(\bm{P})n_{1\textrm{S}}(t)/n_{1\textrm{S},\textrm{eq}}$,
which follows from assuming that the momentum density of the particles, 
$f_{\text{1S}}(\bm{P})$, has its equilibrium form, $f_{\text{1S},\textrm{eq}}(\bm{P})$, which is the Maxwell--Boltzmann distribution of the total momentum, up to a momentum-independent  constant.\footnote{
The constant is $n_{1\textrm{S}}(t)/n_{1\textrm{S},\textrm{eq}}$,  which follows from requiring that the momentum integral of the momentum density of the particles gives the particle number density up to a degeneracy factor counting internal degrees of freedom.}

\begin{figure}[ht]
    \centering
    \includegraphics[scale=0.74]{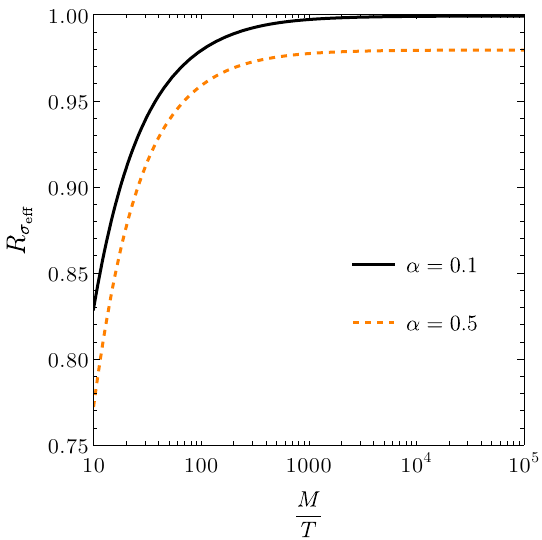}
    \caption{Ratios of thermally averaged effective cross section in the laboratory frame with recoil corrections, cf.~\eqref{Cross_section_eff},
      and the corresponding thermally averaged effective cross section without recoil corrections plotted as a function of $M/T$ for couplings $\alpha=0.1$ (black solid line) and $\alpha=0.5$ (orange dashed line).
      }
    \label{fig:bsd_abelian_plots2}
\end{figure}

For sufficiently large annihilation rates, $\Gamma^{\text{1S}}_{\textrm{ann}} \gg H$, and neglecting bound-state to bound-state transitions, the coupled equations reduce to a single evolution equation~\cite{Ellis:2015vaa}
\begin{equation}
    (\partial_t + 3H) n = - \frac{1}{2} \left\langle \left(\sigma_{\hbox{\scriptsize eff}} v_{\hbox{\tiny M\o l}}\right)_{\textrm{lab}} \right\rangle_{\textrm{lab}} (n^2-n^2_{\textrm{eq}}) \, ,
    \label{Boltzmann_eq_eff}
\end{equation}
where the thermally averaged effective cross section in the laboratory frame is defined as 
 \begin{equation}
   \left\langle \left(\sigma_{\hbox{\scriptsize eff}} v_{\hbox{\tiny M\o l}}\right)_{\textrm{lab}} \right\rangle_{\textrm{lab}}  =
   \left\langle \left(\sigma_{\hbox{\scriptsize ann}} v_{\hbox{\tiny M\o l}}\right)_{\textrm{lab}} \right\rangle_{\textrm{lab}}
   + \sum_{n} \left\langle   (\sigma^n_{\textrm{bsf}} \, v_{\hbox{\tiny M\o l}})_{\text{lab}} \right\rangle_{\text{lab}} \, \frac{\langle (\Gamma_{\textrm{ann}}^n)_{\text{lab}}\rangle_{\text{lab}} }{\langle (\Gamma_{\textrm{ann}}^n)_{\text{lab}} \rangle_{\text{lab}}
     +\langle (\Gamma_{\textrm{bsd}}^n)_{\text{lab}} \rangle_{\text{lab}}} \, .
    \label{Cross_section_eff}
 \end{equation}
In figure~\ref{fig:bsd_abelian_plots2}, 
we show the impact of the center-of-mass motion recoil corrections to the effective cross section by plotting the ratio of~\eqref{Cross_section_eff} and the corresponding thermally averaged effective cross section without recoil corrections. 
The ratio decreases for increasing temperatures, and for large temperatures there is a reduction in the thermally averaged effective cross section due to the recoil corrections  by about 15-25\% for both choices of coupling that we display.
Hence, the contribution of the recoil corrections to the effective cross section, and, therefore, the Boltzmann equations, is significant for increasing $T$ even for small couplings.

\begin{figure}[ht]
    \centering
    \includegraphics[scale=0.75]{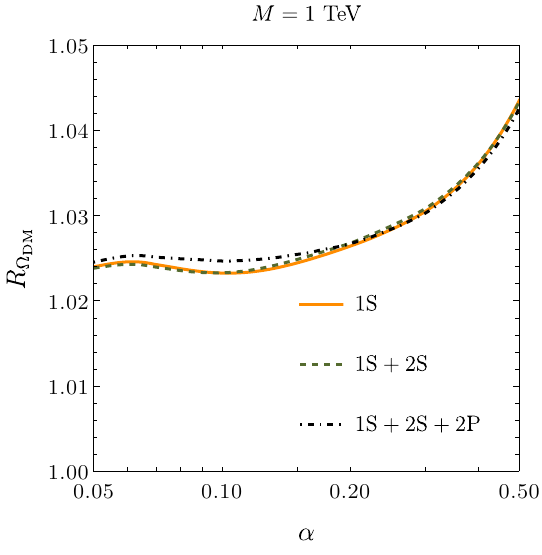}
    \hspace{1cm}
    \includegraphics[scale=0.74]{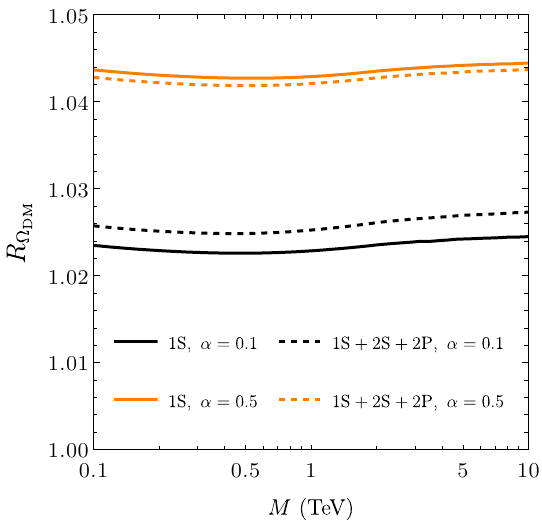}
    \caption{(Left) Ratio of the present dark matter relic abundance in QED$_\text{DM}$ as seen from the laboratory frame including center-of-mass recoil corrections
      and the relic abundance computed without recoil corrections plotted as a function of the coupling $\alpha$ for fixed dark matter mass $M=1$ TeV.
      The orange solid line is for the ratio obtained by only considering the 1S bound state,
      the green dashed line by including the 2S state,
      the black dash-dotted line by including 2P states beyond the no-transition approximation.
      (Right) Ratio of the present dark matter relic abundance as a function of the dark matter mass for two benchmark values of the coupling $\alpha$.
      Black solid and orange solid lines are for $\alpha=0.1$ and $\alpha=0.5$, respectively, and obtained by considering only the 1S state.
      Black dashed and orange dashed lines are for $\alpha=0.1$ and $\alpha=0.5$, respectively, and obtained by including the excited states 2S and 2P beyond the no-transition approximation.}
    \label{fig:energy_density}
\end{figure}

Equation~\eqref{Boltzmann_eq_eff} can be recast in terms of the yield $Y\equiv n/s$, $s$ being the entropy density, and solved numerically.
The solution can be related to the present-day dark matter relic density $\Omega_{\textrm{DM}} = M s_0 Y_0/\rho_{\textrm{crit},0}$,
where $Y_0, s_0$ and $\rho_{\textrm{crit},0}$ are the present yield, entropy density and critical density, respectively.
Taking the values of $s_0$ and  $\rho_{\textrm{crit},0}$ from~\cite{ParticleDataGroup:2022pth}, one obtains $\Omega_{\textrm{DM}} h^2=(M/\textrm{GeV)}\, Y_0 /(3.645 \times 10^{-9})$, where $h$ is the reduced Hubble constant.
Eventually, this value can be compared with the observed dark matter energy density $\Omega_{\hbox{\tiny DM}} h^2 = 0.1200 \pm 0.0012$~\cite{Planck:2018nkj} to determine coupling and mass $M$ of the dark matter model.
The temperature-dependent relativistic degrees of freedom entering the Hubble rate in eq.~\eqref{Boltzmann_eq_eff} are assumed to be those of the Standard Model with the addition of the dark photon~\cite{vonHarling:2014kha}.

In the two panels of figure~\ref{fig:energy_density}, we show the ratio of the present dark matter energy density, $R_{\Omega_\textrm{DM}}$, obtained with center-of-mass recoil effects in the laboratory frame and the one obtained without recoil effects
plotted as a function of $\alpha$ (left panel) and $M$ (right panel).
The left plot shows the ratio for a fixed dark matter mass of 1 TeV.
For a wide range of couplings from 0.05 to 0.5, we observe that when considering the evolution of dark matter unbound pairs and only the ground state 1S, the ratio $R_{\Omega_\textrm{DM}}$ (orange solid line) is larger than one,
reflecting the fact that the dark matter relic abundance is less depleted in the laboratory frame due to the inclusion of center-of-mass recoil effects.
For values of the coupling up to $\alpha=0.2$, the recoil correction stays constant around 2.5\%.
For stronger couplings the recoil correction starts increasing, eventually reaching the maximal value of about 4.5\% for $\alpha=0.5$.
Including the contribution from the first excited state 2S in the effective cross section~\eqref{Cross_section_eff}, the ratio $R_{\Omega_\textrm{DM}}$, now represented by the green dashed line, is nearly the same as the ratio obtained from only the 1S state. 
We complete the study of the evolution of the dark matter energy density by adding bound-state to bound-state transitions into the coupled Boltzmann equations in~\eqref{coupled_boltzmann_eqs} (we call this beyond the \emph{no-transition approximation}).
We include only transitions between the ground state and the first excited states 2P (quantum numbers $\ell =1$ and $m=0,\pm 1$).
The coupled evolution equations can be recast into a single effective Boltzmann equation as in~\eqref{Boltzmann_eq_eff} introducing  a more general expression of the effective cross section that has been derived in ref.~\cite{Garny:2021qsr}.
Therefore, in total, we account for the Sommerfeld enhanced annihilation of scattering states, the decay of the 1S and 2S states,\footnote{
We neglect the decay of the three 2P states, since it happens at order $1/M^4$ in the center-of-mass frame.}
the bound-state formation and dissociation of the 1S, 2S and 2P states, and also the bound-state transitions 1S $\leftrightarrow$ 2P.
This setup has been recently studied in~\cite{Biondini:2023zcz}, now we extend it by taking into account the contribution from the recoil.
In the left panel of figure~\ref{fig:energy_density}, the present dark matter energy density with recoil correction and all the above effects included is represented by the black dash-dotted line.
The recoil correction amounts to an effect between 2.5\% and 4.5\%.
The dark matter energy abundance is monotonically increasing with increasing $\alpha$, since the effective cross section~\eqref{Cross_section_eff}, as well as its generalized version accounting for bound-state effects beyond the no-transition approximation,
are decreasing functions with increasing coupling, see figure~\ref{fig:bsd_abelian_plots2}.
Moreover, the recoil correction on the dark matter relic abundance seems to be independent of whether considering only the ground state or adding higher excited states and also transitions among them for the whole range of considered values for $\alpha$.
Hence, one can quantify the recoil correction on the relic energy density, to a good degree of precision, already including only the ground state in the evolution equations.

The right plot in figure~\ref{fig:energy_density} shows $R_{\Omega_\textrm{DM}}$ for a wide range of dark matter masses from 0.1 TeV to 10 TeV.
In the considered range, the recoil effect on the energy density is almost independent of the dark matter mass $M$.
The black solid and orange solid lines include only the ground state contribution to the evolution equations for the specific values $\alpha=0.1$ and $\alpha=0.5$, respectively.
The black dashed and orange dashed lines also include bound-state effects from the 2S and 2P states.

We conclude that the correction due to the motion of the center of mass of the heavy dark matter pairs is above the 1\% accuracy of the present measurement of the dark matter energy density,
with values ranging between 2.5\% and 4.5\% for the considered values of $\alpha$ from $0.05$ to $0.5$.
In the laboratory frame, the recoil leads to less depletion of the energy density due to a decreased effective cross section,
and is independent of the particular value of the dark matter mass and the inclusion of bound-state effects from higher excited states.
We obtain similar results for the recoil corrections to the dark matter relic abundance in the dark non-abelian model.
In the unbroken non-abelian gauge theories SU(2) and SU(3),
for coupling $\alpha(2M)=0.03$ and at one-loop running, we get a small correction to the relic density coming from recoil corrections of about 2\%.

\section{Conclusions and outlook}
\label{sec:concl}
A novel particle remains a viable and well motivated option to explain the compelling evidence of dark matter in the universe. 
In such a scenario, the production of dark particles occurs in the early stages of the universe evolution, when the universe is a hot and dense medium. 
Through thermal freeze-out, 
the dynamics of the heavy dark matter particle-antiparticle pairs 
eventually leads to the presently observed relic density.  
In this work, we have computed the leading order effects due to the relative
motion between the thermal plasma and the center-of-mass of the dark matter pairs for the annihilation and bound-state formation cross sections, and for the bound-state dissociation and transition widths. 
The leading-order recoil corrections for bound-state formation and dissociation in a thermal bath are computed here for the first time
in a field theoretical framework.
We have used non-relativistic effective field theories to 
exploit the hierarchy of energy scales typical of the problem and systematically factorize  
high-energy from low-energy contributions, and in-vacuum from thermal effects.

As prototypical dark matter models, we have considered a QED-like dark sector made of Dirac fermions and dark photons, and the corresponding non-abelian version featuring an SU($N$) gauge group. 
For the annihilation cross section and bound-state decay, we have derived the recoil corrections in the laboratory frame by computing the contributions of the dimension 8, center-of-mass dependent,
four-fermion operators and inspecting the Lorentz transformation properties of the fermion-antifermion wavefunctions.
The main original results of this work can be found in eqs.~\eqref{bsf_xsection_final_result} and \eqref{bsf_xsection_final_result_non_abelian}
for the bound-state formation cross section, and in eqs.~\eqref{bsd_width_final_result} and \eqref{bsd_width_final_result_non_abelian} for the bound-state dissociation width in the abelian and non-abelian model, respectively. 
One can also obtain the bound-state formation cross section at finite temperature in the laboratory frame by mapping the corresponding expression in the center-of-mass frame similarly
to the case of the in-vacuum annihilation cross section (see eqs.~\eqref{NR_hard_cross_section_lab_frame} and \eqref{Lorentz_boost2}). 
The same consideration holds for the dissociation width.  
Expressions for bound-state to bound-state transition widths are given in eqs.~\eqref{de-excitation_width} and \eqref{excitation_width}. 

The scale associated with the center-of-mass momentum of a heavy dark matter pair in kinetic equilibrium is of the order of $\sqrt{MT} \ll M$ for non-relativistic particles.  
In the paper, we plot the various cross sections and widths as functions of $M/T$ and the coupling. 
Recoil corrections are more important at larger temperatures. 
Around the freeze-out temperature, we observe that the center-of-mass motion affects the cross sections and widths at most at about 15\%--25\%.  
Upon solving the Boltzmann equations and including the recoil corrections,  we find that their impact on the relic density ranges between 2.5\% and 4.5\% for the masses and couplings considered in this work. 
The effect is  larger than the experimental accuracy of the dark matter energy density. 

Although the paper focuses on the dark matter freeze-out, its results for the cross sections and widths have a much wider range of applicability. 
The considered SU(3) dark matter model is QCD, henceforth its results apply immediately 
to the dissociation and formation of heavy quarkonium moving relatively to the quark gluon plasma formed in high-energy heavy-ion collisions.  
In particular, under the assumed hierarchy of scales and for a weakly-coupled plasma, we have computed the leading non-relativistic recoil corrections to quarkonium 
annihilation, formation and gluodissociation.
With respect to earlier computations, we have presented here 
for the first time the complete set of leading-order recoil corrections in the temperature and recoil energy over $M$ ratio.

Finally, we have obtained the bound-state to bound-state transition widths 
in the case of a bound state moving with respect to a thermal bath.
Those expressions may add to the precision spectroscopy at the base of atomic clocks, 
if the atomic clock is not at rest and interacts with a thermal environment.

\section*{Acknowledgments}
The work of S.B. is supported by the Swiss National Science Foundation (SNSF) under the Ambizione grant PZ00P2\_185783.
N.B., G.Q. and A.V. acknowledge support from the DFG (Deutsche Forschungsgemeinschaft, German Research Foundation) cluster of excellence ``ORIGINS''
under Germany's Excellence Strategy -  EXC-2094-390783311. 

\appendix
\numberwithin{equation}{section}
\section{Lorentz transformations}
\label{sec:app_A}
Let $\bm{p}_1$ and $\bm{p}_2$, and $E_1$ and $E_2$ be the momenta and energies, respectively, of two particles in a reference frame $S$,
and $\bm{p}_1'$ and $\bm{p}_2'$, and $E_1'$ and $E_2'$ the momenta and energies, respectively, of the two particles in a reference frame $S'$ moving with respect to $S$ with velocity $\bm{v}$. 
The Lorentz transformations relating momenta and energies in the two reference frames are
\begin{eqnarray}
    &&\bm{p}'_1 = \bm{p}_1 + ( \gamma -1) (\bm{p}_1 \cdot \bm{v}) \frac{\bm{v}}{\bm{v}^2}-  \gamma E_1 \bm{v} \, , 
    \label{LT_COM_lab_1}
    \\
    && \bm{p}'_2 = \bm{p}_2 + ( \gamma -1) (\bm{p}_2 \cdot \bm{v}) \frac{\bm{v}}{\bm{v}^2}-  \gamma E_2 \bm{v}
     \label{LT_COM_lab_2}
    \, , 
    \\
    && E_1' =  \gamma (E_1 - \bm{p}_1 \cdot \bm{v}) \, , 
     \label{LT_COM_lab_3}
    \\
     && E_2' =  \gamma (E_2 - \bm{p}_2 \cdot \bm{v}) \, ,
      \label{LT_COM_lab_4}
\end{eqnarray}
where $\gamma = 1/\sqrt{1-\bm{v}^2}$ is the Lorentz factor.
Since the relative momenta of the pairs in the two reference frames are defined as 
\begin{eqnarray}
    \bm{p} \equiv  \frac{\bm{p}_1-\bm{p}_2}{2} \, , \qquad\qquad  \bm{p}' \equiv \frac{\bm{p}'_1-\bm{p}'_2}{2} \, ,
    \label{p1p2_to_prel}
\end{eqnarray}
and the total momenta as
\begin{eqnarray}
    \bm{P} \equiv \bm{p}_1  + \bm{p}_2 \, , \qquad\qquad  \bm{P}' \equiv \bm{p}'_1 + \bm{p}'_2 \, ,
    \label{p1p2_to_Ptot}
\end{eqnarray}
it follows that the Lorentz transformations relating them read
\begin{equation}
   \bm{p}'  = \bm{p}  + (\gamma-1)  (\bm{p} \cdot \bm{v}) \frac{\bm{v}}{\bm{v}^2} -  \frac{\gamma}{2} (E_1-E_2) \bm{v} \, ,
   \label{LT_prel_COM_lab}
\end{equation}
and 
\begin{equation}
   \bm{P}'  = \bm{P}  + (\gamma-1)  (\bm{P} \cdot \bm{v}) \frac{\bm{v}}{\bm{v}^2} -  \gamma (E_1+E_2) \bm{v} \,.
   \label{LT_Ptot_COM_lab}
\end{equation}
In the following, we consider the special case where the reference frame $S'$ is the center-of-mass frame (cm) of the two particles.
We call then $S$ the laboratory frame (lab).

The total momentum in the laboratory frame is 
\begin{equation}
  \bm{P}_{\textrm{lab}}  = (\bm{p}_1)_{\textrm{lab}} + (\bm{p}_2)_{\textrm{lab}} \, ,
\end{equation}
whereas in the center-of-mass frame it is, by definition, 
\begin{equation}
  \bm{P}_{\textrm{cm}}  = (\bm{p}_1)_{\textrm{cm}} + (\bm{p}_2)_{\textrm{cm}} = \bm{0} \, .
    \label{p1p2_to_Pcm}
\end{equation}
Using eq.~\eqref{LT_Ptot_COM_lab} in~\eqref{p1p2_to_Pcm}, we get 
\begin{eqnarray}
   &&  \bm{0} =  \bm{P}_{\textrm{cm}} = \bm{P}_{\textrm{lab}} + ( \gamma-1)  (\bm{P}_{\textrm{lab}} \cdot \bm{v}) \frac{\bm{v}}{\bm{v}^2} - \gamma ((E_1)_{\textrm{lab}} +(E_2)_{\textrm{lab}}) \bm{v}   \, .
   \label{change_frame_Ptot}
\end{eqnarray}
This equality fixes $\bm{v}$ as a function of the center-of-mass momentum and energy of the pair in the laboratory frame.
Its solution reads
\begin{equation}
  \bm{v} =  \frac{\bm{P}_{\textrm{lab}}}{(E_1)_{\textrm{lab}}+(E_2)_{\textrm{lab}}} \,.
  \label{vel_cm_LO}
\end{equation}
If the particles have the same mass $M$ and are non relativistic, which implies $E_1 \approx M$ and $E_2\approx M$, then we get $\bm{v} \approx \bm{P}_{\textrm{lab}}/(2M)$.
This is the value of $\bm{v}$ used in the main body of the paper to compute the leading relativistic corrections to the various observables.

Furthermore, the condition~\eqref{p1p2_to_Pcm} fixes the energies of the two particles in the center-of-mass frame to be equal:
\begin{equation}
(E_1)_{\textrm{cm}} = (E_2)_{\textrm{cm}}. 
    \label{E1E2_to_cm}
\end{equation}
Using eqs.~\eqref{LT_COM_lab_3} and~\eqref{LT_COM_lab_4} in~\eqref{E1E2_to_cm}, we obtain a relation between the relative momentum in the laboratory frame, $\bm{p}_{\textrm{lab}}$,
the velocity $\bm{v}$ and the energy difference in the laboratory frame $(E_1)_{\textrm{lab}}-(E_2)_{\textrm{lab}}$
\begin{equation}
    \bm{p}_{\textrm{lab}} \cdot \bm{v} = \frac{(E_1)_{\textrm{lab}}-(E_2)_{\textrm{lab}}}{2} \, .
    \label{scalar_product_p_v}
\end{equation}
Then, by trading $(E_1)_{\textrm{lab}}-(E_2)_{\textrm{lab}}$ for $2\,\bm{p}_{\textrm{lab}} \cdot \bm{v}$, the Lorentz transformation \eqref{LT_prel_COM_lab} can be rewritten as 
\begin{eqnarray}
   \bm{p}_{\textrm{cm}} &=& \bm{p}_{\textrm{lab}} + \frac{1-\gamma}{\gamma \,\bm{v}^2}  (\bm{p}_{\textrm{lab}} \cdot \bm{v})\bm{v} \, .
   \label{LT_prel_COM_lab_v2}
\end{eqnarray}
Selecting the component of the relative momentum along the direction of $\bm{v}$, eq.~\eqref{LT_prel_COM_lab_v2} implies
\begin{eqnarray}
    \bm{p}_{\textrm{lab}} \cdot \bm{v} = \gamma \, \bm{p}_{\textrm{cm}} \cdot \bm{v} \, ,
\end{eqnarray}
which shows that the relative momentum component parallel to $\bm{v}$ gets larger by a factor $\gamma$ in the laboratory frame with respect to the center-of-mass frame.
Only the momentum component along $\bm{v}$ gets modified.
This can be made explicit by decomposing the momentum $\bm{p}$ into a component parallel to $\bm{v}$, $\bm{p}_{\parallel} \equiv (\bm{p}\cdot\bm{v})\bm{v}/\bm{v}^2$,
and a component orthogonal to it, $\bm{p}_{\perp} \equiv \bm{p} - \bm{p}_{\parallel}$, and rewriting accordingly eq.~\eqref{LT_prel_COM_lab_v2}:
\begin{equation}
  \bm{p}_{\textrm{cm}} = (\bm{p}_{\perp})_{\textrm{lab}} + \frac{(\bm{p}_{\parallel})_{\textrm{lab}}}{\gamma}\,.
\label{pcmplab}
\end{equation}
The square of the relative momentum changes from one frame to the other as 
\begin{eqnarray}
  | \bm{p}_{\textrm{cm}}|^2 = | (\bm{p}_{\perp})_{\textrm{lab}}|^2 + \frac{|(\bm{p}_{\parallel})_{\textrm{lab}}|^2}{\gamma^2} \, .
\end{eqnarray}
From eq.~\eqref{pcmplab} it follows that the momentum volume element gets also larger by a factor $\gamma$ in the laboratory frame with respect to the center-of-mass frame:
\begin{equation}
  d^3p_{\textrm{cm}} = \frac{d^3p_{\textrm{lab}}}{\gamma} \,.
\label{d3pgamma}
\end{equation}

The total energy of the two particles is $E_{\textrm{cm}}\equiv (E_1)_{\textrm{cm}} + (E_2)_{\textrm{cm}}$ in the center-of-mass frame and $E_{\textrm{lab}} \equiv (E_1)_{\textrm{lab}} + (E_2)_{\textrm{lab}}$ in the laboratory frame.
From~\eqref{LT_COM_lab_3} and~\eqref{LT_COM_lab_4} it follows that
\begin{equation}
  E_{\textrm{cm}} = \gamma(E_{\textrm{lab}} - \bm{P}_{\textrm{lab}}\cdot\bm{v}) = \frac{E_{\textrm{lab}}}{\gamma}\,,
\end{equation}
where in the last equality we have used eq.~\eqref{vel_cm_LO}, i.e. $\bm{v} =  \bm{P}_{\textrm{lab}}/E_{\textrm{lab}}$, which implies $1/\gamma^2 = 1 -  (\bm{P}_{\textrm{lab}}/E_{\textrm{lab}})^2$.
While the center-of-mass energy increases by a factor $\gamma$ in the laboratory frame with respect to the center-of-mass frame,
the opposite happens to the energy difference of two two-particle states, $\Delta E$, for a suitable choice of the center-of-mass frame.
The reason is that the Lorentz factor $\gamma$ depends on the total energy of the pair and therefore it changes by $\Delta \gamma = - \gamma^3\,\bm{P}^2_{\textrm{lab}}\, (\Delta E)_{\textrm{lab}}/E_{\textrm{lab}}^3$ from one state to the other.
Fixing the center-of-mass frame to be just the center-of-mass frame of one chosen state, and computing the relative velocity $\bm{v}$ and the Lorentz factor with respect to it, we get
\begin{equation}
  (\Delta E)_{\textrm{cm}} = \gamma (\Delta E)_{\textrm{lab}}\,.
\label{DeltaEgamma}  
\end{equation}
Since $\Delta E$ may be understood as a frequency, the above relation expresses the Lorentz dilation of the time intervals measured from transition frequencies in the laboratory frame with respect to the center-of-mass frame.

Similarly to the relative momentum, we may decompose the relative distance, $\bm{r}$, between the two particles into a component parallel to $\bm{v}$, $\bm{r}_{\parallel} \equiv (\bm{r}\cdot\bm{v})\bm{v}/\bm{v}^2$,
and a component orthogonal to it, $\bm{r}_{\perp} \equiv \bm{r} - \bm{r}_{\parallel}$.
The Lorentz transformation of $\bm{r}$ reads 
\begin{equation}
  \bm{r}_{\textrm{cm}} = \bm{r}_{\textrm{lab}} + \frac{\gamma-1}{\bm{v}^2}  (\bm{r}_{\textrm{lab}} \cdot \bm{v})\bm{v} = (\bm{r}_{\perp})_{\textrm{lab}} + \gamma (\bm{r}_{\parallel})_{\textrm{lab}}\,,
\label{rcmrlab}
\end{equation}
where we understand $\bm{r}_{\textrm{lab}}$ as determined from the coordinates of the two particles taken at the same time in the laboratory frame.\footnote{
The difference between this condition and eq.~\eqref{E1E2_to_cm} is at the origin of the contraction of the distance along the motion direction in the laboratory frame in eq.~\eqref{rcmrlab}
and the dilation of the relative momentum along the motion direction in the laboratory frame in eq.~\eqref{pcmplab}.
}
The square of the relative distance changes from one frame to the other as 
\begin{eqnarray}
  | \bm{r}_{\textrm{cm}}|^2 = | (\bm{r}_{\perp})_{\textrm{lab}}|^2 + \gamma^2 |(\bm{r}_{\parallel})_{\textrm{lab}}|^2  \, .
\end{eqnarray}
From eq.~\eqref{rcmrlab} it also follows that the volume element gets contracted by a factor $1/\gamma$ in the laboratory frame with respect to the center-of-mass frame:
\begin{equation}
  d^3r_{\textrm{cm}} = \gamma \, d^3r_{\textrm{lab}} \,.
\end{equation}

In quantum mechanics a Lorentz transformation may be represented by a unitary transformation $U(\bm{v})$.
The explicit form of the transformation is not relevant here, but its action on a generic discrete energy eigenstate $|n\rangle$, scattering energy eigenstate $|\bm{p}\rangle$, and on the relative distance operator $\bm{r}$ is
\begin{eqnarray}
&&  U^\dagger(\bm{v})\, |n\rangle_{\textrm{cm}} = |n\rangle_{\textrm{lab}} \,,
  \label{Udagn}  \\
  &&  U^\dagger(\bm{v})\, |\bm{p}\rangle_{\textrm{cm}}  = N(\bm{v})\,|\bm{p}\rangle_{\textrm{lab}} \,, \qquad |N(\bm{v})|^2 = \gamma\,,
  \label{Udagp}  \\
  &&  U^\dagger(\bm{v})\, \bm{r}_{\textrm{cm}} \,U(\bm{v}) =  \bm{r}_{\textrm{lab}} + \frac{\gamma-1}{\bm{v}^2} (\bm{r}_{\textrm{lab}}\cdot \bm{v})\bm{v}\,.
  \label{UdagrU}
\end{eqnarray}  
Equation~\eqref{Udagn} follows from the invariance of the normalization of discrete states under Lorentz transformations: $\langle n| m \rangle_{\textrm{lab}} \!= \langle n| m \rangle_{\textrm{cm}} = \delta_{nm}$.
The notation $\langle n| A |m \rangle_{\textrm{lab}}$ ($\langle n| A |m \rangle_{\textrm{cm}}$) means that the observable $A$, the  bra and the ket are in the laboratory (center-of-mass) frame;
the same notation applies to scattering states.
Equation~\eqref{Udagp} follows from the fact that the non-relativistic normalization of scattering states is not Lorentz invariant.
The normalization factor $N(\bm{v})$ is then necessary to keep both $|\bm{p}\rangle_{\textrm{cm}}$ and $|\bm{p}\rangle_{\textrm{lab}}$ non-relativistically normalized:
$\langle \bm{p}| \bm{q} \rangle_{\textrm{lab}} = \delta^3(\bm{p}_{\textrm{lab}} - \bm{q}_{\textrm{lab}})$ and $\langle \bm{p}| \bm{q} \rangle_{\textrm{cm}} = \delta^3(\bm{p}_{\textrm{cm}}  - \bm{q}_{\textrm{cm}})$.
It can be computed from 
\begin{eqnarray}
  |N(\bm{v})|^2 \,\langle \bm{p}|\bm{q}\rangle_{\textrm{lab}} &=& \langle \bm{p}|\bm{q}\rangle_{\textrm{cm}} =  \delta^3(\bm{p}_{\textrm{cm}}  - \bm{q}_{\textrm{cm}})
  = \delta^3\left( \bm{p}_{\textrm{lab}} - \bm{q}_{\textrm{lab}}  + \frac{1-\gamma}{\gamma \,\bm{v}^2}  ((\bm{p}_{\textrm{lab}} - \bm{q}_{\textrm{lab}}) \cdot \bm{v})\bm{v}\right)
  \nonumber\\
  &=& \gamma \, \delta^3(\bm{p}_{\textrm{lab}} - \bm{q}_{\textrm{lab}}) \,,
\label{Nvderivation}
\end{eqnarray}
where we have used eq.~\eqref{LT_prel_COM_lab_v2}.\footnote{
The last equality follows from
\begin{eqnarray*}
&& \delta^3\left( \bm{p}_{\textrm{lab}} - \bm{q}_{\textrm{lab}}  + \frac{1-\gamma}{\gamma \,\bm{v}^2}  ((\bm{p}_{\textrm{lab}} - \bm{q}_{\textrm{lab}}) \cdot \bm{v})\bm{v}\right) \\
&& = \sum_{n=0}^\infty \frac{1}{n!} \left(  \frac{1-\gamma}{\gamma \,\bm{v}^2}  ((\bm{p}_{\textrm{lab}} - \bm{q}_{\textrm{lab}}) \cdot \bm{v}) (\bm{v}\cdot \bm{\nabla}_{\bm{p}_{\textrm{lab}}})\right)^n\delta^3(\bm{p}_{\textrm{lab}} - \bm{q}_{\textrm{lab}}) \\
  && = \sum_{n=0}^\infty \frac{1}{n!} \left(\frac{1-\gamma}{\gamma \,\bm{v}^2}\right)^n n!\, (\bm{v}^2)^n \, (-1)^n \, \delta^3(\bm{p}_{\textrm{lab}} - \bm{q}_{\textrm{lab}})
  = \frac{1}{1-(\gamma-1)/\gamma} \delta^3(\bm{p}_{\textrm{lab}} - \bm{q}_{\textrm{lab}})
  =  \gamma \, \delta^3(\bm{p}_{\textrm{lab}} - \bm{q}_{\textrm{lab}})   \,.
\end{eqnarray*}
\label{footnote:Nv}
}
From eq.~\eqref{Nvderivation}, it follows $|N(\bm{v})|^2 = \gamma$.
Equation~\eqref{UdagrU} expresses at the operator level the transformation \eqref{rcmrlab}.

The relevant quantum-mechanical matrix elements appearing in this work may be boosted in the different reference frames using eqs.~\eqref{Udagn}-\eqref{UdagrU}.
Let us first consider the bound-state wavefunction at the origin:
\begin{eqnarray}
  \left( |\Psi_{n\ell m}(\bm{0})|^2 \right)_{\textrm{cm}}  &=& \langle n\ell m| \delta^3(\bm{r}) | n\ell m\rangle_{\textrm{cm}} = \langle n\ell m|U(\bm{v}) \, U^\dagger(\bm{v})  \delta^3(\bm{r}) U(\bm{v}) \, U^\dagger(\bm{v})| n\ell m\rangle_{\textrm{cm}}
  \nonumber\\
  &=& \langle n\ell m| \delta^3 \left( \bm{r} + (\gamma-1)   (\bm{r}\cdot \bm{v})\bm{v}/\bm{v}^2 \right) | n\ell m\rangle_{\textrm{lab}}
  \nonumber\\
  &=& \int d^3r_{\textrm{lab}}  \, \delta^3 \left( \bm{r}_{\textrm{lab}} + (\gamma-1)   (\bm{r}_{\textrm{lab}}\cdot \bm{v})\bm{v}/\bm{v}^2 \right) \, \left( |\Psi_{n\ell m}(\bm{r}_{\textrm{lab}})|^2 \right)_{\textrm{lab}}
  \nonumber\\
  &=& \int d^3r_{\textrm{lab}}  \, \frac{1}{\gamma} \delta^3 \left( \bm{r}_{\textrm{lab}} \right) \, \left( |\Psi_{n\ell m}(\bm{r}_{\textrm{lab}})|^2 \right)_{\textrm{lab}} = \frac{\left( |\Psi_{n\ell m}(\bm{0})|^2 \right)_{\textrm{lab}}}{\gamma}\,,
\label{Psinorigin}
\end{eqnarray}
where we have specified the principal, orbital and magnetic quantum numbers.
In the last line, we have rewritten the delta function using an argument similar to the one in footnote~\ref{footnote:Nv}.
Instead, due to the different normalization, for the scattering state wavefunction at the origin, we obtain
\begin{eqnarray}
  \left( |\Psi_{\bm{p}\ell}(\bm{0})|^2 \right)_{\textrm{cm}}  &=& \langle \bm{p}  \ell| \delta^3(\bm{r}) | \bm{p} \ell\rangle_{\textrm{cm}} = \langle \bm{p} \ell |U(\bm{v}) \, U^\dagger(\bm{v})  \delta^3(\bm{r}) U(\bm{v}) \, U^\dagger(\bm{v})| \bm{p} \ell\rangle_{\textrm{cm}}
  \nonumber\\
  &=&  |N(\bm{v})|^2 \langle \bm{p}\ell| \delta^3 \left( \bm{r} + (\gamma-1)   (\bm{r}\cdot \bm{v})\bm{v}/\bm{v}^2 \right) | \bm{p}\ell\rangle_{\textrm{lab}}
  \nonumber\\
  &=&  \gamma \int d^3r_{\textrm{lab}}  \, \delta^3 \left( \bm{r}_{\textrm{lab}} + (\gamma-1)   (\bm{r}_{\textrm{lab}}\cdot \bm{v})\bm{v}/\bm{v}^2 \right) \, \left( |\Psi_{\bm{p}\ell}(\bm{r}_{\textrm{lab}})|^2 \right)_{\textrm{lab}}
  \nonumber\\
  &=& \gamma \int d^3r_{\textrm{lab}}  \, \frac{1}{\gamma} \delta^3 \left( \bm{r}_{\textrm{lab}} \right) \, \left( |\Psi_{\bm{p}\ell}(\bm{r}_{\textrm{lab}})|^2 \right)_{\textrm{lab}} = \left( |\Psi_{\bm{p}\ell}(\bm{0})|^2 \right)_{\textrm{lab}}\,,
\label{Psiporigin}
\end{eqnarray}
where we have specified the partial wave in order to use the same notation that can be found in the main text of the paper.
Dipole matrix elements between generic bound states are related in the different reference frames by
\begin{eqnarray}
\langle n|\bm{r}|m\rangle_{\textrm{cm}} &=& \langle n|U(\bm{v}) \, U^\dagger(\bm{v}) \bm{r}   U(\bm{v}) \, U^\dagger(\bm{v}) |m\rangle_{\textrm{cm}} = \langle n|  \bm{r} + (\gamma-1)   (\bm{r}\cdot \bm{v})\bm{v}/\bm{v}^2  |m\rangle_{\textrm{lab}}
\nonumber\\
&\approx& \langle n|  \bm{r} |m\rangle_{\textrm{lab}}  +  \frac{1}{2}   \langle n|(\bm{r}\cdot \bm{v})\bm{v}|m\rangle_{\textrm{lab}},
\label{dipolenm}
\end{eqnarray}  
where in the last line we have expanded for small $\bm{v}$ and retained only terms up to order $\bm{v}^2$.
Finally, dipole matrix elements between a generic bound and scattering state in different reference frames are related by
\begin{eqnarray}
\langle n|\bm{r}| \bm{p} \rangle_{\textrm{cm}} &=& \langle n|U(\bm{v}) \, U^\dagger(\bm{v}) \bm{r}   U(\bm{v}) \, U^\dagger(\bm{v}) |\bm{p} \rangle_{\textrm{cm}}=  N(\bm{v}) \langle n|  \bm{r} + (\gamma-1)   (\bm{r}\cdot \bm{v})\bm{v}/\bm{v}^2  |\bm{p}\rangle_{\textrm{lab}}
  \nonumber\\
&\approx&  \left(1 + \frac{\bm{v}^2}{4} \right)   \langle n|  \bm{r} |\bm{p} \rangle_{\textrm{lab}}  +  \frac{1}{2}   \langle n|(\bm{r}\cdot \bm{v})\bm{v}| \bm{p} \rangle_{\textrm{lab}},
\label{dipolenp}
\end{eqnarray}  
where in the last line we have expanded in $\bm{v}$ up to order $\bm{v}^2$.

\section{Thermal averages in the laboratory and center-of-mass frame}
\label{sec:app_B}
Let us consider a generic cross section involving the scattering of two incoming particles with center-of-mass momentum
$\bm{P}_{\textrm{lab}} = (\bm{p}_1)_{\textrm{lab}} + (\bm{p}_2)_{\textrm{lab}}$ and relative momentum $\bm{p}_{\textrm{lab}} = ((\bm{p}_1)_{\textrm{lab}} - (\bm{p}_2)_{\textrm{lab}})/2$
times M\o ller velocity in the laboratory frame, $(\sigma \, v_{\hbox{\scriptsize M\o l}})_{\textrm{lab}}(\bm{p}_{\textrm{lab}},\bm{P}_{\textrm{lab}})$.
The \emph{thermal average in the laboratory frame} is defined as
\begin{equation}
\left\langle \left(\sigma\, v_{\hbox{\tiny M\o l}}\right)_{\textrm{lab}} \right\rangle_{\textrm{lab}}
= \frac{\displaystyle \int \frac{d^3 (p_1)_{\textrm{lab}}}{(2 \pi)^3}  \,\frac{d^3 (p_2)_{\textrm{lab}}}{(2 \pi)^3} \, e^{-\frac{(E_1)_{\textrm{lab}}}{T}} \, e^{-\frac{(E_2)_{\textrm{lab}}}{T}} \, (\sigma \, v_{\hbox{\scriptsize M\o l}})_{\textrm{lab}}(\bm{p}_{\textrm{lab}},\bm{P}_{\textrm{lab}})}
{\displaystyle \int \frac{d^3 (p_1)_{\textrm{lab}}}{(2 \pi)^3}  \, \frac{d^3 (p_2)_{\textrm{lab}}}{(2 \pi)^3} \, e^{-\frac{(E_1)_{\textrm{lab}}}{T}} \, e^{-\frac{(E_2)_{\textrm{lab}}}{T}}}\,,
\label{sigmalablab}
\end{equation}
where  $e^{-(E_1)_{\textrm{lab}}/T}$ and $e^{-(E_2)_{\textrm{lab}}/T}$ are the Maxwell--Boltzmann distributions of the incoming particles in the laboratory frame, i.e. the frame where the bath is at rest, 
and $((E_1)_{\textrm{lab}},(\bm{p}_1)_{\textrm{lab}})$, $((E_2)_{\textrm{lab}},(\bm{p}_2)_{\textrm{lab}})$ are the four-momenta of the two incoming particles.
For unbound particles of mass $M$ on mass shell, $(E_1)_{\textrm{lab}} = \sqrt{(\bm{p}_1)_{\textrm{lab}}^2+M^2}$ and $(E_2)_{\textrm{lab}} = \sqrt{(\bm{p}_2)_{\textrm{lab}}^2+M^2}$.
We average over Maxwell--Boltzmann distributions because we assume $M/T \gg 1$.

The numerator on the right-hand side of \eqref{sigmalablab}, 
\begin{equation}
\int \frac{d^3 (p_1)_{\textrm{lab}}}{(2 \pi)^3}  \,\frac{d^3 (p_2)_{\textrm{lab}}}{(2 \pi)^3} \, e^{-\frac{(E_1)_{\textrm{lab}}}{T}} \, e^{-\frac{(E_2)_{\textrm{lab}}}{T}} \, (\sigma \, v_{\hbox{\scriptsize M\o l}})_{\textrm{lab}}(\bm{p}_{\textrm{lab}},\bm{P}_{\textrm{lab}})\,,
\label{sigmalablabnum}
\end{equation}
is Lorentz invariant~\cite{Gondolo:1990dk}, which is the reason for the use of the M\o ller velocity.
For each pair of momenta $\bm{p}_{\textrm{lab}}$ and $\bm{P}_{\textrm{lab}}$ in the laboratory frame, we may identify a center-of-mass frame
such that with respect to it the particles move with relative momentum $\bm{p}_{\textrm{cm}}$, center-of-mass momentum $\bm{P}_{\textrm{cm}} = \bm{0}$, and the laboratory frame moves with relative velocity $-\bm{v}$.
From eqs.~\eqref{LT_prel_COM_lab} and~\eqref{LT_Ptot_COM_lab}, it follows that the two pairs of kinematical variables are related by the Lorentz transformations
\begin{equation}
\begin{aligned}
  &  \bm{p}_{\textrm{lab}}  = \bm{p}_{\textrm{cm}}   + (\gamma-1)  (\bm{p}_{\textrm{cm}}  \cdot \bm{v}) \frac{\bm{v}}{\bm{v}^2} \,,\\
  &  \bm{P}_{\textrm{lab}}  =  \gamma  E_{\textrm{cm}} \bm{v}   \,.
     \label{change_frame_2}
\end{aligned}
\end{equation}
The energy $E_{\textrm{cm}} = ((E_1)_{\textrm{lab}}+(E_2)_{\textrm{lab}})/\gamma = 2 \sqrt{\bm{p}_{\textrm{cm}}^2+M^2}$ is the total energy of the two particles in the center-of-mass frame.
The Jacobian of the transformation from the kinematical variables $(\bm{p})_{\textrm{lab}}$ and $(\bm{P})_{\textrm{lab}}$ to $\bm{p}_{\textrm{cm}}$ and $\bm{v}$ is 
\begin{equation}
\gamma^6  E_{\textrm{cm}}^3 \left(  1 - \frac{ (\bm{p}_{\textrm{cm}}\cdot \bm{v})^2}{(E_{\textrm{cm}}/2)^2} \right).
\label{Jacobian}
\end{equation}  
The factor $\displaystyle \gamma^2  \left(  1 - \frac{ (\bm{p}_{\textrm{cm}}\cdot \bm{v})^2}{(E_{\textrm{cm}}/2)^2} \right)$ cancels against the transformation factor of the M\o ller velocity, see eq.~\eqref{MolLor},
so that the integral \eqref{sigmalablabnum} can be eventually written in terms of the center-of-mass kinematical variables $\bm{p}_{\textrm{cm}}$ and $\bm{v}$ as 
\begin{equation}
  \frac{1}{(2 \pi)^3} \int_{|\bm{v}|\le 1} d^3 v \, \frac{d^3 p_{\textrm{cm}}}{(2 \pi)^3} \, \gamma^4  E_{\textrm{cm}}^3 \, e^{-\frac{\gamma E_{\textrm{cm}}}{T}} \, (\sigma \, v_{\hbox{\scriptsize M\o l}})_{\textrm{cm}}(\bm{p}_{\textrm{cm}},\bm{v})\,.
\label{sigmacmcmnum}
\end{equation}
The result agrees with an analogous expression that can be found in ref.~\cite{Arcadi:2019oxh}.
Note that $e^{-\gamma E_{\textrm{cm}}/T}$ may be also rewritten as $e^{-|(p_1)_{\textrm{cm}}^\mu u_\mu|/T} e^{-|(p_2)_{\textrm{cm}}^\mu u_\mu|/T}$, where $e^{-|p^\mu u_\mu|/T}$ is the small $T$ limit of the particle distribution in the moving thermal bath defined in~\eqref{bose_fkt}.

The \emph{thermal average in the center-of-mass frame} is defined as
\begin{equation}
\left\langle \left(\sigma\, v_{\hbox{\tiny M\o l}}\right)_{\textrm{cm}} \right\rangle_{\textrm{cm}}
= \frac{\displaystyle \int_{|\bm{v}|\le 1} d^3 v  \, \frac{d^3 p_{\textrm{cm}}}{(2 \pi)^3} \, \gamma^4  E_{\textrm{cm}}^3 \, e^{-\frac{\gamma E_{\textrm{cm}}}{T}} \, (\sigma \, v_{\hbox{\scriptsize M\o l}})_{\textrm{cm}}(\bm{p}_{\textrm{cm}},\bm{v})}
{\displaystyle \int_{|\bm{v}|\le 1} d^3 v \, \frac{d^3 p_{\textrm{cm}}}{(2 \pi)^3}  \, \gamma^4  E_{\textrm{cm}}^3 \, e^{-\frac{\gamma E_{\textrm{cm}}}{T}} }\,.
\label{sigmacmcm}
\end{equation}
While the numerators in the right-hand sides of eqs.~\eqref{sigmalablab} and~\eqref{sigmacmcm} are Lorentz invariant, the denominators, which are up to a degeneracy factor the products of the particle number densities at equilibrium, are not.
Hence the conversion factor from $\left\langle \left(\sigma\, v_{\hbox{\tiny M\o l}}\right)_{\textrm{lab}} \right\rangle_{\textrm{lab}}$ to $\left\langle \left(\sigma\, v_{\hbox{\tiny M\o l}}\right)_{\textrm{cm}} \right\rangle_{\textrm{cm}}$
is given by the ratio of the particle number densities at equilibrium in the two frames.
An explicit calculation gives
\begin{equation}
\left\langle \left(\sigma\, v_{\hbox{\tiny M\o l}}\right)_{\textrm{lab}} \right\rangle_{\textrm{lab}}
= \frac{1}{2} \left( 1+\frac{K_1^2(M/T)}{K_2^2(M/T)} \right) \left\langle \left(\sigma\, v_{\hbox{\tiny M\o l}}\right)_{\textrm{cm}} \right\rangle_{\textrm{cm}}\,,
\label{sigmalabsigmacm}
\end{equation}
where $K_i$ are modified Bessel functions of the second kind.
The above expression was first derived in ref.~\cite{Gondolo:1990dk}.

As an application, we consider the thermal average of the annihilation cross section derived in eq.~\eqref{NR_hard_cross_section_lab_frame}.
In the laboratory frame at order $T/M$, it reads
\begin{equation}
\begin{aligned}
  \left\langle \left(\sigma_{\hbox{\scriptsize ann}} v_{\hbox{\tiny M\o l}}\right)_{\textrm{lab}} \right\rangle_{\textrm{lab}} &=
  \left[ 1 -  \frac{8\pi^3}{(MT)^3}\int \frac{d^3p_{\textrm{lab}}}{(2\pi)^3}\, \frac{d^3P_{\textrm{lab}}}{(2\pi)^3}
    e^{-\frac{\bm{p}_{\textrm{lab}}^2}{MT}}e^{-\frac{\bm{P}_{\textrm{lab}}^2}{4MT}}\,\frac{\bm{P}_{\textrm{lab}}^2}{4M^2} \right] \sigma^{\hbox{\tiny NR}}_{\hbox{\scriptsize ann}}v^{(0)}_{\hbox{\scriptsize rel}}
\\
  &=
  \left(1-\frac{3T}{2M}\right) \sigma^{\hbox{\tiny NR}}_{\hbox{\scriptsize ann}}v^{(0)}_{\hbox{\scriptsize rel}}\,,
\label{ann_xsection_average}
\end{aligned}
\end{equation}
where we have expanded $(E_1)_{\textrm{lab}} + (E_2)_{\textrm{lab}} = 2M + \bm{p}_{\textrm{lab}}^2/M + \bm{P}_{\textrm{lab}}^2/(4M) + \dots$~.
The result is consistent with eq.~\eqref{sigmalabsigmacm} in the limit $M\gg T$, as a consequence of 
$\sigma^{\hbox{\tiny NR}}_{\hbox{\scriptsize ann}}v^{(0)}_{\hbox{\scriptsize rel}}$ being independent of the momenta.
Concerning the absolute accuracy of the thermal average \eqref{ann_xsection_average} and the other thermal averages considered in this work,
we remark that corrections proportional to the center-of-mass momentum $\bm{P}$ due to the motion of the dark fermion-antifermion pair in the laboratory frame, 
or equivalently corrections due to the motion of the thermal bath in the center-of-mass reference frame, give corrections to the thermal averages of relative order $T/M$ at low temperatures. 
These are of the same order as the corrections due to the relative momentum $\bm{p}$ that are of relative order $\bm{p}^2/M^2$,  which have not been considered here.

Similarly, the thermal average of a bound-state decay width in the laboratory frame is defined as 
\begin{equation}
\left\langle (\Gamma)_{\textrm{lab}} \right\rangle_{\textrm{lab}} = 
\frac{\displaystyle \int \frac{d^3 P_{\textrm{lab}}}{(2 \pi)^3} \, e^{-\frac{(E_n)_{\textrm{lab}}}{T}} \, (\Gamma)_{\textrm{lab}}(\bm{P}_{\textrm{lab}})}
{\displaystyle \int \frac{d^3 P_{\textrm{lab}}}{(2 \pi)^3} \, e^{-\frac{(E_n)_{\textrm{lab}}}{T}} }\,,
\label{Gammalablab}
\end{equation}
and related to the thermal average of the decay width in the center-of-mass frame through
\begin{equation}
  \left\langle (\Gamma)_{\textrm{lab}} \right\rangle_{\textrm{lab}}
= \frac{\displaystyle \int_{|\bm{v}|\le 1} d^3 v \, \gamma^5 \, e^{-\frac{\gamma (E_n)_{\textrm{cm}}}{T}} \, \frac{(\Gamma)_{\textrm{cm}}(\bm{v})}{\gamma}}
{\displaystyle \int_{|\bm{v}|\le 1} d^3 v \, \gamma^5 \, e^{-\frac{\gamma (E_n)_{\textrm{cm}}}{T}}}
= \left\langle \frac{(\Gamma)_{\textrm{cm}}}{\gamma} \right\rangle_{\textrm{cm}} \,,
\label{GammalabGammacm}
\end{equation}
where $E_n$ is the energy of the bound state.
In the center-of-mass frame, it is given in eq.~\eqref{Encm}.
In the laboratory frame, it may be computed either from $\gamma (E_n)_{\textrm{cm}}$, which amounts at boosting the energy from the center-of-mass frame,
or directly from the bound-state potentials and kinetic energy corrections listed in footnote~\ref{footnotepotential}.
The result is the same and reads
\begin{equation}
(E_n)_{\textrm{lab}} = 2M - \frac{M\alpha^2}{4n^2} + \frac{\bm{P}_{\textrm{lab}}^2}{4M} + \frac{M\alpha^2}{4n^2} \frac{\bm{P}_{\textrm{lab}}^2}{8M^2} - \frac{\bm{P}_{\textrm{lab}}^4}{64M^3}\,.
\end{equation}
The first two terms drop out in the thermal average, as they do not depend on the momentum; 
the last two terms are suppressed with respect to $\bm{P}_{\textrm{lab}}^2/(4M)$ by $E/M$ or $T/M$.
The first equality in eq.~\eqref{GammalabGammacm} follows from the Lorentz transformation $\bm{P}_{\textrm{lab}}  =  \gamma  (E_n)_{\textrm{cm}} \bm{v}$, whose Jacobian is $\gamma^5(E_n)_{\textrm{cm}}^3$,
from $(E_n)_{\textrm{lab}} = \gamma (E_n)_{\textrm{cm}}$ and  $(\Gamma)_{\textrm{lab}}(\bm{P}_{\textrm{lab}}) = (\Gamma)_{\textrm{cm}}(\bm{v})/\gamma$.
Note that $(E_n)_{\textrm{cm}}$ is a constant that does not depend on the integration variables.

As an application, we consider the thermal average of the annihilation widths computed in eq.~\eqref{optical_cross_section3} and following.
If we neglect thermal corrections affecting the annihilation width in the center-of-mass frame (see footnote~\ref{foot:annthermal}),
then $(\Gamma_{\textrm{ann}})_{\textrm{cm}}$ is $\bm{v}$ independent and factorizes outside the integral, leading to
\begin{equation}
\begin{aligned}
  \left\langle (\Gamma_{\textrm{ann}})_{\textrm{lab}} \right\rangle_{\textrm{lab}} &=
  \frac{\displaystyle \int_{|\bm{v}|\le 1} d^3 v \, \gamma^4 \, e^{-\frac{\gamma (E_n)_{\textrm{cm}}}{T}}}  {\displaystyle \int_{|\bm{v}|\le 1} d^3 v \, \gamma^5 \, e^{-\frac{\gamma (E_n)_{\textrm{cm}}}{T}}}\, (\Gamma_{\textrm{ann}})_{\textrm{cm}} 
  =  \frac{K_1( (E_n)_{\textrm{cm}} /T)}{K_2( (E_n)_{\textrm{cm}} /T)} \, (\Gamma_{\textrm{ann}})_{\textrm{cm}} \\
&\approx  \left(1-\frac{3T}{2 (E_n)_{\textrm{cm}}}\right)  \, (\Gamma_{\textrm{ann}})_{\textrm{cm}} \,.
\end{aligned}    
\end{equation}

Finally, we summarize the steps that we followed in this work to compute consistently thermal averages in the laboratory frame at first order in the center-of-mass momentum.

\emph{(i)} Thermal averages of cross sections and decay widths in the laboratory frame are defined as in eqs.~\eqref{sigmalablab} and~\eqref{Gammalablab}, respectively.

\emph{(ii)} Since matrix elements are most easily computed in the center-of-mass frame, see for instance ref.~\cite{Biondini:2023zcz}, 
they are first boosted in the center-of-mass frame according to the transformation formulas derived in appendix~\ref{sec:app_A}.
Then the relative momentum in the center-of-mass frame is re-expressed in terms of the momentum in the laboratory frame by means of the Lorentz transformation~\eqref{pcmplab}, as this is our integration variable.
We may follow the same procedure with the energy differences, or compute them directly in the laboratory frame.

\emph{(iii)} All expressions inside the thermal average integrals are expanded in powers of $\bm{P}$ in accordance with the power counting.
In particular, the Maxwell--Boltzmann distributions for scattering states entering the cross section thermal averages are expanded as
\begin{equation}
  e^{-\frac{(E_1)_{\textrm{lab}}}{T}} \, e^{-\frac{(E_2)_{\textrm{lab}}}{T}} = e^{-\frac{2M}{T}} e^{-\frac{\bm{p}_{\textrm{lab}}^2}{MT}} e^{-\frac{\bm{P}_{\textrm{lab}}^2}{4MT}}
  \left(  1 + \frac{\bm{P}_{\textrm{lab}}^2\bm{p}_{\textrm{lab}}^2}{8M^3T} + \frac{\bm{P}_{\textrm{lab}}^4}{64M^3T} +  \frac{(\bm{P}_{\textrm{lab}}\cdot \bm{p}_{\textrm{lab}})^2}{4M^3T} + \dots  \right)\,,
\end{equation}
and the Maxwell--Boltzmann distribution for bound states entering the decay width thermal averages is expanded as
\begin{equation}
 e^{-\frac{(E_n)_{\textrm{lab}}}{T}} = e^{-\frac{2M}{T}} e^{\frac{M\alpha^2}{4n^2T}} e^{-\frac{\bm{P}_{\textrm{lab}}^2}{4MT}}
  \left(  1 -  \frac{M\alpha^2}{4n^2} \frac{\bm{P}_{\textrm{lab}}^2}{8M^2T} + \frac{\bm{P}_{\textrm{lab}}^4}{64M^3T} + \dots \right) \,,
\end{equation}
where we display only the relevant terms, i.e.~those that depend on the center-of-mass momentum and contribute to the thermal average at order $T/M$ or $E/M$ according to the counting $P_{\textrm{lab}} \sim p_{\textrm{lab}} \sim \sqrt{MT}$.
Constant factors, like $e^{-2M/T}$ and $e^{M\alpha^2/(4n^2T)}$, drop out in the thermal averages.
Also the number densities in the denominators of the thermal averages are expanded in powers of $T/M$ or $E/M$ up to first order.

\section{Dipole matrix elements: general expressions}
\label{sec:app_C}
The analytic expression for the dipole matrix element $\langle n|\bm{r}|\bm{p}\rangle$ in a (non-)abelian model derived in ref.~\cite{Biondini:2023zcz} in the center-of-mass frame 
holds for a coordinate system in which the relative momentum $\bm{p}$ of the dark matter unbound pair is chosen along the $z$-direction. 
In this work, we put the center-of-mass momentum $\bm{P}$ in the laboratory frame (or equivalently the thermal bath velocity $\bm{v}$ in the center-of-mass frame) along the $z$-direction, $\bm{P}=P\bm{e}_z$. 
The relative distance and momentum in spherical coordinates are given by $\bm{r}= (\cos{\phi}\sin{\theta},\sin{\phi}\sin{\theta},\cos{\theta})r$ and $\bm{p}= (\cos{\phi_{p}}\sin{\theta_{p}},\sin{\phi_{p}}\sin{\theta_{p}},\cos{\theta_{p}})p$,
respectively. 
Moreover $p=Mv^{(0)}_{\textrm{rel}}/2$ and $a_0=2/(M\alpha)$, such that $\alpha/v^{(0)}_{\textrm{rel}}=(a_0p)^{-1}$.

The Coulomb wavefunction for a dark matter bound state $|n\rangle \equiv |n,\ell,m\rangle$, with quantum numbers $n$, $\ell$ and $m$, reads
\begin{equation}
\Psi_{n\ell m}(\bm{r}) = \langle\bm{r}|n\ell m \rangle = R_{n \ell}(r)Y_{\ell}^{m}(\Omega) \,,
\label{boundwave}
\end{equation}
with $Y_{\ell}^{m}(\Omega)$ being spherical harmonics and the radial functions given by
\begin{equation}
\begin{aligned}
  R_{n \ell}(r) = \frac{1}{(2\ell+1)!}\sqrt{\left(\frac{2}{na_0}\right)^3\frac{(n+\ell)!}{2n(n-\ell-1)!}}
  \left(\frac{2r}{na_0}\right)^\ell e^{-\frac{r}{na_0}}~_{1}F_{1}\left(\ell+1-n,2\ell+2,\frac{2r}{na_0}\right) \,.
\end{aligned}
\end{equation}
The scattering wavefunction for a dark matter unbound state $|\bm{p}\rangle$, where $\bm{p}$ points into an arbitrary direction, can be expanded into partial waves $\Psi_{\bm{p} \ell}(\bm{r})=\langle\bm{r}|\bm{p}\ell\rangle$ as  
\begin{eqnarray}
  \Psi_{\bm{p}}(\bm{r}) = \sum_{\ell=0}^{\infty} \Psi_{\bm{p} \ell}(\bm{r}) 
&=& \sqrt{\frac{2\pi/(a_0p)}{1-e^{-2\pi/(a_0p)}}}\sum_{\ell=0}^{\infty} i^\ell \frac{(2pr)^{\ell}}{(2\ell)!}P_{\ell}(\bm{e}_r\cdot \bm{e}_p)e^{ipr} \nonumber \\
&&\times ~_{1}F_{1}\left(\ell+1-\frac{i}{a_0p},2\ell+2,-2ipr\right)\prod \limits_{\kappa=1}^{\ell}\sqrt{\kappa^{2} + (a_0p)^{-2}}~,
\label{coulomb_wave}
\end{eqnarray}
where $\bm{e}_r=\bm{r}/r$, $\bm{e}_p=\bm{p}/p$, $_{1}F_{1}\left(  a,b,c\right)$ is the confluent hypergeometric function and $P_\ell(x)$ are the Legendre polynomials. It holds that
\begin{eqnarray}
P_{\ell}(\bm{e}_r\cdot \bm{e}_p) &=& P_{\ell}(\cos{\theta})P_{\ell}(\cos{\theta_p}) \\
&& + ~2\sum_{m=1}^{\ell}\frac{(\ell - m)!}{(\ell + m)!}P^m_{\ell}(\cos{\theta})P^m_{\ell}(\cos{\theta_p})\cos{(m(\phi - \phi_p))} \, ,
\label{Legendre_expand}
\end{eqnarray}
where $P^m_\ell(x)$ are the associated Legendre polynomials. 
The matrix element $\langle n\ell m|\bm{r}|\bm{p}\rangle$ in the center-of-mass frame is then  
\begin{eqnarray}
&&\langle n\ell m|\bm{r}|\bm{p}\rangle = \sum\limits_{\ell'=\ell\pm 1, \ell'\ge 0}\int d^3r\,\bm{r}\,\Psi_{n\ell m}^*(\bm{r})\Psi_{\bm{p} \ell'}(\bm{r}) \nonumber\\
&&~= N_{n\ell m}(p)X^1_{n\ell}(p)G^1_{n\ell}(p) \nonumber\\
  &&\hspace{0.5cm}\times \Bigg\{(\ell+1)P_{\ell+1}(\cos{\theta_p})\left[\ell \left(\delta_{m,1}-\frac{\delta_{m,-1}}{\ell(\ell+1)}\right)\bm{e}_x - i\ell \left(\delta_{m,1}+\frac{\delta_{m,-1}}{\ell(\ell+1)}\right)\bm{e}_y + 2\delta_{m,0}\bm{e}_z\right]
  \nonumber\\
&&\hspace{1cm} + \sum_{m'=1}^{\ell+1}(\ell-m'+1)P^{m'}_{\ell+1}(\cos{\theta_p})\left[(\ell -m')(C^{x}_{\ell m m'}\bm{e}_x -iC^{y}_{\ell m m'}\bm{e}_y) +2C^{z}_{\ell m m'}\bm{e}_z \right] \nonumber\\
&&\hspace{1cm} - \sum_{m'=1}^{\ell+1}P^{m'}_{\ell+1}(\cos{\theta_p})\left[\tilde{C}^{x}_{\ell m m'}\bm{e}_x +i\tilde{C}^{y}_{\ell m m'}\bm{e}_y \right] \Bigg\} \nonumber\\
&&~~~+ N_{n\ell m}(p)X^2_{n\ell}(p)G^2_{n\ell}(p)\times \Bigg\{\ell P_{\ell-1}(\cos{\theta_p}) \nonumber\\
&&\hspace{1.3cm}\times \left[-(\ell +1) \left(\delta_{m,1}-\frac{\delta_{m,-1}}{\ell(\ell+1)}\right)\bm{e}_x + i(\ell +1) \left(\delta_{m,1}+\frac{\delta_{m,-1}}{\ell(\ell+1)}\right)\bm{e}_y + 2\delta_{m,0}\bm{e}_z\right] \nonumber\\
&&\hspace{1.cm} + \sum_{m'=1}^{\ell-1}(\ell+m')P^{m'}_{\ell-1}(\cos{\theta_p})\left[(\ell + m' +1)(-C^{x}_{\ell m m'}\bm{e}_x +iC^{y}_{\ell m m'}\bm{e}_y) +2C^{z}_{\ell m m'}\bm{e}_z \right] \nonumber\\
&&\hspace{1cm} + \sum_{m'=1}^{\ell-1}P^{m'}_{\ell-1}(\cos{\theta_p})\left[\tilde{C}^{x}_{\ell m m'}\bm{e}_x +i\tilde{C}^{y}_{\ell m m'}\bm{e}_y \right] \Bigg\} \, ,
\label{matrixelement}
\end{eqnarray}
where $\bm{e}_x$, $\bm{e}_y$ and $\bm{e}_z$ are the unit vectors in Cartesian coordinates.
Due to the selection rule for electric-dipole transitions, only the partial waves $\ell'=\ell \pm 1$ give a non-vanishing contribution.
In \eqref{matrixelement}, we have defined the following prefactors
\begin{eqnarray}
&&N_{n\ell m}(p) =  \frac{i^{\ell+3}2^{\ell}(-1)^{n-\ell+m}}{(2\ell+1)!}\sqrt{\left(\frac{2}{na_0}\right)^3\frac{(n+\ell)!}{2n(n-\ell-1)!}}\left(\frac{2}{na_0}\right)^\ell \sqrt{\frac{2\pi/(a_0p)}{1-e^{-2\pi/(a_0p)}}} \nonumber
\\
  &&\hspace{2cm}\times \sqrt{\frac{\pi}{2\ell+1}\frac{(\ell-m)!}{(\ell+m)!}}\frac{e^{-\frac{2}{a_0p}\left(1+i(\ell+1-n)a_0 p\right)\arctan{(na_0 p)}}}
     {p^\ell \left(1+(na_0p)^{-2}\right)^\ell} \, ,\\
  &&X^1_{n\ell}(p) = \frac{ie^{-2i\arctan{\left(na_0p\right)}}}{p^4\left(1+(na_0p)^{-2}\right)^2}
     \prod \limits_{\kappa=1}^{\ell+1}\sqrt{\kappa^{2} + (a_0p)^{-2}} \, , \\
  &&X^2_{n\ell}(p) = \frac{a_0 n \ell (2\ell+1)}{2p^3\left(1+(na_0p)^{-2}\right)}
     \prod \limits_{\kappa=1}^{\ell-1}\sqrt{\kappa^{2} + (a_0p)^{-2}} \, ,
\end{eqnarray}
whereas $G^1_{n\ell}(p)$ and $G^2_{n\ell}(p)$ are combinations of hypergeometric functions, $_2F_1\left(a,b,c,d\right)$, 
\begin{eqnarray}
&&G^1_{n\ell}(p) = ~_{2}F_{1}\left(\ell+2-\frac{i}{a_0p},\ell+1-n,2\ell+2,\frac{4ina_0p}{\left(1+ina_0p\right)^2}\right) \nonumber \\
&&\phantom{xxxx} -e^{4i\arctan{(na_0p})}~_{2}F_{1}\left(\ell-\frac{i}{a_0p},\ell+1-n,2\ell+2,\frac{4ina_0p}{\left(1+ina_0p\right)^2}\right) \, ,
\\
&&G^2_{n\ell}(p) = ~_{2}F_{1}\left(\ell-\frac{i}{a_0p},\ell+1-n,2\ell,\frac{4ina_0p}{\left(1+ina_0p\right)^2}\right) \nonumber
\\
&& \phantom{xxxx}-e^{4i\arctan{(na_0p})}~_{2}F_{1}\left(\ell-\frac{i}{a_0p},\ell-1-n,2\ell,\frac{4ina_0p}{\left(1+ina_0p\right)^2}\right) \, .
\end{eqnarray}
The constants $C^{x/y/z}_{\ell m m'}$, $\tilde{C}^{x/y}_{\ell m m'}$ in~\eqref{matrixelement} are defined as:
\begin{eqnarray}
&&C^{x}_{\ell m m'} = e^{-im'\phi_p}\delta_{m',m-1} + e^{im'\phi_p}\delta_{m',-(m+1)}(-1)^{m'+1}\frac{(\ell -m'-1)!}{(\ell+m'+1)!} \, ,\\
&&C^{y}_{\ell m m'} = e^{-im'\phi_p}\delta_{m',m-1} - e^{im'\phi_p}\delta_{m',-(m+1)}(-1)^{m'+1}\frac{(\ell -m'-1)!}{(\ell+m'+1)!} \, , \\
&&C^{z}_{\ell m m'} = e^{-im'\phi_p}\delta_{m',m} + e^{im'\phi_p}\delta_{m',-m}(-1)^{m'}\frac{(\ell -m')!}{(\ell+m')!} \, , \\
&&\tilde{C}^{x}_{\ell m m'} = e^{-im'\phi_p}\delta_{m',m+1} + e^{im'\phi_p}\delta_{m',-(m-1)}(-1)^{m'-1}\frac{(\ell -m'+1)!}{(\ell+m'-1)!} \, , \\
&&\tilde{C}^{x}_{\ell m m'} = e^{-im'\phi_p}\delta_{m',m+1} - e^{im'\phi_p}\delta_{m',-(m-1)}(-1)^{m'-1}\frac{(\ell -m'+1)!}{(\ell+m'-1)!} \, .
\end{eqnarray}
Equation~\eqref{matrixelement} reduces to the expression given in ref.~\cite{Biondini:2023zcz} for
polar angle $\theta_p=0$, which corresponds to putting the relative momentum along the $z$-direction.
Moreover, for the particular case of $n$S-states, for which $\ell=m=0$, the squared matrix elements are
\begin{equation}
\begin{aligned}
|\langle n\textrm{S}|\bm{r}|\bm{p}\rangle|^2 &= |\langle n\textrm{S}|\bm{r}|\bm{p}=p\bm{e}_z\rangle|^2 \, , \\
|\langle n\textrm{S}|z|\bm{p}\rangle|^2 &= |\langle n\textrm{S}|\bm{r}|\bm{p}\rangle|^2 \cos^2{(\theta_p)} \, .
\end{aligned}
\end{equation}
For excited states with non-vanishing orbital angular momentum, it holds that
\begin{equation}
\begin{aligned}
\sum_{m=-\ell}^{\ell} |\langle n\ell m|\bm{r}|\bm{p}\rangle|^2 &= \sum_{m=-\ell}^{\ell} |\langle n\ell m|\bm{r}|\bm{p}=p\bm{e}_z\rangle|^2 \, .
\end{aligned}
\end{equation}
We provide some analytic expressions for the squared matrix elements in the center-of-mass frame for the particular bound states 1S, 2S and 2P used in this work:
\begin{eqnarray}
  &&|\langle 1  \textrm{S}|z|\bm{p}\rangle|^{2} = \frac{2^9\pi^2a_0^4 }{p(1+(a_0p)^2)^5}\frac{e^{-\frac{4}{a_0p}\arctan(a_{0}p)}}{1-e^{-\frac{2\pi}{a_0p}}} \cos^2{(\theta_p)} , 
  \\
  &&|\langle 2  \textrm{S}|z|\bm{p}\rangle|^{2} = \frac{2^{18}\pi^2a_0^4\left(1+(a_0p)^2\right)}{p\left(1+(2a_0p)^2\right)^{6}}\frac{e^{-\frac{4}{a_0p}\arctan{\left(2a_0p\right)}}}{1-e^{-\frac{2\pi}{a_0p}}}\cos^2{(\theta_p)} , 
  \\
  && |\langle 2\textrm{P}_{m=0}|\bm{r}|\bm{p}\rangle|^{2}= \left[4(1+(a_0p)^{-2})(3\cos^2{(\theta_p)}+1) + 1+(2a_0p)^{-2} \right.  \nonumber
  \\
  &&~~~ \left.- 4\sqrt{1+(a_0p)^{-2}}\sqrt{1+(2a_0p)^{-2}}(3\cos^2{(\theta_p)}-1) \right]\frac{2^4\pi^2e^{-\frac{4}{a_0p}\arctan{\left(2a_0p\right)}}}{3^2 a_0^8 p^{13}(1+(2a_0p)^{-2})^7\left(1-e^{-\frac{2\pi}{a_0p}}\right)} , \nonumber
  \\ 
  \\
  &&|\langle 2\textrm{P}_{m=0}|z|\bm{p}\rangle|^{2}= \left(\sqrt{1+(2a_0p)^{-2}} - 2\sqrt{1+(a_0p)^{-2}}(3\cos^2{(\theta_p)}-1) \right)^2 \nonumber \\
  &&\hspace{6 cm}\times \frac{2^4\pi^2e^{-\frac{4}{a_0p}\arctan{\left(2a_0p\right)}}}{3^2 a_0^8 p^{13}(1+(2a_0p)^{-2})^7\left(1-e^{-\frac{2\pi}{a_0p}}\right)} ,
  \\
  &&|\langle 2\textrm{P}_{m=\pm 1}|\bm{r}|\bm{p}\rangle|^{2}= \left[\frac{1}{2}(1+(a_0p)^{-2})(20-12\cos^2{(\theta_p)}) + 1+(2a_0p)^{-2} \right. \nonumber \\
  &&~~~ \left.- 2\sqrt{1+(a_0p)^{-2}}\sqrt{1+(2a_0p)^{-2}}(1-3\cos^2{(\theta_p)}) \right]\frac{2^4\pi^2e^{-\frac{4}{a_0p}\arctan{\left(2a_0p\right)}}}{3^2 a_0^8 p^{13}(1+(2a_0p)^{-2})^7\left(1-e^{-\frac{2\pi}{a_0p}}\right)} , \nonumber
  \\
  \\
&&|\langle 2\textrm{P}_{m=\pm 1}|z|\bm{p}\rangle|^{2}= 18\cos^2{(\theta_p)}\sin^2{(\theta_p)}\left(1+(a_0p)^{-2}\right) \nonumber \\
  &&\hspace{6cm} \times \frac{2^4\pi^2e^{-\frac{4}{a_0p}\arctan{\left(2a_0p\right)}}}{3^2 a_0^8 p^{13}(1+(2a_0p)^{-2})^7\left(1-e^{-\frac{2\pi}{a_0p}}\right)} .
\end{eqnarray}

The dipole matrix element in a non-abelian SU($N$) model in the center-of-mass frame, $\langle n\ell m|\bm{r}|\bm{p}\rangle^{[\textbf{adj}]}$,
where  $|n\ell m\rangle$ is the bound state of a color-singlet dark matter pair and $|\bm{p}\rangle^{[\textbf{adj}]}$ is the color-adjoint state with arbitrary relative momentum vector $\bm{p}$,
has the same analytic structure as the abelian equivalent in~\eqref{matrixelement}; 
they only differ in the factors $N_{n\ell m}(p)$, $X^{1,2}_{n\ell}(p)$ and $G^{1,2}_{n\ell}(p)$, which now read
\begin{eqnarray}
  &&N^{\scalebox{0.65}{\textrm{SU}($N$)}}_{n\ell m}(p) =
  \frac{i^{\ell+3}2^{\ell}(-1)^{n-\ell+m}}{(2\ell+1)!}\sqrt{\left(\frac{2}{na_0}\right)^3\frac{(n+\ell)!}{2n(n-\ell-1)!}}\left(\frac{2}{na_0}\right)^\ell \sqrt{\frac{2\pi/((N^2-1)a_0p)}{e^{\frac{2\pi}{(N^2-1)a_0p}}-1}}  \nonumber
\\
  &&\hspace{0.2cm}\times \sqrt{\frac{\pi}{2\ell+1}\frac{(\ell-m)!}{(\ell+m)!}}\frac{e^{\frac{2}{(N^2-1)a_0p}\left[1-i(\ell+1-n)(N^2-1)a_0 p\right]\arctan{(na_0 p)}}}
     {p^\ell \left(1+(na_0p)^{-2}\right)^\ell}, \, \\
&&X^{1,\scalebox{0.65}{\textrm{SU}($N$)}}_{n\ell}(p) = \frac{ie^{-2i\arctan{\left(na_0p\right)}}}{p^4\left(1+(na_0p)^{-2}\right)^2}
     \prod \limits_{\kappa=1}^{\ell+1}\sqrt{\kappa^{2} + ((N^2-1)a_0p)^{-2}} \, , \\
&&X^{2,\scalebox{0.65}{\textrm{SU}($N$)}}_{n\ell}(p) = \frac{a_0 n \ell (2\ell+1)}{2p^3\left(1+(na_0p)^{-2}\right)}
     \prod \limits_{\kappa=1}^{\ell-1}\sqrt{\kappa^{2} + ((N^2-1)a_0p)^{-2}} \, ,
\end{eqnarray}
and
\begin{align}
&G^{1,\scalebox{0.65}{\textrm{SU}($N$)}}_{n\ell}(p) = \left(1+\frac{iN}{2C_Fa_0p}\right)\,_{2}F_{1}\left(\ell+2+\frac{i}{(N^2-1)a_0p},\ell+1-n,2\ell+2,\frac{4ina_0p}{\left(1+ina_0p\right)^2}\right) \nonumber\\
& -\frac{iN}{C_Fa_0p}e^{2i\arctan{(na_0p})}\,_{2}F_{1}\left(\ell+1+\frac{i}{(N^2-1)a_0p},\ell+1-n,2\ell+2,
  \frac{4ina_0p}{\left(1+ina_0p\right)^2}\right) \nonumber\\
& -\left(1-\frac{iN}{2C_Fa_0p}\right)e^{4i\arctan{(na_0p})}\,_{2}F_{1}\left(\ell+\frac{i}{(N^2-1)a_0p},
  \ell+1-n,2\ell+2,\frac{4ina_0p}{\left(1+ina_0p\right)^2}\right)  \, ,
\end{align}
\begin{align}
&G^{2,\scalebox{0.65}{\textrm{SU}($N$)}}_{n\ell}(p) = \left(1-\frac{N n}{2C_F}\right)\,_{2}F_{1}\left(\ell+\frac{i}{(N^2-1)a_0p},\ell+1-n,2\ell,
    \frac{4ina_0p}{\left(1+ina_0p\right)^2}\right)\nonumber\\
&+N\frac{n}{C_F}e^{2i\arctan{(na_0p})}\,_{2}F_{1}\left(\ell+\frac{i}{(N^2-1)a_0p},\ell-n,2\ell,
  \frac{4ina_0p}{\left(1+ina_0p\right)^2}\right) \nonumber\\
&-\left(1+\frac{N n}{2C_F}\right)e^{4i\arctan{(na_0p})}\,_{2}F_{1}\left(\ell+\frac{i}{(N^2-1)a_0p},\ell-1-n,2\ell,
  \frac{4ina_0p}{\left(1+ina_0p\right)^2}\right) \, ,
\end{align}
with $a_0=2/(C_FM\alpha)$.

\bibliographystyle{JHEP.bst}
\bibliography{DMcom2.bib}

\end{document}